\documentclass[longauth]{aa}

\usepackage{csquotes}
\usepackage[switch, modulo]{lineno}
\renewcommand{\linenumbers}[0]{}
\usepackage{graphicx}
\usepackage{natbib}
\usepackage{scalerel}
\usepackage{lastpage}
\usepackage{dcolumn} 
\usepackage{tablefootnote} 
\usepackage{multirow}
\usepackage{threeparttable}
\usepackage[table]{xcolor}
\usepackage{tikz}
\usepackage{natbib}
\usepackage{tikz}
\usepackage{euclid}

\usetikzlibrary{shapes.geometric, arrows, positioning}
\tikzstyle{startstop} = [rectangle, rounded corners, text centered, draw=black, align=center]
\tikzstyle{process} = [rectangle, text width=4cm, minimum height=1cm, text centered, draw=black, align=center]
\tikzstyle{arrow} = [thick,->,>=stealth]
\usepackage{xcolor}
\usepackage{txfonts}
\usepackage[pdfencoding=auto,psdextra]{hyperref}
\hypersetup{
    colorlinks=true,
    linkcolor=blue,
    filecolor=magenta,      
    urlcolor=blue,
    citecolor=blue
}
\makeatletter
\renewcommand*\aa@pageof{, page \thepage{} of \pageref*{LastPage}}
\makeatother

\usepackage[utf8]{inputenc}

\usepackage[switch, modulo]{lineno}

\usepackage[nameinlink,capitalise]{cleveref}
\crefname{section}{Sect.}{Sects.}
\Crefname{section}{Section}{Sections}
\crefname{figure}{Fig.}{Figs.}
\Crefname{figure}{Figure}{Figures}
\crefname{equation}{Eq.}{Eqs.}
\Crefname{equation}{Equation}{Equations}
\crefname{table}{Table}{Tables}
\crefname{appendix}{Appendix}{Appendices}

\usepackage{orcidlink} 
\newcommand{\orcid}[1]{\orcidlink{#1}}

\newcommand*{\gaia}{\textit{Gaia}\xspace}

\begin{document}

%
%
\title{Euclid Quick Data Release (Q1)}
\subtitle{\Euclid spectroscopy of quasars. 2. Physical properties from spectral fitting}    

\author{Euclid Collaboration: J.~Calhau\orcid{0000-0003-1803-6899}\thanks{\email{joao.feiocalhau@inaf.it}}\inst{\ref{aff1}}
\and G.~Calderone\orcid{0000-0002-7738-5389}\inst{\ref{aff2}}
\and A.~Feltre\orcid{0000-0001-6865-2871}\inst{\ref{aff3}}
\and M.~Scialpi\orcid{0009-0006-5100-4986}\inst{\ref{aff4},\ref{aff5},\ref{aff3}}
\and V.~Allevato\orcid{0000-0001-7232-5152}\inst{\ref{aff1}}
\and H.~Landt\orcid{0000-0001-8391-6900}\inst{\ref{aff6}}
\and F.~Ricci\orcid{0000-0001-5742-5980}\inst{\ref{aff7},\ref{aff8}}
\and Y.~Fu\orcid{0000-0002-0759-0504}\inst{\ref{aff9},\ref{aff10}}
\and L.~Spinoglio\orcid{0000-0001-8840-1551}\inst{\ref{aff11}}
\and F.~Shankar\orcid{0000-0001-8973-5051}\inst{\ref{aff12}}
\and L.~Nicastro\orcid{0000-0001-8534-6788}\inst{\ref{aff13}}
\and A.~Viitanen\orcid{0000-0001-9383-786X}\inst{\ref{aff14},\ref{aff15},\ref{aff8}}
\and G.~Zamorani\orcid{0000-0002-2318-301X}\inst{\ref{aff13}}
\and M.~Mezcua\orcid{0000-0003-4440-259X}\inst{\ref{aff16},\ref{aff17}}
\and F.~La~Franca\orcid{0000-0002-1239-2721}\inst{\ref{aff7},\ref{aff8}}
\and D.~Stern\orcid{0000-0003-2686-9241}\inst{\ref{aff18}}
\and E.~Lusso\orcid{0000-0003-0083-1157}\inst{\ref{aff4},\ref{aff3}}
\and J.~Wolf\orcid{0000-0003-0643-7935}\inst{\ref{aff19},\ref{aff20}}
\and A.~Paulino-Afonso\orcid{0000-0002-0943-0694}\inst{\ref{aff21},\ref{aff22}}
\and S.~Andreon\orcid{0000-0002-2041-8784}\inst{\ref{aff23}}
\and N.~Auricchio\orcid{0000-0003-4444-8651}\inst{\ref{aff13}}
\and C.~Baccigalupi\orcid{0000-0002-8211-1630}\inst{\ref{aff24},\ref{aff2},\ref{aff25},\ref{aff26}}
\and M.~Baldi\orcid{0000-0003-4145-1943}\inst{\ref{aff27},\ref{aff13},\ref{aff28}}
\and S.~Bardelli\orcid{0000-0002-8900-0298}\inst{\ref{aff13}}
\and P.~Battaglia\orcid{0000-0002-7337-5909}\inst{\ref{aff13}}
\and A.~Biviano\orcid{0000-0002-0857-0732}\inst{\ref{aff2},\ref{aff24}}
\and M.~Bolzonella\orcid{0000-0003-3278-4607}\inst{\ref{aff13}}
\and E.~Branchini\orcid{0000-0002-0808-6908}\inst{\ref{aff29},\ref{aff30},\ref{aff23}}
\and M.~Brescia\orcid{0000-0001-9506-5680}\inst{\ref{aff31},\ref{aff1}}
\and S.~Camera\orcid{0000-0003-3399-3574}\inst{\ref{aff32},\ref{aff33},\ref{aff34}}
\and V.~Capobianco\orcid{0000-0002-3309-7692}\inst{\ref{aff34}}
\and C.~Carbone\orcid{0000-0003-0125-3563}\inst{\ref{aff35}}
\and J.~Carretero\orcid{0000-0002-3130-0204}\inst{\ref{aff36},\ref{aff37}}
\and M.~Castellano\orcid{0000-0001-9875-8263}\inst{\ref{aff8}}
\and G.~Castignani\orcid{0000-0001-6831-0687}\inst{\ref{aff13}}
\and S.~Cavuoti\orcid{0000-0002-3787-4196}\inst{\ref{aff1},\ref{aff38}}
\and K.~C.~Chambers\orcid{0000-0001-6965-7789}\inst{\ref{aff39}}
\and A.~Cimatti\inst{\ref{aff40}}
\and C.~Colodro-Conde\inst{\ref{aff41}}
\and G.~Congedo\orcid{0000-0003-2508-0046}\inst{\ref{aff42}}
\and C.~J.~Conselice\orcid{0000-0003-1949-7638}\inst{\ref{aff43}}
\and L.~Conversi\orcid{0000-0002-6710-8476}\inst{\ref{aff44},\ref{aff45}}
\and Y.~Copin\orcid{0000-0002-5317-7518}\inst{\ref{aff46}}
\and F.~Courbin\orcid{0000-0003-0758-6510}\inst{\ref{aff47},\ref{aff48},\ref{aff49}}
\and H.~M.~Courtois\orcid{0000-0003-0509-1776}\inst{\ref{aff50}}
\and M.~Cropper\orcid{0000-0003-4571-9468}\inst{\ref{aff51}}
\and H.~Degaudenzi\orcid{0000-0002-5887-6799}\inst{\ref{aff14}}
\and G.~De~Lucia\orcid{0000-0002-6220-9104}\inst{\ref{aff2}}
\and H.~Dole\orcid{0000-0002-9767-3839}\inst{\ref{aff52}}
\and F.~Dubath\orcid{0000-0002-6533-2810}\inst{\ref{aff14}}
\and X.~Dupac\inst{\ref{aff45}}
\and S.~Dusini\orcid{0000-0002-1128-0664}\inst{\ref{aff53}}
\and A.~Ealet\orcid{0000-0003-3070-014X}\inst{\ref{aff46}}
\and S.~Escoffier\orcid{0000-0002-2847-7498}\inst{\ref{aff54}}
\and M.~Farina\orcid{0000-0002-3089-7846}\inst{\ref{aff11}}
\and S.~Ferriol\inst{\ref{aff46}}
\and F.~Finelli\orcid{0000-0002-6694-3269}\inst{\ref{aff13},\ref{aff55}}
\and S.~Fotopoulou\orcid{0000-0002-9686-254X}\inst{\ref{aff56}}
\and N.~Fourmanoit\orcid{0009-0005-6816-6925}\inst{\ref{aff54}}
\and M.~Frailis\orcid{0000-0002-7400-2135}\inst{\ref{aff2}}
\and M.~Fumana\orcid{0000-0001-6787-5950}\inst{\ref{aff35}}
\and L.~Gabarra\orcid{0000-0002-8486-8856}\inst{\ref{aff57}}
\and S.~Galeotta\orcid{0000-0002-3748-5115}\inst{\ref{aff2}}
\and K.~George\orcid{0000-0002-1734-8455}\inst{\ref{aff58}}
\and B.~Gillis\orcid{0000-0002-4478-1270}\inst{\ref{aff42}}
\and C.~Giocoli\orcid{0000-0002-9590-7961}\inst{\ref{aff13},\ref{aff28}}
\and J.~Gracia-Carpio\orcid{0000-0003-4689-3134}\inst{\ref{aff59}}
\and A.~Grazian\orcid{0000-0002-5688-0663}\inst{\ref{aff60}}
\and F.~Grupp\inst{\ref{aff59},\ref{aff61}}
\and W.~G.~Hartley\inst{\ref{aff14}}
\and S.~V.~H.~Haugan\orcid{0000-0001-9648-7260}\inst{\ref{aff62}}
\and S.~Hemmati\orcid{0000-0003-2226-5395}\inst{\ref{aff63}}
\and W.~Holmes\orcid{0009-0007-8554-4646}\inst{\ref{aff18}}
\and I.~M.~Hook\orcid{0000-0002-2960-978X}\inst{\ref{aff64}}
\and F.~Hormuth\inst{\ref{aff65}}
\and A.~Hornstrup\orcid{0000-0002-3363-0936}\inst{\ref{aff66},\ref{aff67}}
\and M.~Huertas-Company\orcid{0000-0002-1416-8483}\inst{\ref{aff41},\ref{aff68},\ref{aff69}}
\and K.~Jahnke\orcid{0000-0003-3804-2137}\inst{\ref{aff19}}
\and M.~Jhabvala\inst{\ref{aff70}}
\and B.~Joachimi\orcid{0000-0001-7494-1303}\inst{\ref{aff71}}
\and S.~Kermiche\orcid{0000-0002-0302-5735}\inst{\ref{aff54}}
\and A.~Kiessling\orcid{0000-0002-2590-1273}\inst{\ref{aff18}}
\and B.~Kubik\orcid{0009-0006-5823-4880}\inst{\ref{aff46}}
\and M.~K\"ummel\orcid{0000-0003-2791-2117}\inst{\ref{aff61}}
\and M.~Kunz\orcid{0000-0002-3052-7394}\inst{\ref{aff72}}
\and H.~Kurki-Suonio\orcid{0000-0002-4618-3063}\inst{\ref{aff73},\ref{aff74}}
\and A.~M.~C.~Le~Brun\orcid{0000-0002-0936-4594}\inst{\ref{aff75}}
\and V.~Le~Brun\orcid{0000-0002-5027-1939}\inst{\ref{aff76}}
\and S.~Ligori\orcid{0000-0003-4172-4606}\inst{\ref{aff34}}
\and P.~B.~Lilje\orcid{0000-0003-4324-7794}\inst{\ref{aff62}}
\and V.~Lindholm\orcid{0000-0003-2317-5471}\inst{\ref{aff73},\ref{aff74}}
\and I.~Lloro\orcid{0000-0001-5966-1434}\inst{\ref{aff77}}
\and M.~Magliocchetti\orcid{0000-0001-9158-4838}\inst{\ref{aff11}}
\and G.~Mainetti\orcid{0000-0003-2384-2377}\inst{\ref{aff78}}
\and O.~Mansutti\orcid{0000-0001-5758-4658}\inst{\ref{aff2}}
\and O.~Marggraf\orcid{0000-0001-7242-3852}\inst{\ref{aff79}}
\and M.~Martinelli\orcid{0000-0002-6943-7732}\inst{\ref{aff8},\ref{aff80}}
\and N.~Martinet\orcid{0000-0003-2786-7790}\inst{\ref{aff76}}
\and F.~Marulli\orcid{0000-0002-8850-0303}\inst{\ref{aff81},\ref{aff13},\ref{aff28}}
\and R.~J.~Massey\orcid{0000-0002-6085-3780}\inst{\ref{aff82}}
\and N.~Mauri\orcid{0000-0001-8196-1548}\inst{\ref{aff40},\ref{aff28}}
\and E.~Medinaceli\orcid{0000-0002-4040-7783}\inst{\ref{aff13}}
\and S.~Mei\orcid{0000-0002-2849-559X}\inst{\ref{aff83},\ref{aff84}}
\and M.~Meneghetti\orcid{0000-0003-1225-7084}\inst{\ref{aff13},\ref{aff28}}
\and E.~Merlin\orcid{0000-0001-6870-8900}\inst{\ref{aff60}}
\and G.~Meylan\orcid{0000-0001-6503-0209}\inst{\ref{aff85}}
\and P.~Monaco\orcid{0000-0003-2083-7564}\inst{\ref{aff86},\ref{aff2},\ref{aff25},\ref{aff24}}
\and A.~Mora\orcid{0000-0002-1922-8529}\inst{\ref{aff87}}
\and M.~Moresco\orcid{0000-0002-7616-7136}\inst{\ref{aff81},\ref{aff13}}
\and C.~Moretti\orcid{0000-0003-3314-8936}\inst{\ref{aff2},\ref{aff24},\ref{aff25}}
\and L.~Moscardini\orcid{0000-0002-3473-6716}\inst{\ref{aff81},\ref{aff13},\ref{aff28}}
\and C.~Neissner\orcid{0000-0001-8524-4968}\inst{\ref{aff88},\ref{aff37}}
\and S.-M.~Niemi\orcid{0009-0005-0247-0086}\inst{\ref{aff89}}
\and J.~W.~Nightingale\orcid{0000-0002-8987-7401}\inst{\ref{aff90}}
\and C.~Padilla\orcid{0000-0001-7951-0166}\inst{\ref{aff88}}
\and S.~Paltani\orcid{0000-0002-8108-9179}\inst{\ref{aff14}}
\and F.~Pasian\orcid{0000-0002-4869-3227}\inst{\ref{aff2}}
\and K.~Pedersen\inst{\ref{aff91}}
\and W.~J.~Percival\orcid{0000-0002-0644-5727}\inst{\ref{aff92},\ref{aff93},\ref{aff94}}
\and V.~Pettorino\orcid{0000-0002-4203-9320}\inst{\ref{aff89}}
\and A.~Pezzotta\orcid{0000-0003-0726-2268}\inst{\ref{aff23}}
\and S.~Pires\orcid{0000-0002-0249-2104}\inst{\ref{aff95}}
\and G.~Polenta\orcid{0000-0003-4067-9196}\inst{\ref{aff96}}
\and M.~Poncet\inst{\ref{aff97}}
\and L.~A.~Popa\inst{\ref{aff98}}
\and L.~Pozzetti\orcid{0000-0001-7085-0412}\inst{\ref{aff13}}
\and F.~Raison\orcid{0000-0002-7819-6918}\inst{\ref{aff59}}
\and R.~Rebolo\orcid{0000-0003-3767-7085}\inst{\ref{aff41},\ref{aff99},\ref{aff100}}
\and A.~Renzi\orcid{0000-0001-9856-1970}\inst{\ref{aff101},\ref{aff53},\ref{aff13}}
\and J.~Rhodes\orcid{0000-0002-4485-8549}\inst{\ref{aff18}}
\and G.~Riccio\inst{\ref{aff1}}
\and I.~Risso\orcid{0000-0003-2525-7761}\inst{\ref{aff29},\ref{aff30},\ref{aff23}}
\and E.~Romelli\orcid{0000-0003-3069-9222}\inst{\ref{aff2}}
\and M.~Roncarelli\orcid{0000-0001-9587-7822}\inst{\ref{aff13}}
\and B.~Rusholme\orcid{0000-0001-7648-4142}\inst{\ref{aff63}}
\and R.~Saglia\orcid{0000-0003-0378-7032}\inst{\ref{aff61},\ref{aff59}}
\and Z.~Sakr\orcid{0000-0002-4823-3757}\inst{\ref{aff102},\ref{aff103},\ref{aff104}}
\and D.~Sapone\orcid{0000-0001-7089-4503}\inst{\ref{aff105}}
\and B.~Sartoris\orcid{0000-0003-1337-5269}\inst{\ref{aff61},\ref{aff2}}
\and P.~Schneider\orcid{0000-0001-8561-2679}\inst{\ref{aff79}}
\and T.~Schrabback\orcid{0000-0002-6987-7834}\inst{\ref{aff106}}
\and M.~Scodeggio\inst{\ref{aff35}}
\and A.~Secroun\orcid{0000-0003-0505-3710}\inst{\ref{aff54}}
\and E.~Sihvola\orcid{0000-0003-1804-7715}\inst{\ref{aff15}}
\and P.~Simon\inst{\ref{aff79}}
\and C.~Sirignano\orcid{0000-0002-0995-7146}\inst{\ref{aff101},\ref{aff53}}
\and G.~Sirri\orcid{0000-0003-2626-2853}\inst{\ref{aff28}}
\and L.~Stanco\orcid{0000-0002-9706-5104}\inst{\ref{aff53}}
\and C.~Surace\orcid{0000-0003-2592-0113}\inst{\ref{aff76}}
\and P.~Tallada-Cresp\'{i}\orcid{0000-0002-1336-8328}\inst{\ref{aff36},\ref{aff37}}
\and A.~N.~Taylor\inst{\ref{aff42}}
\and H.~I.~Teplitz\orcid{0000-0002-7064-5424}\inst{\ref{aff107}}
\and I.~Tereno\orcid{0000-0002-4537-6218}\inst{\ref{aff108},\ref{aff109}}
\and S.~Toft\orcid{0000-0003-3631-7176}\inst{\ref{aff110},\ref{aff111}}
\and R.~Toledo-Moreo\orcid{0000-0002-2997-4859}\inst{\ref{aff112},\ref{aff113}}
\and F.~Torradeflot\orcid{0000-0003-1160-1517}\inst{\ref{aff37},\ref{aff36}}
\and I.~Tutusaus\orcid{0000-0002-3199-0399}\inst{\ref{aff16},\ref{aff17},\ref{aff103}}
\and J.~Valiviita\orcid{0000-0001-6225-3693}\inst{\ref{aff73},\ref{aff74}}
\and T.~Vassallo\orcid{0000-0001-6512-6358}\inst{\ref{aff2},\ref{aff58}}
\and D.~Vibert\orcid{0009-0008-0607-631X}\inst{\ref{aff76}}
\and Y.~Wang\orcid{0000-0002-4749-2984}\inst{\ref{aff63}}
\and J.~Weller\orcid{0000-0002-8282-2010}\inst{\ref{aff61},\ref{aff59}}
\and A.~Zacchei\orcid{0000-0003-0396-1192}\inst{\ref{aff2},\ref{aff24}}
\and E.~Zucca\orcid{0000-0002-5845-8132}\inst{\ref{aff13}}
\and M.~Ballardini\orcid{0000-0003-4481-3559}\inst{\ref{aff114},\ref{aff115},\ref{aff13}}
\and P.~Bergamini\orcid{0000-0003-1383-9414}\inst{\ref{aff13}}
\and S.~Borgani\orcid{0000-0001-6151-6439}\inst{\ref{aff86},\ref{aff24},\ref{aff2},\ref{aff25},\ref{aff116}}
\and E.~Bozzo\orcid{0000-0002-8201-1525}\inst{\ref{aff14}}
\and C.~Burigana\orcid{0000-0002-3005-5796}\inst{\ref{aff117},\ref{aff55}}
\and R.~Cabanac\orcid{0000-0001-6679-2600}\inst{\ref{aff103}}
\and A.~Cappi\inst{\ref{aff118},\ref{aff13}}
\and T.~Castro\orcid{0000-0002-6292-3228}\inst{\ref{aff2},\ref{aff25},\ref{aff24},\ref{aff116}}
\and J.~A.~Escartin~Vigo\inst{\ref{aff59}}
\and J.~Garc\'ia-Bellido\orcid{0000-0002-9370-8360}\inst{\ref{aff102}}
\and T.~Gasparetto\orcid{0000-0002-7913-4866}\inst{\ref{aff8}}
\and A.~Loureiro\orcid{0000-0002-4371-0876}\inst{\ref{aff119},\ref{aff120}}
\and J.~Macias-Perez\orcid{0000-0002-5385-2763}\inst{\ref{aff121}}
\and R.~Maoli\orcid{0000-0002-6065-3025}\inst{\ref{aff122},\ref{aff8}}
\and R.~B.~Metcalf\orcid{0000-0003-3167-2574}\inst{\ref{aff81},\ref{aff13}}
\and M.~P\"ontinen\orcid{0000-0001-5442-2530}\inst{\ref{aff73}}
\and E.~Sarpa\orcid{0000-0002-1256-655X}\inst{\ref{aff2}}
\and V.~Scottez\orcid{0009-0008-3864-940X}\inst{\ref{aff123},\ref{aff124}}
\and M.~Sereno\orcid{0000-0003-0302-0325}\inst{\ref{aff13},\ref{aff28}}
\and M.~Tenti\orcid{0000-0002-4254-5901}\inst{\ref{aff28}}
\and M.~Tucci\inst{\ref{aff14}}
\and M.~Viel\orcid{0000-0002-2642-5707}\inst{\ref{aff24},\ref{aff2},\ref{aff26},\ref{aff25},\ref{aff116}}
\and M.~Wiesmann\orcid{0009-0000-8199-5860}\inst{\ref{aff62}}
\and J.~A.~Acevedo~Barroso\orcid{0000-0002-9654-1711}\inst{\ref{aff18}}
\and Y.~Akrami\orcid{0000-0002-2407-7956}\inst{\ref{aff102},\ref{aff125}}
\and I.~T.~Andika\orcid{0000-0001-6102-9526}\inst{\ref{aff61}}
\and G.~Angora\orcid{0000-0002-0316-6562}\inst{\ref{aff1},\ref{aff114}}
\and S.~Anselmi\orcid{0000-0002-3579-9583}\inst{\ref{aff53},\ref{aff101},\ref{aff126}}
\and M.~Archidiacono\orcid{0000-0003-4952-9012}\inst{\ref{aff127},\ref{aff128}}
\and G.~Arico\orcid{0000-0002-2802-2928}\inst{\ref{aff28}}
\and F.~Atrio-Barandela\orcid{0000-0002-2130-2513}\inst{\ref{aff129}}
\and E.~Aubourg\orcid{0000-0002-5592-023X}\inst{\ref{aff83},\ref{aff130}}
\and M.~Baes\orcid{0000-0002-3930-2757}\inst{\ref{aff131}}
\and L.~Baumont\orcid{0000-0002-1518-0150}\inst{\ref{aff86},\ref{aff2},\ref{aff24}}
\and L.~Bazzanini\orcid{0000-0003-0727-0137}\inst{\ref{aff114},\ref{aff13}}
\and D.~Bertacca\orcid{0000-0002-2490-7139}\inst{\ref{aff101},\ref{aff60},\ref{aff53}}
\and M.~Bethermin\orcid{0000-0002-3915-2015}\inst{\ref{aff132}}
\and F.~Beutler\orcid{0000-0003-0467-5438}\inst{\ref{aff42}}
\and L.~Bisigello\orcid{0000-0003-0492-4924}\inst{\ref{aff60}}
\and A.~Blanchard\orcid{0000-0001-8555-9003}\inst{\ref{aff103}}
\and L.~Blot\orcid{0000-0002-9622-7167}\inst{\ref{aff133},\ref{aff75}}
\and M.~L.~Brown\orcid{0000-0002-0370-8077}\inst{\ref{aff43}}
\and S.~Bruton\orcid{0000-0002-6503-5218}\inst{\ref{aff134}}
\and A.~Calabro\orcid{0000-0003-2536-1614}\inst{\ref{aff8}}
\and B.~Camacho~Quevedo\orcid{0000-0002-8789-4232}\inst{\ref{aff24},\ref{aff26},\ref{aff2}}
\and F.~Caro\orcid{0009-0003-1053-0507}\inst{\ref{aff8}}
\and C.~S.~Carvalho\inst{\ref{aff109}}
\and F.~Cogato\orcid{0000-0003-4632-6113}\inst{\ref{aff81},\ref{aff13}}
\and T.~E.~Collett\orcid{0000-0001-5564-3140}\inst{\ref{aff135}}
\and S.~Conseil\orcid{0000-0002-3657-4191}\inst{\ref{aff46}}
\and A.~R.~Cooray\orcid{0000-0002-3892-0190}\inst{\ref{aff136}}
\and P.~Corcho-Caballero\orcid{0000-0001-6327-7080}\inst{\ref{aff10}}
\and B.~Csizi\orcid{0000-0003-3227-6581}\inst{\ref{aff106}}
\and O.~Cucciati\orcid{0000-0002-9336-7551}\inst{\ref{aff13}}
\and H.~Dannerbauer\orcid{0000-0001-7147-3575}\inst{\ref{aff41},\ref{aff100}}
\and S.~Davini\orcid{0000-0003-3269-1718}\inst{\ref{aff30}}
\and T.~de~Boer\orcid{0000-0001-5486-2747}\inst{\ref{aff39}}
\and F.~De~Paolis\orcid{0000-0001-6460-7563}\inst{\ref{aff137},\ref{aff138},\ref{aff139}}
\and G.~Desprez\orcid{0000-0001-8325-1742}\inst{\ref{aff10}}
\and A.~D\'iaz-S\'anchez\orcid{0000-0003-0748-4768}\inst{\ref{aff140}}
\and S.~Di~Domizio\orcid{0000-0003-2863-5895}\inst{\ref{aff29},\ref{aff30}}
\and J.~M.~Diego\orcid{0000-0001-9065-3926}\inst{\ref{aff141}}
\and V.~Duret\orcid{0009-0009-0383-4960}\inst{\ref{aff54}}
\and A.~Enia\orcid{0000-0002-0200-2857}\inst{\ref{aff13}}
\and Y.~Fang\orcid{0000-0002-0334-6950}\inst{\ref{aff61}}
\and A.~Farina\orcid{0009-0000-3420-929X}\inst{\ref{aff23},\ref{aff30}}
\and A.~G.~Ferrari\orcid{0009-0005-5266-4110}\inst{\ref{aff28}}
\and A.~Finoguenov\orcid{0000-0002-4606-5403}\inst{\ref{aff73}}
\and F.~Fontanot\orcid{0000-0003-4744-0188}\inst{\ref{aff2},\ref{aff24}}
\and A.~Franco\orcid{0000-0002-4761-366X}\inst{\ref{aff137},\ref{aff138},\ref{aff139}}
\and K.~Ganga\orcid{0000-0001-8159-8208}\inst{\ref{aff83}}
\and R.~Gavazzi\orcid{0000-0002-5540-6935}\inst{\ref{aff76},\ref{aff142}}
\and E.~Gaztanaga\orcid{0000-0001-9632-0815}\inst{\ref{aff16},\ref{aff17},\ref{aff135}}
\and Z.~Ghaffari\orcid{0000-0002-6467-8078}\inst{\ref{aff2},\ref{aff24}}
\and F.~Giacomini\orcid{0000-0002-3129-2814}\inst{\ref{aff28}}
\and F.~Gianotti\orcid{0000-0003-4666-119X}\inst{\ref{aff13}}
\and G.~Gozaliasl\orcid{0000-0002-0236-919X}\inst{\ref{aff143},\ref{aff73}}
\and A.~Gruppuso\orcid{0000-0001-9272-5292}\inst{\ref{aff13},\ref{aff28}}
\and M.~Guidi\orcid{0000-0001-9408-1101}\inst{\ref{aff27},\ref{aff13}}
\and C.~M.~Gutierrez\orcid{0000-0001-7854-783X}\inst{\ref{aff41},\ref{aff100}}
\and A.~Hall\orcid{0000-0002-3139-8651}\inst{\ref{aff42}}
\and N.~A.~Hatch\orcid{0000-0001-5600-0534}\inst{\ref{aff144}}
\and H.~Hildebrandt\orcid{0000-0002-9814-3338}\inst{\ref{aff145}}
\and J.~Hjorth\orcid{0000-0002-4571-2306}\inst{\ref{aff91}}
\and J.~J.~E.~Kajava\orcid{0000-0002-3010-8333}\inst{\ref{aff146},\ref{aff147},\ref{aff148}}
\and Y.~Kang\orcid{0009-0000-8588-7250}\inst{\ref{aff14}}
\and V.~Kansal\orcid{0000-0002-4008-6078}\inst{\ref{aff149},\ref{aff150}}
\and D.~Karagiannis\orcid{0000-0002-4927-0816}\inst{\ref{aff114},\ref{aff151}}
\and J.~Kim\orcid{0000-0003-2776-2761}\inst{\ref{aff57}}
\and C.~C.~Kirkpatrick\inst{\ref{aff15}}
\and A.~Kov\'acs\orcid{0000-0002-5825-579X}\inst{\ref{aff152},\ref{aff153}}
\and I.~Kova{\v{c}}i{\'{c}}\orcid{0000-0001-6751-3263}\inst{\ref{aff131}}
\and K.~Koyama\orcid{0000-0001-6727-6915}\inst{\ref{aff135}}
\and S.~Kruk\orcid{0000-0001-8010-8879}\inst{\ref{aff45}}
\and M.~C.~Lam\orcid{0000-0002-9347-2298}\inst{\ref{aff42}}
\and G.~Leroy\orcid{0009-0004-2523-4425}\inst{\ref{aff6},\ref{aff82}}
\and J.~Lesgourgues\orcid{0000-0001-7627-353X}\inst{\ref{aff154}}
\and L.~Linke\orcid{0000-0002-2622-8113}\inst{\ref{aff106}}
\and S.~J.~Liu\orcid{0000-0001-7680-2139}\inst{\ref{aff11}}
\and X.~Lopez~Lopez\orcid{0009-0008-5194-5908}\inst{\ref{aff13}}
\and G.~Maggio\orcid{0000-0003-4020-4836}\inst{\ref{aff2}}
\and F.~Mannucci\orcid{0000-0002-4803-2381}\inst{\ref{aff3}}
\and F.~R.~Marleau\orcid{0000-0002-1442-2947}\inst{\ref{aff106}}
\and C.~J.~A.~P.~Martins\orcid{0000-0002-4886-9261}\inst{\ref{aff21},\ref{aff22}}
\and M.~Migliaccio\inst{\ref{aff155},\ref{aff156}}
\and M.~Miluzio\inst{\ref{aff45},\ref{aff157}}
\and G.~Morgante\inst{\ref{aff13}}
\and S.~Nadathur\orcid{0000-0001-9070-3102}\inst{\ref{aff135}}
\and K.~Naidoo\orcid{0000-0002-9182-1802}\inst{\ref{aff135},\ref{aff19}}
\and A.~Navarro-Alsina\orcid{0000-0002-3173-2592}\inst{\ref{aff79}}
\and A.~Nersesian\orcid{0000-0001-6843-409X}\inst{\ref{aff158}}
\and S.~Nesseris\orcid{0000-0002-0567-0324}\inst{\ref{aff102}}
\and F.~Oppizzi\orcid{0000-0003-3904-8370}\inst{\ref{aff30}}
\and F.~Pace\orcid{0000-0001-8039-0480}\inst{\ref{aff32},\ref{aff33},\ref{aff34}}
\and D.~Paoletti\orcid{0000-0003-4761-6147}\inst{\ref{aff13},\ref{aff55}}
\and F.~Passalacqua\orcid{0000-0002-8606-4093}\inst{\ref{aff101},\ref{aff53}}
\and K.~Paterson\orcid{0000-0001-8340-3486}\inst{\ref{aff19}}
\and L.~Patrizii\inst{\ref{aff28}}
\and A.~Pisani\orcid{0000-0002-6146-4437}\inst{\ref{aff54}}
\and D.~Potter\orcid{0000-0002-0757-5195}\inst{\ref{aff159}}
\and G.~W.~Pratt\inst{\ref{aff95}}
\and S.~Quai\orcid{0000-0002-0449-8163}\inst{\ref{aff81},\ref{aff13}}
\and M.~Radovich\orcid{0000-0002-3585-866X}\inst{\ref{aff60}}
\and G.~Rodighiero\orcid{0000-0002-9415-2296}\inst{\ref{aff101},\ref{aff60}}
\and K.~Rojas\orcid{0000-0003-1391-6854}\inst{\ref{aff160}}
\and W.~Roster\orcid{0000-0002-9149-6528}\inst{\ref{aff59}}
\and S.~Sacquegna\orcid{0000-0002-8433-6630}\inst{\ref{aff161}}
\and M.~Sahl\'en\orcid{0000-0003-0973-4804}\inst{\ref{aff162}}
\and D.~B.~Sanders\orcid{0000-0002-1233-9998}\inst{\ref{aff39}}
\and A.~Schneider\orcid{0000-0001-7055-8104}\inst{\ref{aff159}}
\and D.~Sciotti\orcid{0009-0008-4519-2620}\inst{\ref{aff8},\ref{aff80}}
\and E.~Sellentin\inst{\ref{aff163},\ref{aff9}}
\and S.~Serjeant\orcid{0000-0002-0517-7943}\inst{\ref{aff164}}
\and J.~G.~Sorce\orcid{0000-0002-2307-2432}\inst{\ref{aff165},\ref{aff52}}
\and M.~Talia\orcid{0000-0003-4352-2063}\inst{\ref{aff81},\ref{aff13}}
\and K.~Tanidis\orcid{0000-0001-9843-5130}\inst{\ref{aff166}}
\and C.~Tao\orcid{0000-0001-7961-8177}\inst{\ref{aff54}}
\and F.~Tarsitano\orcid{0000-0002-5919-0238}\inst{\ref{aff167},\ref{aff168},\ref{aff14}}
\and G.~Testera\orcid{0000-0003-2970-766X}\inst{\ref{aff30}}
\and R.~Teyssier\orcid{0000-0001-7689-0933}\inst{\ref{aff169}}
\and S.~Tosi\orcid{0000-0002-7275-9193}\inst{\ref{aff29},\ref{aff23},\ref{aff30}}
\and A.~Troja\orcid{0000-0003-0239-4595}\inst{\ref{aff2}}
\and C.~Uhlemann\orcid{0000-0001-7831-1579}\inst{\ref{aff170},\ref{aff90}}
\and C.~Valieri\inst{\ref{aff28}}
\and A.~Venhola\orcid{0000-0001-6071-4564}\inst{\ref{aff171}}
\and G.~Verza\orcid{0000-0002-1886-8348}\thanks{Deceased}\inst{\ref{aff172},\ref{aff173}}
\and S.~Vinciguerra\orcid{0009-0005-4018-3184}\inst{\ref{aff76}}
\and M.~von~Wietersheim-Kramsta\orcid{0000-0003-4986-5091}\inst{\ref{aff82},\ref{aff6}}
\and L.~Wang\orcid{0000-0002-6736-9158}\inst{\ref{aff174},\ref{aff10}}
\and A.~H.~Wright\orcid{0000-0001-7363-7932}\inst{\ref{aff145}}}
										   
\institute{INAF-Osservatorio Astronomico di Capodimonte, Via Moiariello 16, 80131 Napoli, Italy\label{aff1}
\and
INAF-Osservatorio Astronomico di Trieste, Via G. B. Tiepolo 11, 34143 Trieste, Italy\label{aff2}
\and
INAF-Osservatorio Astrofisico di Arcetri, Largo E. Fermi 5, 50125, Firenze, Italy\label{aff3}
\and
Dipartimento di Fisica e Astronomia, Universit\`{a} di Firenze, via G. Sansone 1, 50019 Sesto Fiorentino, Firenze, Italy\label{aff4}
\and
University of Trento, Via Sommarive 14, I-38123 Trento, Italy\label{aff5}
\and
Department of Physics, Centre for Extragalactic Astronomy, Durham University, South Road, Durham, DH1 3LE, UK\label{aff6}
\and
Department of Mathematics and Physics, Roma Tre University, Via della Vasca Navale 84, 00146 Rome, Italy\label{aff7}
\and
INAF-Osservatorio Astronomico di Roma, Via Frascati 33, 00078 Monteporzio Catone, Italy\label{aff8}
\and
Leiden Observatory, Leiden University, Einsteinweg 55, 2333 CC Leiden, The Netherlands\label{aff9}
\and
Kapteyn Astronomical Institute, University of Groningen, PO Box 800, 9700 AV Groningen, The Netherlands\label{aff10}
\and
INAF-Istituto di Astrofisica e Planetologia Spaziali, via del Fosso del Cavaliere, 100, 00100 Roma, Italy\label{aff11}
\and
School of Physics \& Astronomy, University of Southampton, Highfield Campus, Southampton SO17 1BJ, UK\label{aff12}
\and
INAF-Osservatorio di Astrofisica e Scienza dello Spazio di Bologna, Via Piero Gobetti 93/3, 40129 Bologna, Italy\label{aff13}
\and
Department of Astronomy, University of Geneva, ch. d'Ecogia 16, 1290 Versoix, Switzerland\label{aff14}
\and
Department of Physics and Helsinki Institute of Physics, Gustaf H\"allstr\"omin katu 2, University of Helsinki, 00014 Helsinki, Finland\label{aff15}
\and
Institute of Space Sciences (ICE, CSIC), Campus UAB, Carrer de Can Magrans, s/n, 08193 Barcelona, Spain\label{aff16}
\and
Institut d'Estudis Espacials de Catalunya (IEEC),  Edifici RDIT, Campus UPC, 08860 Castelldefels, Barcelona, Spain\label{aff17}
\and
Jet Propulsion Laboratory, California Institute of Technology, 4800 Oak Grove Drive, Pasadena, CA, 91109, USA\label{aff18}
\and
Max-Planck-Institut f\"ur Astronomie, K\"onigstuhl 17, 69117 Heidelberg, Germany\label{aff19}
\and
European Southern Observatory, Karl-Schwarzschild-Str.~2, 85748 Garching, Germany\label{aff20}
\and
Centro de Astrof\'{\i}sica da Universidade do Porto, Rua das Estrelas, 4150-762 Porto, Portugal\label{aff21}
\and
Instituto de Astrof\'isica e Ci\^encias do Espa\c{c}o, Universidade do Porto, CAUP, Rua das Estrelas, PT4150-762 Porto, Portugal\label{aff22}
\and
INAF-Osservatorio Astronomico di Brera, Via Brera 28, 20122 Milano, Italy\label{aff23}
\and
IFPU, Institute for Fundamental Physics of the Universe, via Beirut 2, 34151 Trieste, Italy\label{aff24}
\and
INFN, Sezione di Trieste, Via Valerio 2, 34127 Trieste TS, Italy\label{aff25}
\and
SISSA, International School for Advanced Studies, Via Bonomea 265, 34136 Trieste TS, Italy\label{aff26}
\and
Dipartimento di Fisica e Astronomia, Universit\`a di Bologna, Via Gobetti 93/2, 40129 Bologna, Italy\label{aff27}
\and
INFN-Sezione di Bologna, Viale Berti Pichat 6/2, 40127 Bologna, Italy\label{aff28}
\and
Dipartimento di Fisica, Universit\`a di Genova, Via Dodecaneso 33, 16146, Genova, Italy\label{aff29}
\and
INFN-Sezione di Genova, Via Dodecaneso 33, 16146, Genova, Italy\label{aff30}
\and
Department of Physics "E. Pancini", University Federico II, Via Cinthia 6, 80126, Napoli, Italy\label{aff31}
\and
Dipartimento di Fisica, Universit\`a degli Studi di Torino, Via P. Giuria 1, 10125 Torino, Italy\label{aff32}
\and
INFN-Sezione di Torino, Via P. Giuria 1, 10125 Torino, Italy\label{aff33}
\and
INAF-Osservatorio Astrofisico di Torino, Via Osservatorio 20, 10025 Pino Torinese (TO), Italy\label{aff34}
\and
INAF-IASF Milano, Via Alfonso Corti 12, 20133 Milano, Italy\label{aff35}
\and
Centro de Investigaciones Energ\'eticas, Medioambientales y Tecnol\'ogicas (CIEMAT), Avenida Complutense 40, 28040 Madrid, Spain\label{aff36}
\and
Port d'Informaci\'{o} Cient\'{i}fica, Campus UAB, C. Albareda s/n, 08193 Bellaterra (Barcelona), Spain\label{aff37}
\and
INFN -- Sezione di Napoli, Via Cinthia 6, 80126, Napoli, Italy\label{aff38}
\and
Institute for Astronomy, University of Hawaii, 2680 Woodlawn Drive, Honolulu, HI 96822, USA\label{aff39}
\and
Dipartimento di Fisica e Astronomia "Augusto Righi" - Alma Mater Studiorum Universit\`a di Bologna, Viale Berti Pichat 6/2, 40127 Bologna, Italy\label{aff40}
\and
Instituto de Astrof\'{\i}sica de Canarias, E-38205 La Laguna, Tenerife, Spain\label{aff41}
\and
Institute for Astronomy, University of Edinburgh, Royal Observatory, Blackford Hill, Edinburgh EH9 3HJ, UK\label{aff42}
\and
Jodrell Bank Centre for Astrophysics, Department of Physics and Astronomy, University of Manchester, Oxford Road, Manchester M13 9PL, UK\label{aff43}
\and
European Space Agency/ESRIN, Largo Galileo Galilei 1, 00044 Frascati, Roma, Italy\label{aff44}
\and
ESAC/ESA, Camino Bajo del Castillo, s/n., Urb. Villafranca del Castillo, 28692 Villanueva de la Ca\~nada, Madrid, Spain\label{aff45}
\and
Universit\'e Claude Bernard Lyon 1, CNRS/IN2P3, IP2I Lyon, UMR 5822, Villeurbanne, F-69100, France\label{aff46}
\and
Institut de Ci\`{e}ncies del Cosmos (ICCUB), Universitat de Barcelona (IEEC-UB), Mart\'{i} i Franqu\`{e}s 1, 08028 Barcelona, Spain\label{aff47}
\and
Instituci\'o Catalana de Recerca i Estudis Avan\c{c}ats (ICREA), Passeig de Llu\'{\i}s Companys 23, 08010 Barcelona, Spain\label{aff48}
\and
Institut de Ciencies de l'Espai (IEEC-CSIC), Campus UAB, Carrer de Can Magrans, s/n Cerdanyola del Vall\'es, 08193 Barcelona, Spain\label{aff49}
\and
UCB Lyon 1, CNRS/IN2P3, IUF, IP2I Lyon, 4 rue Enrico Fermi, 69622 Villeurbanne, France\label{aff50}
\and
Mullard Space Science Laboratory, University College London, Holmbury St Mary, Dorking, Surrey RH5 6NT, UK\label{aff51}
\and
Universit\'e Paris-Saclay, CNRS, Institut d'astrophysique spatiale, 91405, Orsay, France\label{aff52}
\and
INFN-Padova, Via Marzolo 8, 35131 Padova, Italy\label{aff53}
\and
Aix-Marseille Universit\'e, CNRS/IN2P3, CPPM, Marseille, France\label{aff54}
\and
INFN-Bologna, Via Irnerio 46, 40126 Bologna, Italy\label{aff55}
\and
School of Physics, HH Wills Physics Laboratory, University of Bristol, Tyndall Avenue, Bristol, BS8 1TL, UK\label{aff56}
\and
Department of Physics, University of Oxford, Keble Road, Oxford OX1 3RH, UK\label{aff57}
\and
University Observatory, LMU Faculty of Physics, Scheinerstr.~1, 81679 Munich, Germany\label{aff58}
\and
Max Planck Institute for Extraterrestrial Physics, Giessenbachstr. 1, 85748 Garching, Germany\label{aff59}
\and
INAF-Osservatorio Astronomico di Padova, Via dell'Osservatorio 5, 35122 Padova, Italy\label{aff60}
\and
Universit\"ats-Sternwarte M\"unchen, Fakult\"at f\"ur Physik, Ludwig-Maximilians-Universit\"at M\"unchen, Scheinerstr.~1, 81679 M\"unchen, Germany\label{aff61}
\and
Institute of Theoretical Astrophysics, University of Oslo, P.O. Box 1029 Blindern, 0315 Oslo, Norway\label{aff62}
\and
Caltech/IPAC, 1200 E. California Blvd., Pasadena, CA 91125, USA\label{aff63}
\and
Department of Physics, Lancaster University, Lancaster, LA1 4YB, UK\label{aff64}
\and
Felix Hormuth Engineering, Goethestr. 17, 69181 Leimen, Germany\label{aff65}
\and
Technical University of Denmark, Elektrovej 327, 2800 Kgs. Lyngby, Denmark\label{aff66}
\and
Cosmic Dawn Center (DAWN), Denmark\label{aff67}
\and
Universit\'e PSL, Observatoire de Paris, Sorbonne Universit\'e, CNRS, LERMA, 75014, Paris, France\label{aff68}
\and
Universit\'e Paris-Cit\'e, 5 Rue Thomas Mann, 75013, Paris, France\label{aff69}
\and
NASA Goddard Space Flight Center, Greenbelt, MD 20771, USA\label{aff70}
\and
Department of Physics and Astronomy, University College London, Gower Street, London WC1E 6BT, UK\label{aff71}
\and
Universit\'e de Gen\`eve, D\'epartement de Physique Th\'eorique and Centre for Astroparticle Physics, 24 quai Ernest-Ansermet, CH-1211 Gen\`eve 4, Switzerland\label{aff72}
\and
Department of Physics, P.O. Box 64, University of Helsinki, 00014 Helsinki, Finland\label{aff73}
\and
Helsinki Institute of Physics, Gustaf H{\"a}llstr{\"o}min katu 2, University of Helsinki, 00014 Helsinki, Finland\label{aff74}
\and
Laboratoire d'etude de l'Univers et des phenomenes eXtremes, Observatoire de Paris, Universit\'e PSL, Sorbonne Universit\'e, CNRS, 92190 Meudon, France\label{aff75}
\and
Aix-Marseille Universit\'e, CNRS, CNES, LAM, Marseille, France\label{aff76}
\and
SKAO, Jodrell Bank, Lower Withington, Macclesfield SK11 9FT, UK\label{aff77}
\and
Centre de Calcul de l'IN2P3/CNRS, 21 avenue Pierre de Coubertin 69627 Villeurbanne Cedex, France\label{aff78}
\and
Universit\"at Bonn, Argelander-Institut f\"ur Astronomie, Auf dem H\"ugel 71, 53121 Bonn, Germany\label{aff79}
\and
INFN-Sezione di Roma, Piazzale Aldo Moro, 2 - c/o Dipartimento di Fisica, Edificio G. Marconi, 00185 Roma, Italy\label{aff80}
\and
Dipartimento di Fisica e Astronomia "Augusto Righi" - Alma Mater Studiorum Universit\`a di Bologna, via Piero Gobetti 93/2, 40129 Bologna, Italy\label{aff81}
\and
Department of Physics, Institute for Computational Cosmology, Durham University, South Road, Durham, DH1 3LE, UK\label{aff82}
\and
Universit\'e Paris Cit\'e, CNRS, Astroparticule et Cosmologie, 75013 Paris, France\label{aff83}
\and
CNRS-UCB International Research Laboratory, Centre Pierre Bin\'etruy, IRL2007, CPB-IN2P3, Berkeley, USA\label{aff84}
\and
Institute of Physics, Laboratory of Astrophysics, Ecole Polytechnique F\'ed\'erale de Lausanne (EPFL), Observatoire de Sauverny, 1290 Versoix, Switzerland\label{aff85}
\and
Dipartimento di Fisica - Sezione di Astronomia, Universit\`a di Trieste, Via Tiepolo 11, 34131 Trieste, Italy\label{aff86}
\and
Telespazio UK S.L. for European Space Agency (ESA), Camino bajo del Castillo, s/n, Urbanizacion Villafranca del Castillo, Villanueva de la Ca\~nada, 28692 Madrid, Spain\label{aff87}
\and
Institut de F\'{i}sica d'Altes Energies (IFAE), The Barcelona Institute of Science and Technology, Campus UAB, 08193 Bellaterra (Barcelona), Spain\label{aff88}
\and
European Space Agency/ESTEC, Keplerlaan 1, 2201 AZ Noordwijk, The Netherlands\label{aff89}
\and
School of Mathematics, Statistics and Physics, Newcastle University, Herschel Building, Newcastle-upon-Tyne, NE1 7RU, UK\label{aff90}
\and
DARK, Niels Bohr Institute, University of Copenhagen, Jagtvej 155, 2200 Copenhagen, Denmark\label{aff91}
\and
Waterloo Centre for Astrophysics, University of Waterloo, Waterloo, Ontario N2L 3G1, Canada\label{aff92}
\and
Department of Physics and Astronomy, University of Waterloo, Waterloo, Ontario N2L 3G1, Canada\label{aff93}
\and
Perimeter Institute for Theoretical Physics, Waterloo, Ontario N2L 2Y5, Canada\label{aff94}
\and
Universit\'e Paris-Saclay, Universit\'e Paris Cit\'e, CEA, CNRS, AIM, 91191, Gif-sur-Yvette, France\label{aff95}
\and
Space Science Data Center, Italian Space Agency, via del Politecnico snc, 00133 Roma, Italy\label{aff96}
\and
Centre National d'Etudes Spatiales -- Centre spatial de Toulouse, 18 avenue Edouard Belin, 31401 Toulouse Cedex 9, France\label{aff97}
\and
Institute of Space Science, Str. Atomistilor, nr. 409 M\u{a}gurele, Ilfov, 077125, Romania\label{aff98}
\and
Consejo Superior de Investigaciones Cientificas, Calle Serrano 117, 28006 Madrid, Spain\label{aff99}
\and
Universidad de La Laguna, Dpto. Astrof\'\i sica, E-38206 La Laguna, Tenerife, Spain\label{aff100}
\and
Dipartimento di Fisica e Astronomia "G. Galilei", Universit\`a di Padova, Via Marzolo 8, 35131 Padova, Italy\label{aff101}
\and
Instituto de F\'isica Te\'orica UAM-CSIC, Campus de Cantoblanco, 28049 Madrid, Spain\label{aff102}
\and
Institut de Recherche en Astrophysique et Plan\'etologie (IRAP), Universit\'e de Toulouse, CNRS, UPS, CNES, 14 Av. Edouard Belin, 31400 Toulouse, France\label{aff103}
\and
Universit\'e St Joseph; Faculty of Sciences, Beirut, Lebanon\label{aff104}
\and
Departamento de F\'isica, FCFM, Universidad de Chile, Blanco Encalada 2008, Santiago, Chile\label{aff105}
\and
Universit\"at Innsbruck, Institut f\"ur Astro- und Teilchenphysik, Technikerstr. 25/8, 6020 Innsbruck, Austria\label{aff106}
\and
Infrared Processing and Analysis Center, California Institute of Technology, Pasadena, CA 91125, USA\label{aff107}
\and
Departamento de F\'isica, Faculdade de Ci\^encias, Universidade de Lisboa, Edif\'icio C8, Campo Grande, PT1749-016 Lisboa, Portugal\label{aff108}
\and
Instituto de Astrof\'isica e Ci\^encias do Espa\c{c}o, Faculdade de Ci\^encias, Universidade de Lisboa, Tapada da Ajuda, 1349-018 Lisboa, Portugal\label{aff109}
\and
Cosmic Dawn Center (DAWN)\label{aff110}
\and
Niels Bohr Institute, University of Copenhagen, Jagtvej 128, 2200 Copenhagen, Denmark\label{aff111}
\and
Universidad Polit\'ecnica de Cartagena, Departamento de Electr\'onica y Tecnolog\'ia de Computadoras,  Plaza del Hospital 1, 30202 Cartagena, Spain\label{aff112}
\and
European University of Technology EUt+, European Union\label{aff113}
\and
Dipartimento di Fisica e Scienze della Terra, Universit\`a degli Studi di Ferrara, Via Giuseppe Saragat 1, 44122 Ferrara, Italy\label{aff114}
\and
Istituto Nazionale di Fisica Nucleare, Sezione di Ferrara, Via Giuseppe Saragat 1, 44122 Ferrara, Italy\label{aff115}
\and
ICSC - Centro Nazionale di Ricerca in High Performance Computing, Big Data e Quantum Computing, Via Magnanelli 2, Bologna, Italy\label{aff116}
\and
INAF, Istituto di Radioastronomia, Via Piero Gobetti 101, 40129 Bologna, Italy\label{aff117}
\and
Universit\'e C\^{o}te d'Azur, Observatoire de la C\^{o}te d'Azur, CNRS, Laboratoire Lagrange, Bd de l'Observatoire, CS 34229, 06304 Nice cedex 4, France\label{aff118}
\and
Oskar Klein Centre for Cosmoparticle Physics, Department of Physics, Stockholm University, Stockholm, SE-106 91, Sweden\label{aff119}
\and
Astrophysics Group, Blackett Laboratory, Imperial College London, London SW7 2AZ, UK\label{aff120}
\and
Univ. Grenoble Alpes, CNRS, Grenoble INP, LPSC-IN2P3, 53, Avenue des Martyrs, 38000, Grenoble, France\label{aff121}
\and
Dipartimento di Fisica, Sapienza Universit\`a di Roma, Piazzale Aldo Moro 2, 00185 Roma, Italy\label{aff122}
\and
Institut d'Astrophysique de Paris, 98bis Boulevard Arago, 75014, Paris, France\label{aff123}
\and
ICL, Junia, Universit\'e Catholique de Lille, LITL, 59000 Lille, France\label{aff124}
\and
CERCA/ISO, Department of Physics, Case Western Reserve University, 10900 Euclid Avenue, Cleveland, OH 44106, USA\label{aff125}
\and
Laboratoire Univers et Th\'eorie, Observatoire de Paris, Universit\'e PSL, Universit\'e Paris Cit\'e, CNRS, 92190 Meudon, France\label{aff126}
\and
Dipartimento di Fisica "Aldo Pontremoli", Universit\`a degli Studi di Milano, Via Celoria 16, 20133 Milano, Italy\label{aff127}
\and
INFN-Sezione di Milano, Via Celoria 16, 20133 Milano, Italy\label{aff128}
\and
Departamento de F{\'\i}sica Fundamental. Universidad de Salamanca. Plaza de la Merced s/n. 37008 Salamanca, Spain\label{aff129}
\and
IRFU, CEA, Universit\'e Paris-Saclay 91191 Gif-sur-Yvette Cedex, France\label{aff130}
\and
Universiteit Gent, Department of Physics and Astronomy, Proeftuinstraat 86 N3, 9000 Ghent, Belgium
\label{aff131}
\and
Universit\'e de Strasbourg, CNRS, Observatoire astronomique de Strasbourg, UMR 7550, 67000 Strasbourg, France\label{aff132}
\and
Center for Data-Driven Discovery, Kavli IPMU (WPI), UTIAS, The University of Tokyo, Kashiwa, Chiba 277-8583, Japan\label{aff133}
\and
California Institute of Technology, 1200 E California Blvd, Pasadena, CA 91125, USA\label{aff134}
\and
Institute of Cosmology and Gravitation, University of Portsmouth, Portsmouth PO1 3FX, UK\label{aff135}
\and
Department of Physics \& Astronomy, University of California Irvine, Irvine CA 92697, USA\label{aff136}
\and
Department of Mathematics and Physics E. De Giorgi, University of Salento, Via per Arnesano, CP-I93, 73100, Lecce, Italy\label{aff137}
\and
INFN, Sezione di Lecce, Via per Arnesano, CP-193, 73100, Lecce, Italy\label{aff138}
\and
INAF-Sezione di Lecce, c/o Dipartimento Matematica e Fisica, Via per Arnesano, 73100, Lecce, Italy\label{aff139}
\and
Departamento F\'isica Aplicada, Universidad Polit\'ecnica de Cartagena, Campus Muralla del Mar, 30202 Cartagena, Murcia, Spain\label{aff140}
\and
Instituto de F\'isica de Cantabria, Edificio Juan Jord\'a, Avenida de los Castros, 39005 Santander, Spain\label{aff141}
\and
Institut d'Astrophysique de Paris, UMR 7095, CNRS, and Sorbonne Universit\'e, 98 bis boulevard Arago, 75014 Paris, France\label{aff142}
\and
Department of Computer Science, Aalto University, PO Box 15400, Espoo, FI-00 076, Finland\label{aff143}
\and
School of Physics and Astronomy, University of Nottingham, University Park, Nottingham NG7 2RD, UK\label{aff144}
\and
Ruhr University Bochum, Faculty of Physics and Astronomy, Astronomical Institute (AIRUB), German Centre for Cosmological Lensing (GCCL), 44780 Bochum, Germany\label{aff145}
\and
Department of Physics and Astronomy, Vesilinnantie 5, University of Turku, 20014 Turku, Finland\label{aff146}
\and
Finnish Centre for Astronomy with ESO (FINCA), Quantum, Vesilinnantie 5, University of Turku, 20014 Turku, Finland\label{aff147}
\and
Serco for European Space Agency (ESA), Camino bajo del Castillo, s/n, Urbanizacion Villafranca del Castillo, Villanueva de la Ca\~nada, 28692 Madrid, Spain\label{aff148}
\and
ARC Centre of Excellence for Dark Matter Particle Physics, Melbourne, Australia\label{aff149}
\and
Centre for Astrophysics \& Supercomputing, Swinburne University of Technology,  Hawthorn, Victoria 3122, Australia\label{aff150}
\and
Department of Physics and Astronomy, University of the Western Cape, Bellville, Cape Town, 7535, South Africa\label{aff151}
\and
MTA-CSFK Lend\"ulet Large-Scale Structure Research Group, Konkoly-Thege Mikl\'os \'ut 15-17, H-1121 Budapest, Hungary\label{aff152}
\and
Konkoly Observatory, HUN-REN CSFK, MTA Centre of Excellence, Budapest, Konkoly Thege Mikl\'os {\'u}t 15-17. H-1121, Hungary\label{aff153}
\and
Institute for Theoretical Particle Physics and Cosmology (TTK), RWTH Aachen University, 52056 Aachen, Germany\label{aff154}
\and
Dipartimento di Fisica, Universit\`a di Roma Tor Vergata, Via della Ricerca Scientifica 1, Roma, Italy\label{aff155}
\and
INFN, Sezione di Roma 2, Via della Ricerca Scientifica 1, Roma, Italy\label{aff156}
\and
HE Space for European Space Agency (ESA), Camino bajo del Castillo, s/n, Urbanizacion Villafranca del Castillo, Villanueva de la Ca\~nada, 28692 Madrid, Spain\label{aff157}
\and
STAR Institute, University of Li{\`e}ge, Quartier Agora, All\'ee du six Ao\^ut 19c, 4000 Li\`ege, Belgium\label{aff158}
\and
Department of Astrophysics, University of Zurich, Winterthurerstrasse 190, 8057 Zurich, Switzerland\label{aff159}
\and
University of Applied Sciences and Arts of Northwestern Switzerland, School of Computer Science, 5210 Windisch, Switzerland\label{aff160}
\and
INAF - Osservatorio Astronomico d'Abruzzo, Via Maggini, 64100, Teramo, Italy\label{aff161}
\and
Theoretical astrophysics, Department of Physics and Astronomy, Uppsala University, Box 516, 751 37 Uppsala, Sweden\label{aff162}
\and
Mathematical Institute, University of Leiden, Einsteinweg 55, 2333 CA Leiden, The Netherlands\label{aff163}
\and
School of Physical Sciences, The Open University, Milton Keynes, MK7 6AA, UK\label{aff164}
\and
Univ. Lille, CNRS, Centrale Lille, UMR 9189 CRIStAL, 59000 Lille, France\label{aff165}
\and
Center for Astrophysics and Cosmology, University of Nova Gorica, Nova Gorica, Slovenia\label{aff166}
\and
Kobayashi-Maskawa Institute for the Origin of Particles and the Universe, Nagoya University, Chikusa-ku, Nagoya, 464-8602, Japan\label{aff167}
\and
Institute for Particle Physics and Astrophysics, Dept. of Physics, ETH Zurich, Wolfgang-Pauli-Strasse 27, 8093 Zurich, Switzerland\label{aff168}
\and
Department of Astrophysical Sciences, Peyton Hall, Princeton University, Princeton, NJ 08544, USA\label{aff169}
\and
Fakult\"at f\"ur Physik, Universit\"at Bielefeld, Postfach 100131, 33501 Bielefeld, Germany\label{aff170}
\and
Space physics and astronomy research unit, University of Oulu, Pentti Kaiteran katu 1, FI-90014 Oulu, Finland\label{aff171}
\and
International Centre for Theoretical Physics (ICTP), Strada Costiera 11, 34151 Trieste, Italy\label{aff172}
\and
Center for Computational Astrophysics, Flatiron Institute, 162 5th Avenue, 10010, New York, NY, USA\label{aff173}
\and
SRON Netherlands Institute for Space Research, Landleven 12, 9747 AD, Groningen, The Netherlands\label{aff174}}           

%
%
\abstract
{
A substantial volume of spectroscopic data has become available with the Euclid Quick Data Release (Q1), which represents a unique opportunity to study quasars (QSOs) in the Euclid Deep Fields, provide spectroscopic measurements, and estimates of their physical properties. We present results from a spectroscopic analysis of QSOs with \HE$\leq 22.5$ using Q1. We use external catalogues  from surveys such as the Dark Energy Spectroscopic Instrument (DESI), \gaia, Wide-field Infrared Survey Explorer (WISE), and the QUasars as BRIght beacons for Cosmology in the Southern hemisphere (QUBRICS) for the QSO selection and redshift determination, totalling $5489$ QSOs. We provide measurements of the line fluxes, full widths at half-maximum, and respective uncertainties for the emission lines of \Euclid spectra through the use of the software \texttt{QSFit}. We additionally estimate the QSO continuum slope and present a simple criterion to identify problematic \Euclid spectra based on fitting statistics and signal-to-noise ratio (S/N). Of the total $5489$ QSOs, $5387$ are successfully fitted. We find that 45\% of the total number of successfully fitted spectra are free of problematic artefacts allowing for trustworthy measurements. Our final \Euclid QSO sample spans a redshift range of $0.01 \leq z \leq 4.8$ and shows a median spectral index of $\alpha_{\lambda}\sim -1.17$. Finally, we estimate black hole masses with single epoch methods for $1213$ sources, using the \ion{He}{i}~$\lambda10830$, Pa\,$\beta$, \ha, \hb, and \ion{Mg}{ii} emission lines, obtaining a mean logarithmic mass of $\logten(M_{\rm BH}/\si{\solarmass})\sim8.7$ and mean Eddington ratio of $\lambda_\mathrm{Edd}\sim0.34$. The catalogue with the physical spectral properties of this sample together with the software developed for the analysis is available online to allow scientific exploitation.}


   \titlerunning{ Spectroscopy of QSOs: physical properties fitting }
   \authorrunning{Euclid Collaboration: J.\ Calhau et al.}
   
   \maketitle
%
%
%
%

\section{\label{sc:Intro}Introduction}

Most galaxies in the Universe host supermassive black holes (SMBHs) at their centres \citep[][and references therein]{1963MNRAS.125..169H, 1964ApJ...140..796S, 2013ARA&A..51..511K}. Some of these undergo an active accretion phase during which the infall of material onto these SMBHs produces some of the most powerful emissions observed in the Universe. Galaxies in this phase are referred to as active galactic nuclei (AGN).

A high amount of observational evidence supports the existence of scaling relations between the mass of the black hole (BH) and host galaxy properties, such as, e.g., the velocity dispersion of the bulge, the stellar luminosity, or the stellar mass \citep[][and references therein]{dressler1989observational, 2000ApJ...539L...9F,
2017A&A...598A..51R,2025MNRAS.541.2070S}. Observations further suggest that the star-formation history of host galaxies and the accretion history of the central SMBHs closely track each other across cosmic time \citep[][]{2009ApJ...690...20S, 2014ARA&A..52..415M, 2020MNRAS.493.3341C}. The presence of an AGN is thus believed to be a key driver in galaxy evolution and the central SMBH may even precede the formation of the host galaxy itself \citep{1998A&A...331L...1S,2006MNRAS.370..645B, 2026arXiv260500763M}. As a subset of AGN, quasars (QSOs) stand as the most luminous sources in the Universe \citep{2026arXiv260321976B}. They have been detected up to $z\sim 7.7$ \citep[e.g.][]{2001AJ....122.2833F, 2021ApJ...907L...1W}, including \Euclid data \citep{2026A&A...711A.104Y}, and are expected to be found in the galaxy overdensities of the early Universe \citep{10.1093/mnras/stu101, 2023ApJ...951L...4W, Meyer_2022, 2020A&A...642L...1M}. These discoveries have, among others, led to new insights on the mechanisms of SMBH formation \citep{2020ARA&A..58...27I}, and extended our understanding of the large-scale structure of the Universe \citep[e.g.][]{2023ApJ...951L...4W}, as well as the interplay between host galaxy and central SMBH through cosmic time \citep[][for a review.]{2013ARA&A..51..511K}

In order to get a consistent view of the evolution and interplay of galaxies and AGN, it is crucial to have statistically significant datasets across a broad redshift range. With the advent of the Euclid Quick Data Release \citep[Q1;][]{Q1-TP001, Q1cite}, it is now possible to build such statistical datasets over large portions of the sky \citep{Scaramella-EP1}. 

\Euclid is expected to deliver redshifts for over 25 million galaxies in the redshift range of $z=0.9$--$1.8$ \citep[][]{EuclidSkyOverview} using \ha\ line emission from slitless spectroscopy in the near-infrared (NIR) bands. Pre-launch studies on the potential of \Euclid for the study of QSOs tailored on the \Euclid instrument capabilities found that the reliability of the spectroscopic redshift determination, which is handled by the \Euclid SPE processing function, decreases for wavelengths other than that of \ha\ \citep{EP-Lusso} due to the low S/N or lack of identifiable lines. Additionally, the \ha\, and [\ion{N}{ii}] lines are blended at the spectral resolution of \Euclid, resulting in a relatively high level of scatter, particularly for the full width at half maximum (FWHM) measurements.

Q1 is now publicly available and contains \mbox{$\sim 30$ million} objects, identified through imaging data located in the three Euclid Deep Fields (EDFs) Euclid Deep Field North (EDF-N), South (EDF-S), and Fornax (EDF-F), and encompassing a total of $\rm \sim 63 \, deg^2$ \citep[][]{EuclidSkyOverview, Q1-TP001}. Q1 provides both imaging (in the visible and NIR bands) and spectroscopic data (in the NIR bands), as well as complementary ground-based photometry in the \textit{u, g, r, i,} and \textit{z} bands. So far, almost $230 \, 000$ AGN and AGN candidates have been identified in the Q1 fields alone \citep{Q1-SP027}, illustrating the potential of \Euclid for AGN studies. The identifications were based on photometric selections, such as colour--colour selections, including the use of the above ancillary data, where available, and with the requirement that AGN sources be point-like, and spectroscopic selections from DESI matches (based on line flux, FWHM, and spectral classification from DESI). Additionally, X-ray detections and machine-learning selection techniques were also considered for AGN identification \citep[e.g.][ and references therein for details]{Q1-SP027, Q1-SP003}.

While these existing results even include a multi-wavelength selection of active galaxies in the \Euclid fields \citep[e.g.][F26 hereafter]{Q1-SP068}, we are still missing a targeted spectroscopic study of \Euclid's QSOs. The derivation of AGN properties -- line luminosity, FWHM, velocity offsets, and QSO continuum, and SMBH masses inferred from them -- critically relies on emission line measurements and continuum estimates, which are essential to understand AGN evolution and environmental impact. With the release of Q1 data, we are in a position to exploit the influx of large quantities of newly available \Euclid spectra and provide spectroscopic measurements of these objects. 

In this paper, we present a spectroscopic analysis of the QSOs detected in the Q1, selected from external catalogues, namely DESI DR1 \citep[][]{DESI-DR1} and QUBRICS \citep[][]{2019ApJ...887..268C, 2021MNRAS.506.2471G}, as well as direct visual inspection of the bright QSO \Euclid spectra (F26). We also provide estimates of the QSO continuum and present a simple selection criterion to identify problematic or low-quality \Euclid spectra. 

This paper is organised as follows. Section \ref{sc:Data} presents the data and sample selection. Section \ref{sc:Methods} outlines the  methodology and analysis and describes the final selection of QSOs used in the analysis. Section \ref{sc:Results} presents the results and discussion. Section \ref{sc:Conclusion} summarises our conclusions. For this work, we used a $\Lambda$CDM cosmology with $H_0 = 70 \, {\rm km\,s^{-1}\,Mpc^{-1}}$, $\Omega_{\rm m}=0.3$, and $\Omega_\Lambda = 0.7$. Throughout the figures in this paper we use the notation $\tilde{\mu}$ and $\tilde{\sigma}$ to represent the median and the normalised median absolute deviation (MAD) values, respectively.

\section{\label{sc:Data}Data and sample}

\subsection{\label{sc:Q1}Q1 spectra}
The \Euclid Near-Infrared Spectrometer and Photometer \citep[NISP;][]{Schirmer-EP18, EuclidSkyNISP} is equipped with one blue grism covering the wavelength range \mbox{$\lambda=920$--1250~nm} and three red grisms covering the wavelength range \mbox{$\lambda=1250$--1850~nm}. We show examples of typical \Euclid QSO's spectra in Fig. \ref{fig:Example_spectra}, for both high and low signal-to-noise ratio (S/N) cases (S/N$=47$ and $4$, respectively).\footnote{The \Euclid source identification number (ID) is calculated through the equation detailed in \url{https://euclid.esac.esa.int/dr/q1/dpdd/merdpd/dpcards/mer_finalcatalog.html}. Specifically, a negative sign indicates a negative declination for that source.} For spectroscopy in Q1, only observations in the red grism are available \citep[][]{Q1-TP006}, with a spectral resolution of $R = \lambda / \Delta\lambda  \sim450$. Q1 provides spectra from all sources detected with \HE$\leq 22.5$, a cut-off designed to minimise contamination by spurious detections at fainter magnitudes \citep[][]{ Q1-TP007}. Each spectrum is provided in FITS format and includes a wavelength and normalised flux columns, two columns with a pixel mask and a quality flag, to help in the identification of problematic fluxes and wavelengths, and the variance associated with the flux.

\subsubsection{\label{sc:Euclid_spectra}Spectra preparation}
To prepare our sample for analysis, we applied quality masks to our spectra, as suggested by \citet{Q1-TP007}.
Q1 spectra sometimes feature artefacts due to low S/N or issues during acquisition or data reduction. These issues can arise from factors such as cross-contamination, due to e.g. nearby sources, or transient detection, like cosmic rays \citep[Appendix ~\ref{sc:Appendix_A} and Fig. \ref{fig:Bad_spectra}, as well as][]{Q1-TP006, Q1-TP007}.
To prevent these artefacts from impacting our study, we restricted our analysis to the wavelengths for which the SIR \textsc{MASK} reports values of $0$ (no issues) or $2$ (potential problems due to low S/N) and rejected pixels for which the \textsc{MASK} column reported an odd number (pixel has issues) or a value higher than $64$ (close to spectrum edge).  We decide to include pixels with mask value of $2$ in order to use all potentially significant data, and then apply our own quality checks post-fitting (Sect. \ref{sc:Statistics_Parameters}) to reject truly problematic spectra.
These conditions are relatively strict, but justified in light of the problems revealed by visual inspection of the spectra. For example, a common artefact observed in nearly all spectra is a sudden and sharp decrease in the measured flux at the edges of the spectral range, sometimes resulting in negative flux values, along with an equally unexpected peak in the measured signal. Negative flux values can also sometimes be found throughout a spectrum.
These checks allow us not only to avoid these problematic edge regions of the spectra, but also to reject the wavelengths for which there are problems with the flux measurements or calibration due to spurious pixels with unrealistic high or low flux measurements \citep[][for details]{Q1-TP006}. 

\subsection{\label{sc:External_Cats} QSO selection from \Euclid and external spectroscopic surveys}
For the selection of QSOs in the Q1 spectroscopic data, we make use of a catalogue of visually inspected QSO spectra from \mbox{\Euclid} data (Sect. \ref{sc:Yumings_QSOs}), as well as external catalogues with spectroscopic redshift measurements (Sects. \ref{sc:DESI} and \ref{sc:QUBRICS}).

\subsubsection{\label{sc:Yumings_QSOs}Visually confirmed QSOs from F26}
Our sample incorporates \Euclid QSOs from F26, who selected them using WISE \citep[][]{2010AJ....140.1868W, 2018ApJS..234...23A} and \gaia~DR3 data \citep[][]{GaiaCollaboration_2023_2023A&A...674A...1G, GaiaCollaboration_2023_2023A&A...674A..41G} and subsequently visually inspected their \Euclid NISP spectra to confirm their classification and redshift. 
As detailed in F26, these objects are bright QSOs and an empirical cut of $\HE < 21.5$ is further applied by F26 to the sample in order to assure the highest confidence in the redshift estimation, resulting in a selection of a total of $3468$ sources.

\subsubsection{\label{sc:DESI}DESI}

The Dark Energy Spectroscopic Instrument \citep[DESI;][]{2022AJ....164..207D} collaboration has produced a map of the large-scale structure of the Universe within $0 \leq z \leq 4$, obtaining spectra of stars, galaxies, and QSOs over an area $\rm \approx 14\,000 \, deg^2$ \citep[][]{2024AJ....168...58D}.
The DESI Data Release 1 \citep[DESI DR1;][]{DESI-DR1} includes all data acquired in the first 13 months of the DESI main survey, as well as a reprocessing of the DESI Early Data Release. It includes high-confidence redshifts for 18.7 million targets, including 1.6 million quasars. DESI only overlaps with EDF-N.

The selection of QSOs from the DESI DR1 catalogue\footnote{\url{https://data.desi.lbl.gov/doc/releases/dr1/vac/agnqso/}} and cross-matching with \Euclid's SPE catalogue was done by using the sky coordinates with a radius of $0\farcs5$. In order to restrict our matching results to QSOs, we require that the sources contain the DESI-attributed flag $\rm \texttt{SPECTYPE}=QSO$ \citep{2024AJ....168...58D, DESI-DR1}. These restrictions result in $2304$ selected QSOs, some of which overlap with the other catalogues (Sect. \ref{sc:Final_input_catalogue}).
The main advantage of this sample is that it has spectroscopic coverage over a large wavelength range, from optical (DESI) to NIR (\Euclid). Thus, we can, e.g., estimate black hole masses using different broad emission lines (Sect. \ref{sc:BH_quantities}).

\subsubsection{\label{sc:QUBRICS}QUBRICS}

The QUasars as BRIght beacons for Cosmology in the Southern hemisphere \citep[QUBRICS;][]{2019ApJ...887..268C, 2021MNRAS.506.2471G} is designed to target the brightest QSOs at (but not restricted to) $z>2.5$ 
using photometric data from programmes such as the Dark Energy Survey \citep[DES;][]{2021ApJS..255...20A} and PanSTARRS \citep[][]{2016arXiv161205560C}, among others. 
QUBRICS employs a machine-learning technique for QSO selection, the canonical correlation analysis, achieving a classification precision of approximately $70\%$. More recently, a heuristic reverse selection method \citep[][]{2024A&A...683A..34C} has been implemented, which significantly boosts recall to $90\%$ for QSOs at $z>3$ with only a moderate reduction in precision (down to 60\%). To date, QUBRICS has detected more than $1000$ new QSOs. We matched the \Euclid SPE catalogue with QUBRICS using a radius of $0\farcs5$ and obtain $207$ QSOs.

\subsection{\label{sc:Final_input_catalogue}Parent sample}
\begin{figure}[htbp!]
\centering
\includegraphics[angle=0,width=1.0\hsize]{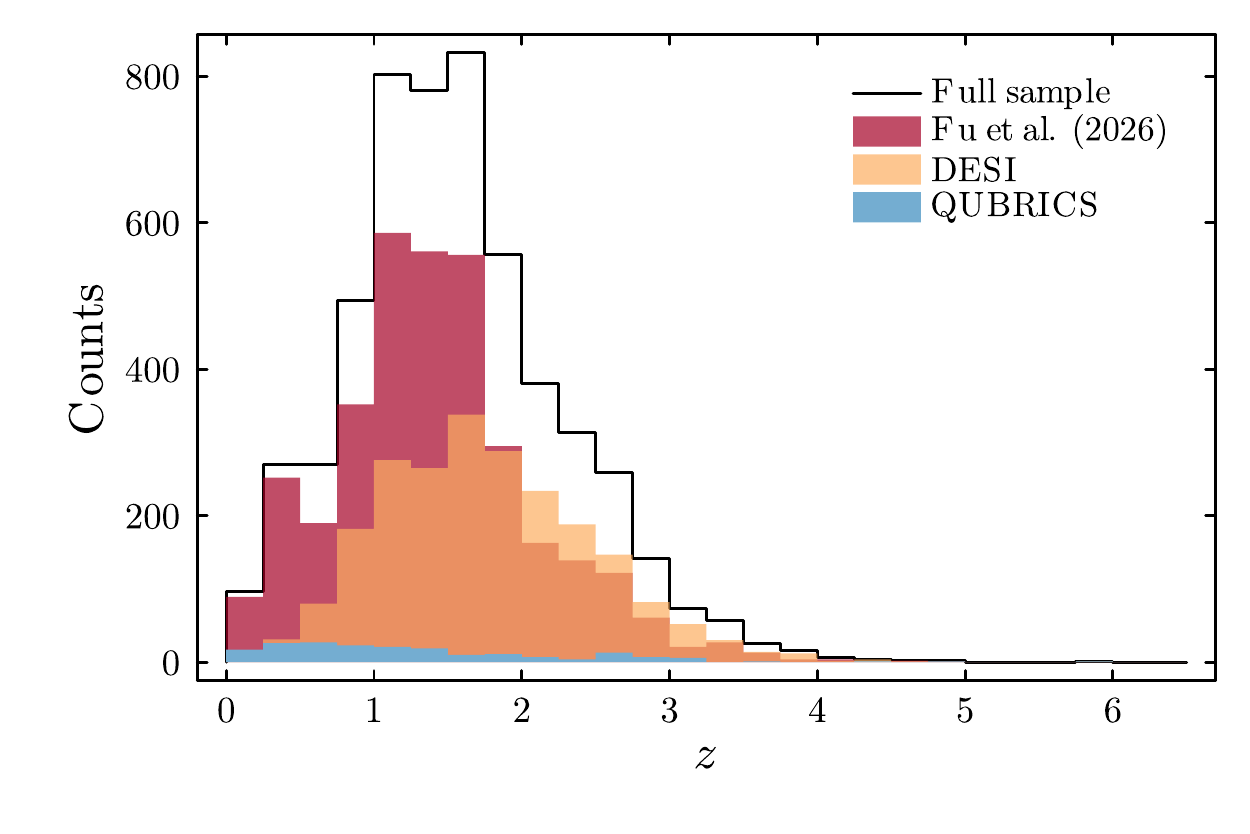}
\caption{Histograms of the redshifts for the final QSO sample considered in this study. The black empty histogram shows the whole sample, with a mean redshift of 1.56. The red, dark yellow, and blue histograms show the sources taken from F26, DESI, and QUBRICS, respectively. For all three source catalogues, the majority of QSOs are found at $z<2$.}
\label{fig:Redshift_dist}
\end{figure}

In total, we make an initial selection of $5489$ unique QSOs across the three EDFs. If a source is present in more than one catalogue, we adopt the redshift from DESI, QUBRICS, or F26, in that priority order. The decision to prioritise DESI redshift solutions over the visual inspection of F26 stems from difficulties in estimating the redshift from \Euclid spectra alone during visual inspection, due to uncertainties linked to, for example, confusion on the line identification when only one line is present, or the presence of artifacts in the spectra. Additionally, DESI's higher spectral resolution allows for a more precise redshift determination, which is desirable for the line fitting process. Approximately 10\% of the parent sample is present in more than one catalogue, with an overlap of 372 (7\%) sources between our DESI selection and F26, 87 (2\%) between F26 and QUBRICS, and 65 (1\%) between QUBRICS and DESI.
Our sample spans a redshift range of $z\in[0.004,6.6]$, with the most extreme values coming from only three QUBRICS sources (Fig.~\ref{fig:Redshift_dist}). 
The highest concentration of QSOs is found at $1 < z < 2$. 
The histogram of the sample roughly follows that of QSOs present in DESI DR1 and F26 and reflects the intrinsic selection biases underlying these catalogues.
The extended tail on the higher redshift range is primarily driven by the visually-confirmed QSOs from F26, although DESI sources are still present up to $z=4$.
We stress that our sample is affected by the biases inherent to the external catalogues used, which is unavoidable. Using different catalogues may change the results presented in this paper. Therefore our results are not necessarily representative of the overall QSO population.

\section{\label{sc:Methods}Data analysis}
\subsection{\label{The QSFIT package}\texttt{QSFit}}
In our analysis, we make use of the software Quasar Spectral Fitting \citep[\texttt{QSFit};][]{2017MNRAS.472.4051C}\footnote{\url{https://gcalderone.github.io/QSFit.jl/}}. Originally developed in the IDL language, it has now been reimplemented in {\tt Julia}\footnote{\url{https://julialang.org/}}, an open source, high-level language with a focus on high performance, and provides a powerful framework for the analysis of AGN and QSO spectra. It enables the measurement of emission line luminosities, FWHM, and QSO continuum, among other quantities \citep[section 2 of][]{2017MNRAS.472.4051C}. 

A key feature of \texttt{QSFit} is that the analysis is done by fitting all spectral components simultaneously. Furthermore, the QSO continuum component is estimated by making use of the entire available spectrum, minimizing contamination from localised features such as emission lines. \texttt{QSFit} allows for the broadening of its original capabilities through user-defined recipes \citep[e.g.][]{2023MNRAS.518..130S}. We make use of this feature to extend \texttt{QSFit}'s original wavelength fitting range of rest-frame optical/ultraviolet (UV) into \Euclid's NIR spectral range.

For the analysis of our \Euclid spectra, we follow the work of \cite{EP-Lusso}, who applied \texttt{QSFit} to \Euclid mock spectra of AGN, laying the groundwork for spectroscopy with NISP by simulating spectroscopic observations.
The following sections provide further details on the fitting process. We refer the interested reader to \cite{2017MNRAS.472.4051C} for a more in-depth description of the \texttt{QSFit} package, its components, and the fitting procedure.

\subsubsection{\label{sc:Fitting}Fitting components}

When fitting a spectrum, \texttt{QSFit} builds a model by combining several distinct components, as illustrated in Fig.~\ref{fig:QSFIT_Ha_Hb_MgII} and detailed below: 

\begin{enumerate}
    
    \item The QSO continuum is estimated over the entire wavelength range of each individual spectrum. The QSO continuum can be modelled using a simple power law, a smoothly-broken power law, or a cut-off power law. For our purposes, we use a simple power law defined as
    \begin{equation}
        L_\lambda = A \left(\frac{\lambda}{\lambda_{\rm 0}}\right)^{\alpha_\lambda}\, ,
    \end{equation}
    where $\rm \lambda_0$ is the reference wavelength and is taken as the median of the entire rest-framed wavelength range, $\alpha_\lambda$ is the spectral index (constrained by the condition $\alpha_\lambda \in[-5,5]$), and $A$ is the luminosity density at $\lambda = \lambda_{\rm 0}$. The choice of using a simple power law is justified by the narrow wavelength window available for fitting in \Euclid spectra from the red grism alone, which in the majority of cases do not allow a more extended parametrization.
    
    \item The Balmer continuum is modelled using the default configuration, where the luminosity density and the Balmer continuum normalization are defined at $300\,\text{nm}$. These quantities are expressed in units of luminosity density $L_{\lambda}$ and continuum luminosity $\lambda L_{\lambda}$, respectively \citep[see details in section 2 of][]{{2017MNRAS.472.4051C}}.
    
    \item The host galaxy template is used to account for the emission of the host galaxy and separate it from the QSO continuum. \texttt{QSFit} includes several different templates for host galaxies, with the default being an elliptical galaxy with an age of $5 \,{\rm Gyr}$ from the SWIRE library \citep[][]{2007ApJ...663...81P}. The choice of using this template is justified by this being the type of galaxy we expect to host luminous AGN like QSOs \citep[e.g. ][]{2004MNRAS.355..196F, 2014MNRAS.440..476F}. Although AGN can be found in late type galaxies at low redshifts, the majority of our sample targets intrinsically bright targets and is found at $z>1$. We tested whether the imposition of a late type host template changed our results and found no significant differences. But the difficulty in accounting for all the variables which are affected by the fitting of the host template makes going over the entirety of the available host templates an impractical endeavour. For this reason we chose to stick with a single template. Since the elliptical 5 Gyr galaxy template becomes hardly detectable at high redshifts we disable the fitting of the host template for sources at $z>2.6$, where the host contribution is expected to become negligible (due to the emission shifting outside \Euclid's spectral range). Including the host galaxy contribution in the analysis often leads to unrealistic QSO continuum estimates (i.e., slopes) for \Euclid spectra. This happens because there is a correlation between the QSO continuum component and the arbitrarily chosen host template, causing the QSO continuum to be negligible in sources where the host contribution overwhelms the QSO continuum component. In an effort to reduce this problem, following \cite{EP-Lusso}, we undertake a `hybrid' fitting procedure, where we fit our sources while both including the contribution of the host galaxy and excluding it. We then take the fit which presents the lowest Bayesian information criterion (BIC) of the two possibilities. We define ${\rm BIC} = \chi^2 + k \ln (n)$, with $\chi^2$ the non-reduced chi-square, $k$ the number of free parameters, and $n$ the number of data points. A total of $448$ from the original $5489$ sources required the fitting of a host template.
    
    \item The main emission lines are modelled with a single Gaussian from which we then estimate the total line luminosity, FWHM, and velocity offset. Although \texttt{QSFit} is able to fit both narrow and broad components of emission lines, we follow \cite{EP-Lusso}, who found that the \Euclid spectra are better modelled by single-component fitting, and consider only a broad component to the emission lines used to estimate BH quantities (\ha, \hb, \ion{Mg}{ii}, Pa\,$\beta$, and \ion{He}{i}).
    
    \item Blended iron lines contaminate some of the measured strong emission lines and in places can form a pseudo-continuum. \texttt{QSFit} uses iron \ion{Fe}{ii} and \ion{Fe}{iii} templates for the UV and optical bands by default \citep{2001ApJS..134....1V, 2004A&A...417..515V}. At \Euclid's NIR rest-frame range, the contribution from iron lines is expected to be minimal \citep{2008ApJS..174..282L}, and we therefore do not include additional iron line templates in our model beyond the optical and UV range. Given \Euclid's spectral resolution ($R\sim450$), we do not fit the narrow components of the iron models and consider only the broad components.
    
    \item Nuisance lines are spectral features that remain unaccounted for by the \texttt{QSFit} emission components described above (e.g. missing iron lines or features such as asymmetric line profiles). They are added in after all other components have been modelled, and we allow the inclusion of a maximum of two nuisance lines per spectrum.

\end{enumerate}

In addition to these points, the spectrum is de-reddened during the analysis by \texttt{QSFit} using the extinction law by \cite{1994ApJ...422..158O}, and we calculate the amount of extinction by using the map model by \cite{1998ApJ...500..525S}.

\noindent

\subsubsection{\label{sc:Statistics_Parameters}Statistics parameters and quality selection}
In addition to the fitting components described above, \texttt{QSFit} also provides a measure of the goodness of the fit (namely, the reduced $\chi^2$) and the total number of points used in the fit ($N_{\rm POINTS}$).
These can serve as simplified selection criteria for identifying (and potentially excluding) problematic spectra. The number of fitted points can be used to identify spectra with substantial wavelength regions failing the pixel mask threshold set in Sect.~\ref{sc:Q1}. The reduced $\chi^2$ is a direct measure of how well \texttt{QSFit} has modelled the spectra and the different components. 

To support these selection processes, we follow \cite{2008ASPC..394..505S}'s definition of the S/N. This method has the advantage of being independent of the instrument used for the measurements, allowing better consistency in comparisons between \Euclid and other observations (such as DESI), while also allowing us to avoid problems inherent to the spectra themselves\footnote{For example, flux or uncertainties in \Euclid spectra files may show negative values.} and the associated uncertainty. The S/N is computed as the ratio of the median of the flux and the noise, with the noise defined as
\begin{multline}
N=(1.482602/\sqrt{6.0})\\
\times \mathrm{median}[\mathrm{abs}(2\,\mathrm{flux}_i-\mathrm{flux}_{i-2}-\mathrm{flux}_{i+2})]\, , 
\end{multline}
for all non-zero pixels $i$ in the spectrum.

To aid in the selection of good spectra, our results catalogue includes independent classification columns on the reliability of some components. These columns are named `<component>\_reliable', where ``<component>'' is placeholder for the fitting components detailed in Sect. \ref{sc:Fitting}, e.g., the QSO continuum, specific line, host galaxy template, etc., included in the fit.

\begin{figure}[htbp!]
\centering
\includegraphics[angle=0,width=1.0\hsize]{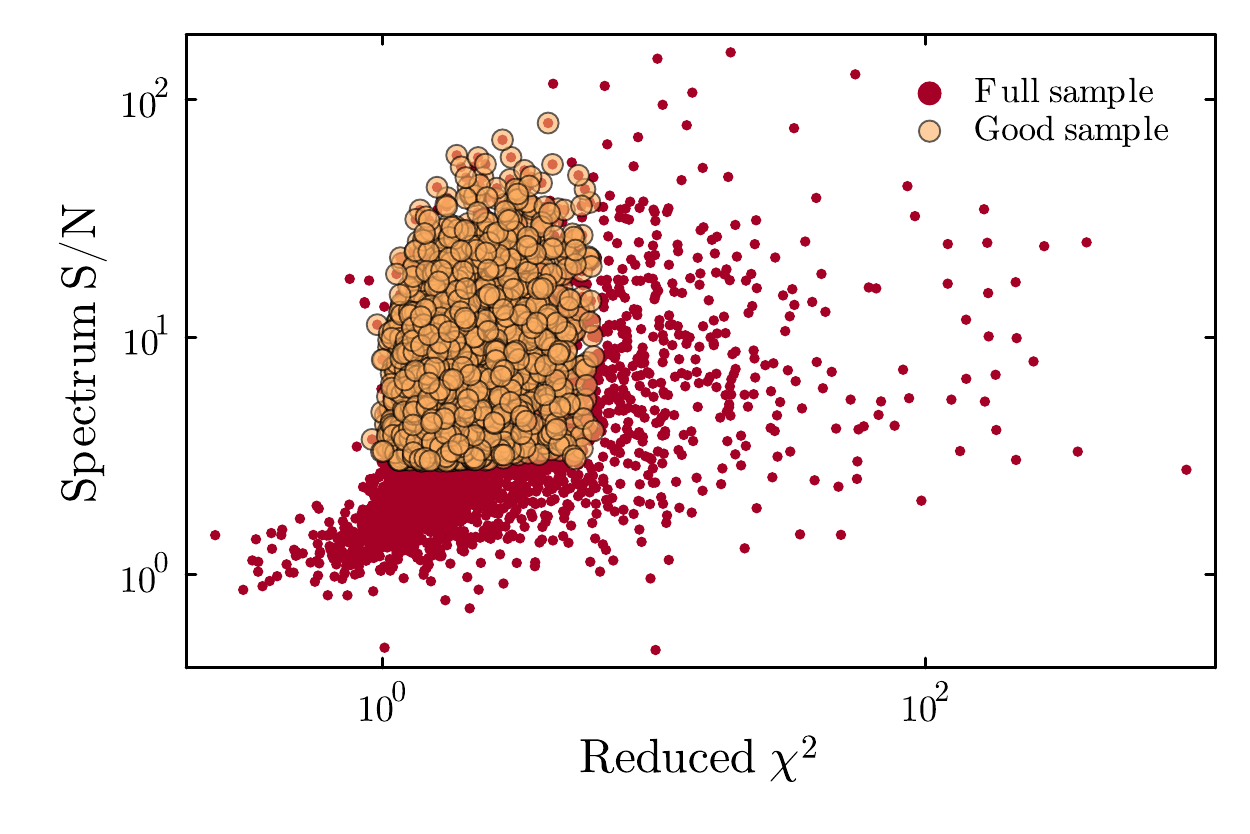}\\
\includegraphics[angle=0,width=1.0\hsize]{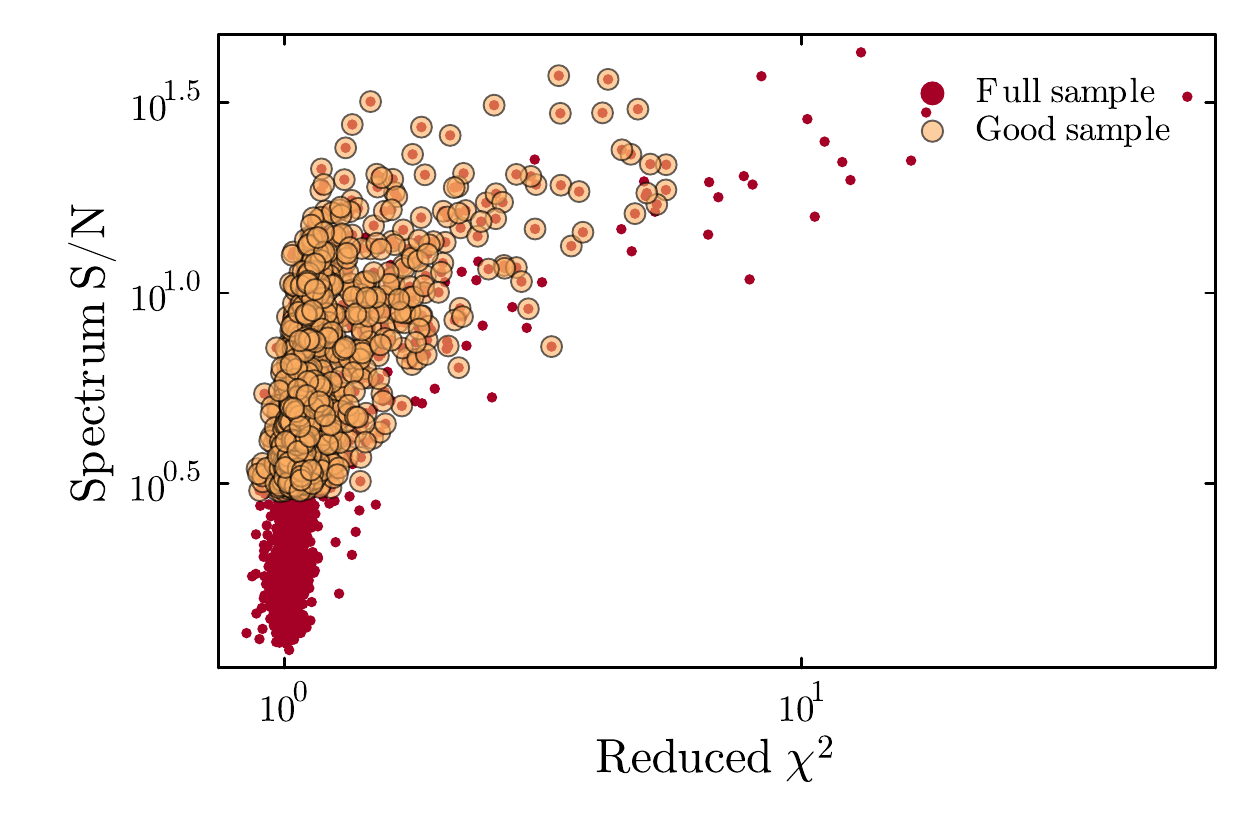}
\caption{\emph{Top panel}: Scatter plot of the $\rm (S/N)_{\rm spectrum}$ as a function the reduced $\rm \chi^2$ statistic (log scale) for the QSOs in this work. Applying the quality cut to the sample results in $\sim 45\%$ of the total selection being considered as `good'. \emph{Bottom panel}: As in top panel but for DESI spectra of QSOs with \ion{Mg}{ii} and \ha\, in DESI and \Euclid, respectively. The quality cut leads to $\rm 57\%$ of the total selection being considered as good.}
\label{fig:Chi2xSNR}
\end{figure}

We flag spectroscopic quantities as unreliable (and therefore untrustworthy) whenever the value measured for the component reached the extremes of the allowed range. Additionally, for certain components (like the luminosity values, which should always be positive), an uncertainty larger than the value measured for the component will be flagged as unreliable. This restriction does not apply to values like the velocity offset of a line or the QSO continuum slope.
The reliability classification is as follows:
\begin{itemize}
    \item A column with a reliability of value `1' indicates a reliable component.
    \item A column with reliability `0' may have untrustworthy measurements. 
    \item A column with reliability `$-1$' indicates that the component has not been fit for this particular source, because the component was not needed for the fitting, e.g., an emission line located outside the available rest-frame wavelength range for the fit.
\end{itemize}
\noindent
In addition to the reliability flag, we adopt an additional selection criterion, referred to in this paper as `quality cut', that selects sources meeting the following conditions: 
\begin{enumerate}
    \item $N_{\rm POINTS}>450$;
    \item $\rm (S/N)_{\rm spectrum}>3$;
    \item $\chi^2_{\mathrm{red}}<6$;
    \item the QSO continuum component must be flagged as `reliable';
    \item the number of points with negative flux values is less than $10\%$ of the total number of available points for fitting.
\end{enumerate}
The restrictions on the lowest number of fitting points is intended to filter out spectra with too many invalid pixels, while not excluding sources where the proximity of a line to the edge of the spectra causes \texttt{QSFit} to ignore that part of the wavelength range~\footnote{The maximum number of points available for fitting after masking is $468$. On average, the sources passing the quality cuts have $461$ points available for fitting.}. To select the values for the S/N and $\rm \chi^2_{\rm red}$ limits, we visually inspected the best-fit models and selected the values that allowed us to discard the most problematic spectra. Table \ref{tab:quality_cut} shows the effects of the selection constraints on the `final sample' (the AGN from the parent sample for which the fitting converged) used in this work. 
\begin{table*}
\caption{
The effects of the adopted `quality cut'. `Parent sample' refers to the initial $5489$ sources obtained from the individual catalogues listed in Sect.~\ref{sc:External_Cats}. `Final sample' refers to the sources of the parent sample that successfully converged while fitting (Sect.~\ref{sc:Final_sample}) and make up the effective number of sources analysed in this work. The `F26', `DESI', and `QUBRICS' subsamples refer to sources from the final sample present in these source catalogues.  We note that, although there are no repeated QSOs in the sample, a source can be in more than one source catalogue.
Approximately $55\%$ of the total fitted spectra for our \Euclid QSOs is rejected by the quality cut. The redshift ($z$) and \HE magnitude columns refer to the sources passing the quality cut. Also shown is the effect of applying the quality cut and reliability flag to the \ha\, and \hb\, lines.}
\centering
\begin{tabular}{ccccc}
\hline
\hline
    \multicolumn{1}{c}{Sample} & sample size & sources passing & $z$ & \HE\\
    \multicolumn{1}{c}{} &    & the quality selections &  & \\
\hline
\hline
    Parent sample & 5489 & -- & $0.004$--$6.6$ & $14.02$--$22.47$ \\
    Final sample  &  5387 &  2410 (44.8\%) & $0.011$--$4.76$ & $15.33$--$22.34$\\
    F26 & 3438 & 2017 (58.9\%) & $0.011$--$4.76$ & $15.33$--$21.39$\\
    DESI  & 2237 & 558 (24.9\%) & $0.12$--$4.12$ & $16.00$--$22.34$\\
    QUBRICS & 196 & 78 (39.8\%) & $0.12$--$3.22$ & $16.00$--$21.85$\\
    \ha\, sources & 2659 & 753 (28\%) & $0.91$--$1.84$ & $16.67$--$22.34$\\
    \hb\, sources & 1884 & 377 (20\%) & $1.57$--$2.83$ & $17.10$--$21.41$\\
\hline
\end{tabular}
\label{tab:quality_cut}
\end{table*}
Figure~\ref{fig:Chi2xSNR} shows the effect of the quality cut on the \Euclid and DESI samples, respectively, through $\rm (S/N)_{\rm spectrum}$ versus reduced $\chi^2$ plots. 
Figure \ref{fig:Bad_spectra} shows examples of two spectrum rejected by the quality cut, specifically due to high $\chi_{\mathrm{red}}^2$ and low $N_{\rm POINTS}$.

Our quality cuts show some degree of arbitrariness because it is not straightforward to predict all the ways in which a spectrum may be problematic. We therefore choose rather conservative cuts in an attempt to obtain a sufficient elimination of problematic spectra. However there still can be individual problematic sources passing all quality checks, as well as non-problematic sources being marked as not good due to a quality condition not being satisfied. Our catalogue includes all sources regardless of whether or not they pass the quality cuts, so the interested reader will have access to all details of even the discarded spectra. We also stress that the reliability flag and the quality cut are independent of each other. Furthermore, it is possible for a source to have an unreliable component (e.g. \hb\,line) while other components (e.g. the QSO continuum or \ha\, line) remain reliable. Therefore, in order to achieve the highest confidence in the results, one should select sources passing both the quality cut and the reliability flag appropriate for the component being inspected (Table.~\ref{tab:quality_cut} for an example of the selection criteria applied to the \ha\, and \hb\, lines).
From Sect. \ref{sc:Results} onwards, we present the results only for the sample subsets that pass both cuts.

\subsection{\label{sc:BH_quantities} Black hole mass estimation}
Under the assumption that the broad-line region cloud motion is virialised around the BH, that its brightness is proportional to its ionization state, and that the emission line width is a proxy for the velocity dispersion of the clouds, the virial product can be used as a proxy for the BH mass, based on the strong correlation between the size of the broad line region and the AGN continuum luminosity \citep[e.g.][]{2004ApJ...613..682P, 2006LNP...693...77P}.
We make use of the relation derived by \cite{Wu_2022} for the \hb\,and \ion{Mg}{ii} lines, which gives the BH mass as
\begin{equation}
    \logten \left(\frac{M_{\rm BH}}{\si{\solarmass}} \right)=a+b \logten  \left(\frac{\lambda L_\lambda}{10^{44}\,{\rm erg\,s^{-1}}}\right) + 2\logten\left(\frac{\rm FWHM}{\rm km\,s^{-1}}\right)\, ,
\end{equation}
where $(a,b)=(0.91, 0.5)$ and $(a,b)=(0.74,0.62)$ for \hb\,and \ion{Mg}{ii}, using the QSO continuum luminosity in the vicinity (at $\lambda=510$~nm and $\lambda=300$~nm, respectively) of these lines and the FWHM of their broad component. Although lines like \hb\,and \ion{Mg}{ii} are commonly used for such estimations, \ha\, can also be used, under the additional assumption that a line luminosity is a proxy for the total ionizing radiation \citep[][and references therein]{2012ApJ...753..125S, 2025A&A...699A.335B}. 
For our BH mass estimates using the \ha\ line, we apply the relation from \cite{2012ApJ...753..125S}
\begin{equation}
\begin{split}
   \logten \left( \frac{M_{\rm BH}}{\si{\solarmass}} \right)=2.216+0.564 \logten \left( \frac{L_{\rm H\alpha}}{10^{44} {\rm erg\,s^{-1}}} \right) \\
    + 1.821 \logten \left( \frac{\rm FWHM_{\rm H\alpha}}{\rm km\,s^{-1}} \right)\, .
\end{split}
\end{equation}

\noindent
To estimate the BH masses of sources with available Pa\,$\beta$, we follow \cite{2026arXiv260413170M} and take the relation from \cite{2015MNRAS.449.1526L}, scaled to a geometrical factor $f=5.5$ \citep[equivalent to a scale factor of $\epsilon=1$, ][and references therein]{Pucha_2025}

\begin{equation}
\begin{split}
   \logten \left( \frac{M_{\rm BH}}{\si{\solarmass}} \right)=7.94+0.872
   \left[ 0.5 \logten \left( \frac{L_{\rm Pa\beta}}{10^{40} {\rm erg\,s^{-1}}} \right) \right. \\
    \left. + 2 \logten \left( \frac{\rm FWHM_{\rm Pa\beta}}{\rm 10^4 \,\, km\,s^{-1}} \right) \right]\, .
\end{split}
\end{equation}

\noindent
Finally, for the \ion{He}{i} line mass estimations, we once again follow \cite{2026arXiv260413170M} and rescale the equation in \cite{2017A&A...598A..51R} from $L_{14-195\,\mathrm{keV}}$ to $L_{\mathrm{\ion{He}{i}}}$. We then use the relations from \cite{2017A&A...598A..51R} and \cite{2022ApJS..261....8R} to obtain the expression for the BH mass using the \ion{He}{i} FWHM as
\begin{equation}
\begin{split}
    \logten \left( \frac{M_\mathrm{BH}}{\si{\solarmass}} \right)=7.86+0.5 \left[\logten \left( \frac{L_{\mathrm{\ion{He}{i}}}}{\mathrm{erg \, s^{-1}}} \right)-39.55 \right]\\
    +2\logten \left(\frac{\rm FWHM_{\rm \ion{He}{i}}}{\rm 10^4 \, km \, s^{-1}} \right)\, .
\end{split}
\end{equation}
The use of single epoch estimators has an intrinsic uncertainty of $\sim$0.3--0.5 dex \citep[][]{2013BASI...41...61S}, which needs to be taken into account in addition to the uncertainties propagated from the quantities used in the estimation. In principle, one would expect the estimations from \hb\, and \ion{Mg}{ii} lines to be more reliable than \ha, Pa\,$\beta$ and \ion{He}{i}, simply because the last three calibrators require the additional assumption that the line luminosity is a proxy for the continuum luminosity. In practice, the assumptions are similar, since the determination of the line luminosity is subject to the same assumptions on the shape and strength of the continuum as $\lambda L_{\lambda}$. We therefore do not expect a particular mass estimator to be more reliable than the rest.
We show examples of the fitting of the lines used to estimate the BH masses in Fig.~\ref{fig:QSFIT_Ha_Hb_MgII}.

\subsection{\label{sc:Lbol}Bolometric luminosities and Eddington ratio estimation}
The bolometric luminosities of sources with \ha, \hb, and \ion{Mg}{ii} emission lines are estimated by following \cite{2017ApJS..228....9K}: we take the continuum luminosities at 510\,\text{nm} and 300\,\text{nm} and apply the corresponding luminosity corrections \citep[respectively $9.26$ and $5.15$, ][]{2006ApJS..166..470R, 2017ApJS..228....9K}. These calculations are based on the assumption that the overall spectral energy distribution (SED) shape is constant for all QSOs. This is supported empirically, but we cannot guarantee it is true for all cases. 

For sources with Pa\,$\beta$ and \ion{He}{i} line emissions, we estimate the bolometric luminosity by translating the Pa\,$\beta$ and \ion{He}{i} luminosities into hard X-ray luminosity and then applying a bolometric correction to obtain the bolometric luminosity.
We use the relations between Pa\,$\beta$ or \ion{He}{i} and the hard X-ray luminosity at $14$--$195 \, \mathrm{keV}$, which can be inferred from figure 10 of \cite{2022ApJS..261....8R} as
\begin{equation}
    \logten(L_\mathrm{line}/\mathrm{erg\,s^{-1}})=\log(L_\mathrm{14-195\, keV}/\mathrm{erg \, s^{-1}})-2.45\, ,
\end{equation}
which we convert to $L_\mathrm{2-10keV}$ assuming $\Gamma = 1.9$ as
\begin{equation}
    \logten(L_\mathrm{line}/\mathrm{erg\,s^{-1}})=\log(L_\mathrm{2-10\, keV}/\mathrm{erg \, s^{-1}})-2.13\, ,
\end{equation}
where $L_\mathrm{line}$ is the luminosity of Pa\,$\beta$ or \ion{He}{i}, and $L_\mathrm{2-10\, keV}$ is the 2--10\,keV luminosity. We then take the 2--10\,keV bolometric correction from \cite{2020A&A...636A..73D} to convert the $L_\mathrm{2-10keV}$ into bolometric $L_{\mathrm{bol}}$.
These steps are necessary because the $510 \, \text{nm}$ and $300 \, \text{nm}$ monochromatic luminosities used in the bolometric luminosities of sources with \ha, \hb, and \ion{Mg}{ii} emission are not only too far from the Pa\,$\beta$ and \ion{He}{i} lines to provide an accurate estimation, but they are also unavailable within the wavelength window of \Euclid for these two lines. The additional number of intermediate steps and assumptions required to estimate the bolometric luminosity for the Pa\,$\beta$ and \ion{He}{i} lines means that these quantities are subject to higher uncertainties \citep[of at least 0.3 dex;][]{2020A&A...636A..73D} than the remaining emission lines, so some care should be taken when comparing measurements from these lines with the remaining three.
We finally estimate the Eddington luminosities as
\begin{equation}
    \frac{L_{\rm Edd}}{{\rm erg \, s^{-1}}} =1.26 \times 10^{38}\left(\frac{M_{\rm BH}}{\si{\solarmass}} \right)
\end{equation}
and with the bolometric luminosities we calculate the Eddington ratios as
\begin{equation}
    \lambda_{\rm Edd}=\frac{L_{\rm bol}}{L_{\rm Edd}}\, .
\end{equation}

\subsection{\label{sec:composite_generation} \Euclid spectral composite analysis}
In order to validate our results and gain insight into the properties of the QSOs in our sample, we also consider a geometric mean composite of \Euclid spectra, and fit it with \texttt{QSFit} (Sect.~\ref{sc:composite_compare}). The geometric mean is chosen over a median or arithmetic mean because it preserves the shape of the continuum, allowing the spectrum to be approximated by power laws. The composite is generated following the same process as \cite{2001AJ....122..549V} and F26, with the only difference that we use a constant sampling resolution of $R=3000$, rather than a constant sampling in wavelength, to preserve the information available at all rest-frame wavelengths. The sampling resolution of $R=3000$ is higher than the \Euclid native one $\sim$~450 to exploit the additional information gained with the dithering in redshift. Also, we chose this sampling resolution value since we found that there is no advantage in using higher values, and the only way to achieve higher resolutions would be to enlarge our sample.
We also consider only the sources up to $z=4$ to avoid increasing the noise on the blue edge of the spectrum due to the low number of sources at higher redshifts. We then use the geometric mean composite for the purposes of comparison with our QSO continuum slopes.
We take the standard deviation divided by the square root of the number of spectra at any given wavelength as the uncertainties for the fit. The composite does not have the same number of sources across its wavelength range, with the highest count around the \ha\, range and the lowest at the edges of the spectra.

The composite spectra is used to identify which spectral lines to use in our analysis of the individual spectra. The transitions are therefore the same for the analyses of the composite and the individual sources; however, there are some key differences:
\begin{itemize}
    \item We include narrow components in the model for the composite, due to the higher resolution from the redshift dithering during its construction.
    \item The QSO continuum for the composite was fitted using a smoothly-broken power law as opposed to the simple power law used in the individual spectra. This is required due to the large wavelength range covered by the composite (spanning rest-frame optical to NIR wavelengths).
    \item The fitting of the composite is done with the contribution of the host galaxy included by default. This is in contrast with the individual source fitting where the fit with the best reduced $\chi^2$ (with or without the inclusion of a host galaxy contribution) is chosen.
\end{itemize}
In any other aspect, the analyses of the composites and the individual spectra are identical.

\subsection{Final sample}
\label{sc:Final_sample}

\begin{figure}[htbp!]
\centering
\includegraphics[angle=0,width=1.0\hsize]{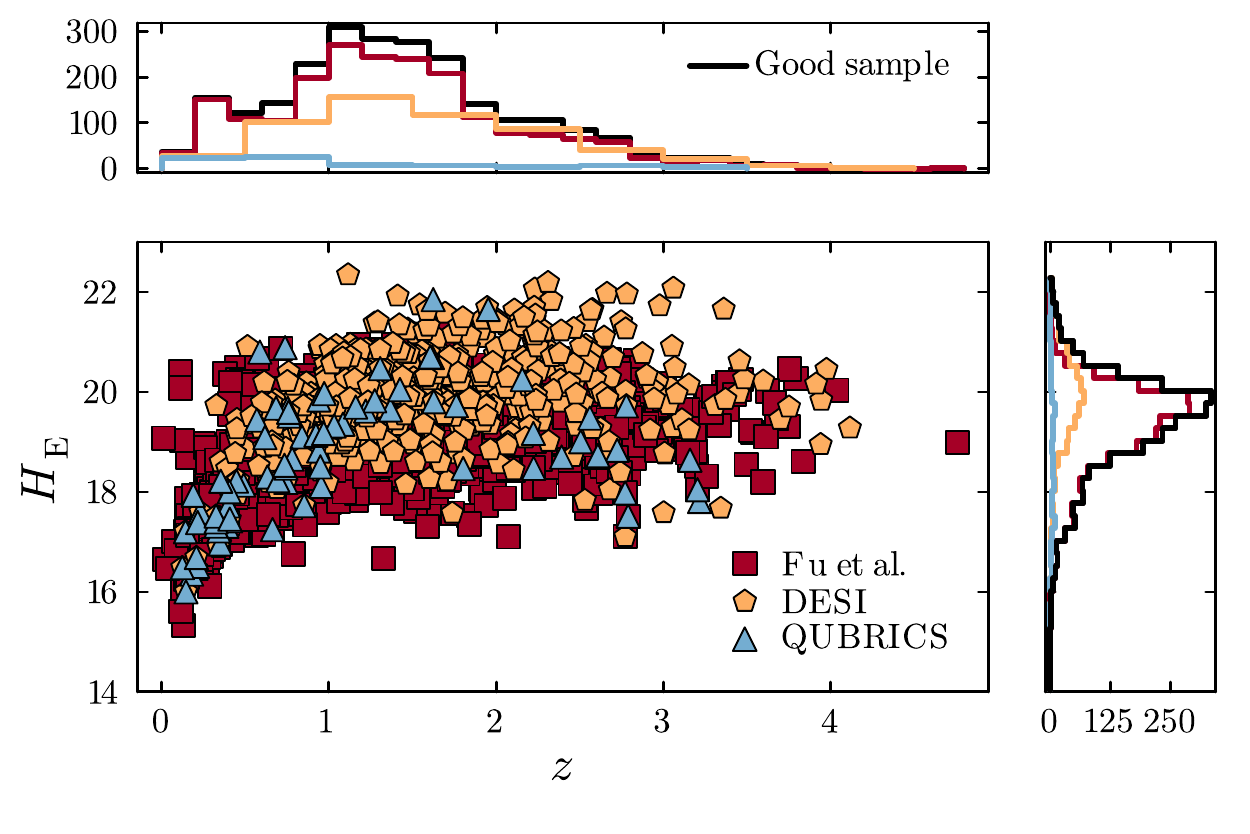}
\caption{The \HE magnitudes of our QSO sample (from \Euclid MER fluxes; \citealt{Q1-TP004}) as a function of redshift. Symbols indicate the parent catalogue: red squares (F26), yellow pentagons (DESI), and blue triangles (QUBRICS). The top and right margin plots show the histograms for the redshift and \HE magnitude of the entire good sample and the sources from each parent catalogue. The highest-redshift QSOs come from F26, while DESI generally includes the faintest sources. If a source appears in multiple catalogues, it is assigned to only one (Sect. \ref{sc:External_Cats}).}
\label{fig:H_mag_redshift_scatter_hist}
\end{figure}

As described in Sect. \ref{sc:External_Cats}, using data from DESI, QUBRICS, and the visually selected catalogue of F26, we compile a sample of $5489$ QSO spectra, which we process using \texttt{QSFit}.
During the analysis, $102$ spectra ($1.9\%$ of the total sample) failed to produce converging fits, resulting in a total of $5387$ analysed QSOs, with $3438$ present in F26, $2237$ in DESI and $196$ in QUBRICS (some sources are found in more than one catalogue).
Of the total number of sources in the parent catalogue, $45\%$ pass the quality cut (Sects. \ref{sc:Euclid_spectra} and \ref{sc:Statistics_Parameters}, as well as Table~\ref{tab:quality_cut}).  
Most of the excluded sources populate the fainter end of the \HE values, where issues with the S/N and artefact contamination are more prevalent. The faintest sources are from the DESI source catalogue while the brightest ones are from F26 (Fig. \ref{fig:H_mag_redshift_scatter_hist} and Table~\ref{tab:quality_cut}).
Applying our quality cut reduces the total number of sources, narrowing the redshift range to $0.01<z<4.76$, with sources from F26 reaching the highest redshifts (Fig.~\ref{fig:H_mag_redshift_scatter_hist}, top margin histogram).
The majority of our good sources ($86\%$) show $\HE<21$, with the average magnitude of the good sample being $\langle \HE \rangle = 19.4 \pm 1.0$. This is primarily driven by two factors: the catalogues used, particularly the visually-confirmed catalogue (F26), in which approximately half of the sample targets have $\HE \leq 21.3$, and limitations of the Q1, which only provides spectra for sources with $\HE<22.5$.
Our good sources include  $58.9\%$ of F26's visually confirmed QSOs. Most of these exclusions are driven by the impositions of our quality cut, with only 3\% being due to rejection by \texttt{QSFit} during the fitting process.
\noindent

\section{\label{sc:Results}Results and discussion}
We divide our results and discussion in the following way: we first discuss the QSO continuum slope of the entire good sample, then proceed to analyse the emission line fitting results, focusing on luminosity and FWHM.
Afterwards, we discuss the bolometric luminosities and BH masses. In the final subsections, we compare the BH masses estimated for the good sample with SDSS and discuss the \Euclid composite slopes. Although our final catalogue includes all sources ($5387$) for which we could fit the data, in the following discussion we consider only the ones that pass the quality and reliability flags ($2410$), as these are the most trustworthy results. Throughout the discussion, the values quoted in reference to our sample refer to the median or mean of a given quantity and their respective median absolute deviations or standard deviations, unless otherwise stated.

\subsection{\label{sc:QSOalpha_Disc} QSO continuum slope}
\begin{figure}[htbp!]
\centering
\includegraphics[angle=0,width=1.0\hsize]{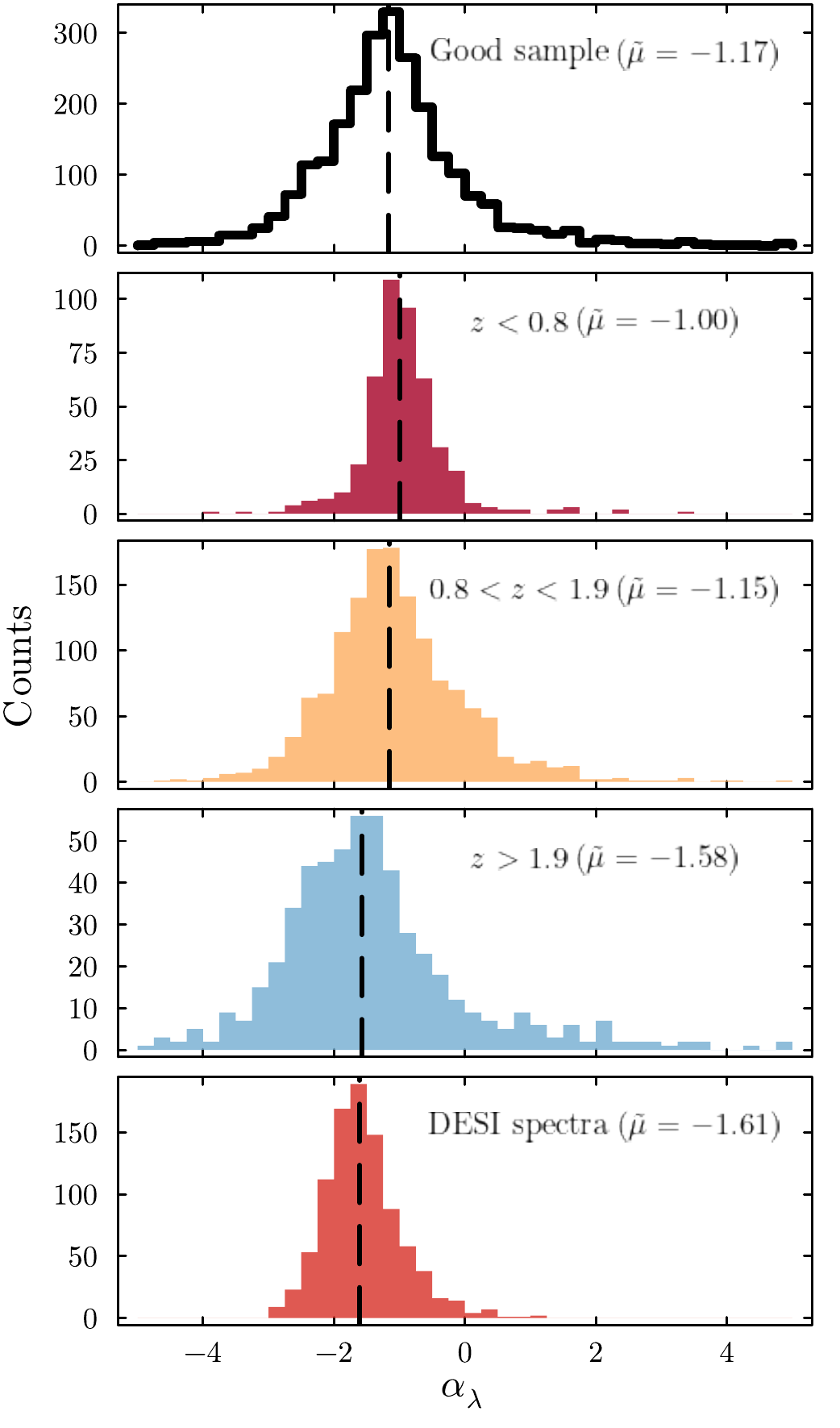}
\caption{Histograms of the QSO continuum slopes $\alpha_{\lambda}$ estimated from \Euclid spectra, in different redshift bins. The bottom panel shows the histogram obtained from DESI spectra.  Vertical lines mark the medians of the values in each bins. In the legend, $\tilde{\mu}$ refers to the median value. Table \ref{tab:QSO_slopes} shows this information as well as the redshift range and average bolometric luminosity of each sample.}
\label{fig:QSOcont_Redshift}
\end{figure}

The histograms of the QSO continuum spectral slopes, as estimated from the \Euclid spectra, are shown in Fig.~\ref{fig:QSOcont_Redshift}. The redshift range, mean bolometric luminosity and wavelength is shown in Table \ref{tab:QSO_slopes}. The black solid line histogram refers to the entire sample.
\begin{table*}
\caption{QSO continuum slopes from this work and from the literature.}
\centering
\begin{tabular}{cccccc}
\hline
\hline
    {Reference} & sample size & rest-frame wavelength $(\text{nm})$ & $\alpha_{\lambda}$ & $\logten{({L_{\mathrm{bol}}/\mathrm{erg\,s^{-1}}})}$& $z$\\
\hline
\hline
    parent \Euclid sample (\Euclid spectra) & $5387$ & $217$--$1830$ & $-1.13$ & $46.1$ & $0.011$--$4.76$ \\
    good \Euclid sample (\Euclid spectra) & $2410$ & $217$--$1830$ & $-1.17$ & $46.2$ & $0.011$--$4.76$ \\
    good \Euclid sample (\Euclid spectra) & $457$ & $695$--$1830$ & $-1.00$ & $46.2$ & $0.011$--$0.8$ \\
    good \Euclid sample (\Euclid spectra) & $1414$ & $431$--$1028$ & $-1.15$ & $46.2$ & $0.8$--$1.9$ \\
    good \Euclid sample (\Euclid spectra) &$539$ &  $217$--$638$ &  $-1.58$ & $46.5$ & $1.9$--$4.76$ \\
    good \Euclid sample (DESI spectra) & $851$ & $103$--$727$ & $-1.61$ & $45.7$ & $0.35$--$2.5$\\
    \cite{2001AJ....122..549V} & -- & $90$--$580$ & $-1.56$ & -- & $0.044$--$4.789$\\
    \cite{Selsing2016} & -- & $100$--$1135$ & $-1.7$ & -- & $1$--$2.1$\\
\hline
\end{tabular}
\label{tab:QSO_slopes}
\end{table*}
The median spectral index is also shown, with $\alpha_{\lambda}= -1.17$ with a normalised MAD of $0.83$.
When considering the subsamples in bins of redshift, i.e. $z<0.8$, $0.8<z<1.9$, and $z>1.9$, the histograms span different values, showing that the measured slopes depend on the rest-frame wavelength range available (which ultimately depends on redshift).  More specifically, we observe steeper slopes with increasing redshifts, with median values in the considered redshift bins of, respectively, $\alpha_{\lambda}=-1.00\pm0.43$, $-1.15\pm0.87$, and $-1.58\pm1.04$ (Fig.~\ref{fig:QSOcont_Redshift}, 
see also Sect.~\ref{sc:composite_compare} for the slopes obtained through the geometric mean composite).

We use a sample of sources with DESI spectra spanning the redshift range $0.35<z<2.5$ (rest-frame wavelength range of $103\, \text{nm}<\lambda<727\,\text{nm}$) to test the QSO continuum slopes obtained from \Euclid spectra. The histogram of the QSO continuum slopes measured on the DESI sample is shown in Fig.~\ref{fig:QSOcont_Redshift} (bottom panel).
The median for the DESI spectra is $\alpha_{\lambda}=-1.61\pm0.51$, which is consistent with the median measured for the \Euclid sample for redshifts $z>1.9$ ($\alpha_{\lambda}=-1.58\pm1.04$). 
The reason for this similarity lies in the similar rest-frame wavelength ranges for DESI and \Euclid constrained to $z>1.9$ (of $217\,\text{nm}< \lambda<661\,\text{nm}$). 
This provides confirmation that the measured slopes depend on the observed rest-frame wavelength range, and explains why our measurements on \Euclid and DESI spectra are in agreement with each other. 

The change of the QSO continuum slope with redshift has also been observed by \cite{Shankar_2016}, when they analysed a sample of SDSS QSOs at $1.0<z<1.2$, also using \texttt{QSFit}. The authors find a similar steepening of the QSO continuum with redshift, but also find a change in the QSO slope depending on the mass of the SMBHs, with the steeper slopes being found for the highest SMBH masses (their Fig. 1). We took the median of our QSO continuum slopes at the redshift range of $0.8<z<1.2$ and found that, similarly to the results of \cite{Shankar_2016}, the QSO slopes become steeper with the BH mass (from $\alpha_{\lambda}\sim-0.9$ to $\alpha_{\lambda}\sim-1.3$). We stress that the change of slope with redshift bins is not evidence of an evolution with time, only a dependence on the wavelength window used to estimate the QSO continuum. Furthermore, at these redshifts, there is the presence of the host galaxy to take into account, which may influence the shape of the measured continuum. We therefore cannot guarantee that the change in the measured slopes are due to changes in the thermal continuum of our QSOs.

We further compare our results with the continuum slopes available in the literature and summarise the results in Table~\ref{tab:QSO_slopes}.
\cite{2001AJ....122..549V}, who built a geometric mean composite covering the rest-frame wavelength range \mbox{$\lambda=90$--580~nm}, find a QSO continuum slope of $\alpha_{\lambda}=-1.56$ for wavelengths between the Ly$\alpha$ and \hb\,lines. This is closer to the median slope obtained from our own sources at $z>1.9$ ($\alpha_{\lambda}=-1.58$, for rest-frame $217\, \text{nm}<\lambda<638 \, \text{nm}$).
We note that, similarly to our measurements, \cite{2001AJ....122..549V} also find a changing QSO continuum slope with rest-frame wavelength probed by observations. Specifically, they find that the composite's spectral index changes sign redward of the $500 \, \text{nm}$ region, from $\alpha_{\lambda}=-1.56$ to $0.45$. This change is attributed to the effects of contamination from the host galaxies and, to a lesser extent, hot dust emission. Our own change of the QSO continuum slope is not so abrupt, as we do not observe a change in the slope sign for sources at lower redshifts, but it is likely that the changes in the slopes observed in our sample are due to similar contamination (see also Sect. \ref{sc:composite_compare}). 
The influence of the host galaxy light is a possibility that is strengthened by the results of \cite{Selsing2016}. They used a composite spectrum generated from data taken with \mbox{X-Shooter} at the VLT for QSOs at $1<z<2.1$ and, unlike the composite of \cite{2001AJ....122..549V}, the work of \cite{Selsing2016} is based on a sample that is assumed to have negligible host galaxy contribution. They find no evolution in their QSO continuum slope, obtaining an average of $\alpha_{\lambda}=-1.7$. Furthermore, the authors also fit the composite with a broken power law, which leads to consistent results of $\alpha_{\lambda}=-1.69$ and $-1.73$ for rest-frame wavelength regions of $\lambda<500 \, \text{nm}$ and $\lambda>500 \, \text{nm}$, respectively. This is similar to the median slope for our $z>1.9$ sources.
While the values reported in \cite{Selsing2016} are derived from a weighted mean composite, the authors test the fitting of the QSO continuum on several different combination methods, including a geometric mean composite spectrum. The QSO continuum slopes obtained by \cite{Selsing2016} are consistent across all methods, including the mean from the individual spectra and so their results should be comparable.

Our results are at odds with the literature in regards to a single power law being a good fit for the entire wavelength range of our sample, as seen through the variation in $\alpha_{\lambda}$ observed with the changing wavelength range. 
However, a proper comparison at longer wavelengths is difficult due to the lack of large quasar samples with published NIR rest-frame spectral data; most quasar studies still focus on the UV and optical ranges.
In this sense, \Euclid represents an improvement and a broadening of the available wavelength range, and this without the Earth's atmospheric absorption windows that plague ground-based NIR observations. Furthermore, for low-redshift sources, our spectral slopes are based almost entirely on the rest-frame NIR wavelength range, which exhibit flatter slopes due to the above mentioned host contribution. This could explain the discrepancies observed between slope measurements based on different rest-frame wavelengths. 

\subsection{\label{sc:overall_line_emission} Emission line properties}
\begin{figure}[htbp!]
\centering
\includegraphics[angle=0,width=1.0\hsize]{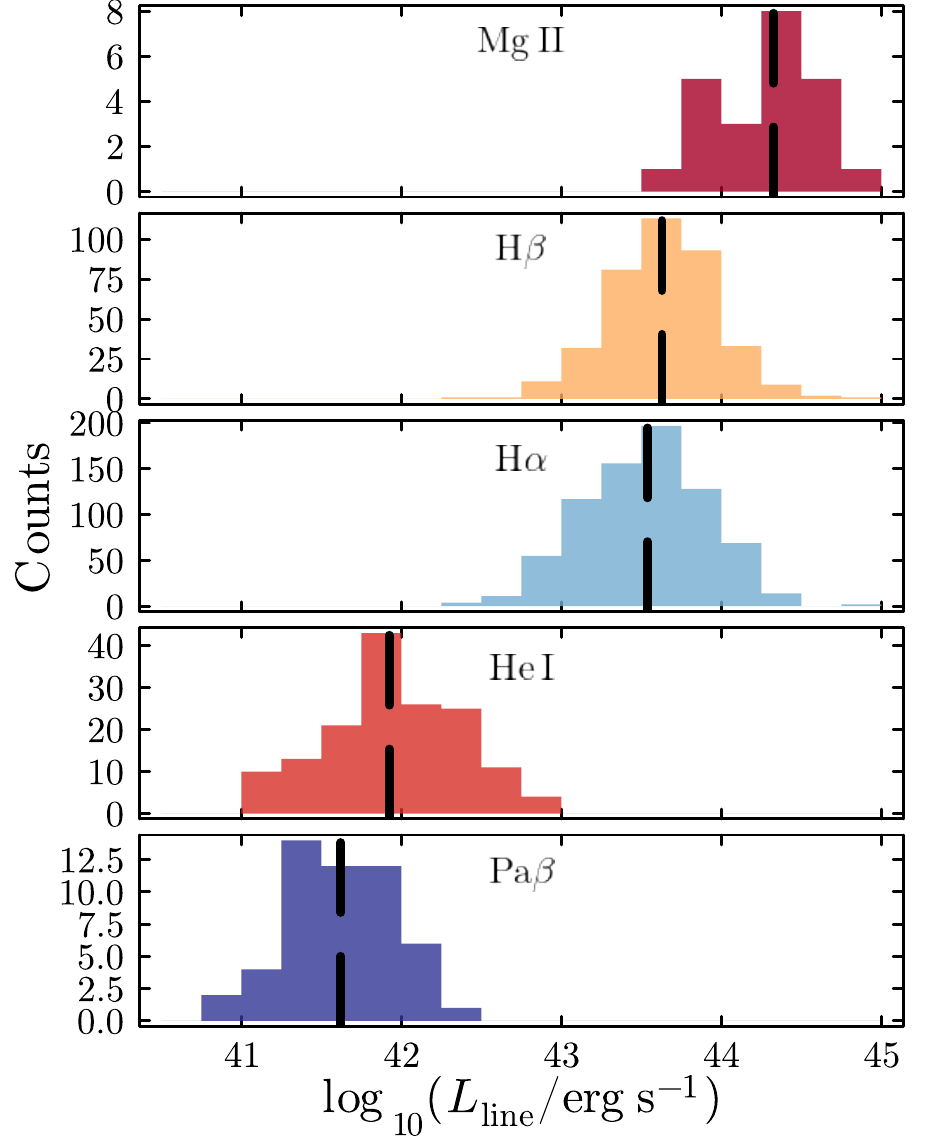}
\caption{Line luminosity histograms for Pa\,$\beta$, \ion{He}{i}, \ha, \hb, and \ion{Mg}{ii}.}
\label{fig:Lines_Lum}
\end{figure}

In this work, we provide measurements of the Pa\,$\beta$, \ion{He}{i},  \ha, \hb, and \ion{Mg}{ii} lines (Sect.~\ref{sc:Fitting} for the fitting procedure) for sources passing both the quality cut and the reliability flag for the respective line (Sect.~\ref{sc:Statistics_Parameters} -- see also Fig.~\ref{fig:QSFIT_Ha_Hb_MgII}, for an example of the fitting of these lines). The results of the fitting are summarised in Table \ref{tab:QSO_quantities_summary}.
\begin{table*}
\caption{Summary of the mean values and respective standard deviations of the measured line luminosities, FWHM, BH masses, and logarithmic Eddington ratios for the sources in our sample. The values are valid only for sources passing both the quality cut and the reliability flag described in Sect.~\ref{sc:Statistics_Parameters}.}
\centering
\begin{tabular}{cccccccccc}
\hline
\hline
    \multicolumn{1}{c}{Emission line} & 
number of sources
    & $\logten(L_{\rm line} / \rm erg \, s^{-1})$ & $\logten(\mathrm{FWHM}/\rm km \, s^{-1})$ & $ \logten(M_{\rm BH} / \si{\solarmass})$ & $\logten(\lambda_{\mathrm{Edd}})$\\
\hline
\hline
    Pa\,$\beta$ & $51$ & $41.6\pm0.3$ & $3.61\pm0.30$ &  $8.0\pm0.6$ & $-1.0\pm0.4$\\
    \ion{He}{i} & $153$ & $41.9\pm0.4$ & $3.45\pm0.25$ &  $8.0\pm0.6$ & $-0.72\pm0.5$\\
    \ha\,& $753$ & $43.5\pm0.4$ & $3.70\pm0.20$ & $8.7\pm0.5$ & $-0.8\pm0.4$\\
    \hb\,& $377$ & $43.6\pm0.3$ & $3.67\pm0.19$ &  $9.0\pm0.4$ & $-0.7\pm0.4$\\
    \ion{Mg}{ii} & $23$ & $44.3\pm0.3$ & $3.54\pm0.14$ & $9.1\pm0.3$ & $-0.4\pm0.3$\\
\hline
\end{tabular}
\label{tab:QSO_quantities_summary}
\end{table*}
Figure~\ref{fig:Lines_Lum} shows the histograms of the luminosities for the five lines considered here.
All luminosity histograms present a Gaussian-like shape. The number counts for \ion{Mg}{ii} and Pa\,$\beta$ lines are markedly lower than the remaining lines. The Pa\,$\beta$ and \ion{He}{i} lines have clearly lower luminosities than \ha, \hb, and \ion{Mg}{ii}, possibly due to the lower redshifts of the respective sources. 

We compared our line luminosities (Table \ref{tab:QSO_quantities_summary}) to the ones obtained in SDSS from \cite{Wu_2022}, who find mean logarithmic line luminosities of $\logten(L_{\mathrm{\ha}}/\mathrm{erg\,s^{-1}})=42.5$, $\logten(L_{\mathrm{\hb}}/\mathrm{erg\,s^{-1}})=42.7 $, and $\logten(L_{\mathrm{\ion{Mg}{ii}}}/\mathrm{erg\,s^{-1}})=43.3$. On average, our \Euclid sources sample the bright end of SDSS AGN. A possible explanation for this discrepancy is that, by filtering out sources with low S/N, we reject lower luminosity sources and thus probe preferentially the bright end of SDSS (see also Sect.~\ref{sc:SMBH_masses_Lbol_discussion}, for a comparison of the bolometric luminosities).

\begin{figure}[htbp!]
\centering
\includegraphics[angle=0,width=1.0\hsize]{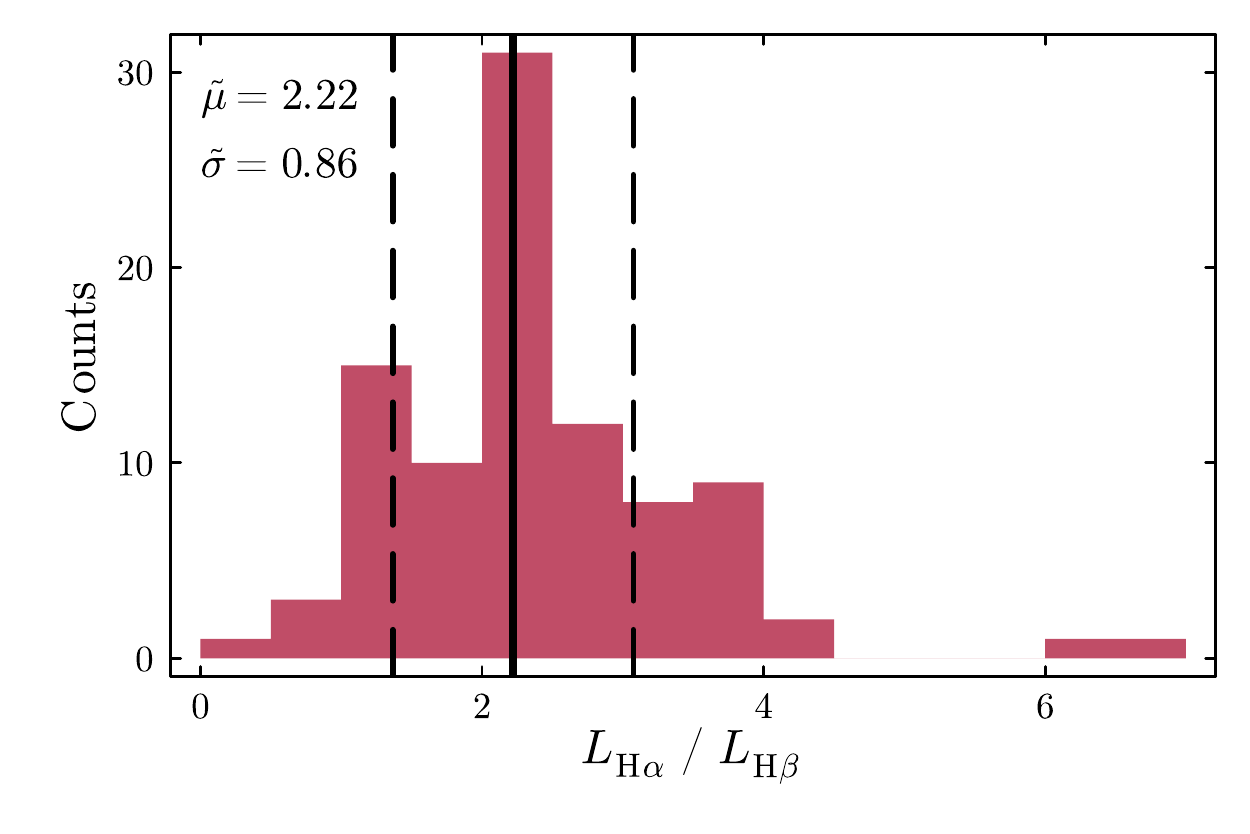}
\caption{Histogram of the ratio between the luminosities of the \ha\, and \hb\,emission lines (the Balmer decrement) from \Euclid spectra, for sources in our sample where both lines are available.}
\label{fig:HaHb_ratio}
\end{figure}

We also compared the relative strengths between the \ha\, and \hb\,lines in our analysis, for sources with both emission lines (93 sources in total), by estimating the ratio between their luminosities. This resulted in a median ratio of $2.22\pm0.86$ and a slightly higher mean ratio of $2.4\pm1.0$ (Fig.~\ref{fig:HaHb_ratio} shows the histogram of this ratio). 
\noindent
This ratio is the Balmer decrement (BD) and both our standard deviation and normalised MAD estimates' of this quantity place it below the values usually accepted for case B recombination \citep[$2.7$--$3.0$; ][]{1971MNRAS.153..471B}. The BD of QSOs is generally expected to be larger than the case B value ($3.1$; see \citealp[]{1977ApJ...215..733O}, \citealp[]{1984PASP...96..393G} and references therein). However, in \cite{2005ApJ...620..629D} and references therein, the BD has been found to be in the range of $2.5$--$5$. We estimated the \ha\,/\hb\,ratio by using an arithmetic mean composite of our sources, resulting in a BD of $\sim 3.04$, much more in line with what is expected from QSOs. 
We acknowledge that our BD values may be biased due to the low spectral resolution, which lead us to use just a single broad component in our fitting. We may also be overestimating of fluxes for the \hb\,line. There have been instances where the \hb\,region has shown an enhancement redwards of the \hb\,line, due to possible contamination by e.g. \ion{Fe}{ii} and \ion{He}{i}\,$\lambda492.2$ (the so-called `red shelf', e.g., \citealp[]{1996ApJS..104...37M}, \citealp[]{2008ApJS..174..282L}), which would also imply an overestimation of our \hb\, BH masses. Although we use an iron template to account for these contributions, the process is not straightforward, due to the difficulty in separating emission from \hb\,from the blended iron lines. Additionally, strong narrow line region outflows may also be a factor, leading to shifted \ion{[O}{iii]} blue wings that lead to further contamination. 
The BD has also been shown to have an anti-correlation with continuum flux \citep[][]{1990ApJ...354..446W, 2004A&A...422..925S}. 
We found no correlation between the BD and the QSO continuum luminosity of our sources. We likewise find no correlation between the BD and the continuum slope.
However, it is worth noting that the QSOs in our sample with both \ha\, and \hb\,emission that pass the quality and reliability flags are few (93), so our statistics suffer from low number counts. The range for the continuum luminosity covered by these sources is also small ($\sim 1$ dex) so it is likely that any correlation that might exist is simply not visible.
Finally, the BD can vary significantly, being prone to variability, sometimes on time-scales of years or months \citep[e.g., figure 16 of][]{2011A&A...528A.130P}, and within single sources, and depends on the geometry of the broad line region and the presence of additional factors like outflows \citep{2010A&A...509A.106S}.

\begin{figure}[htbp!]
\centering
\includegraphics[angle=0,width=1.0\hsize]{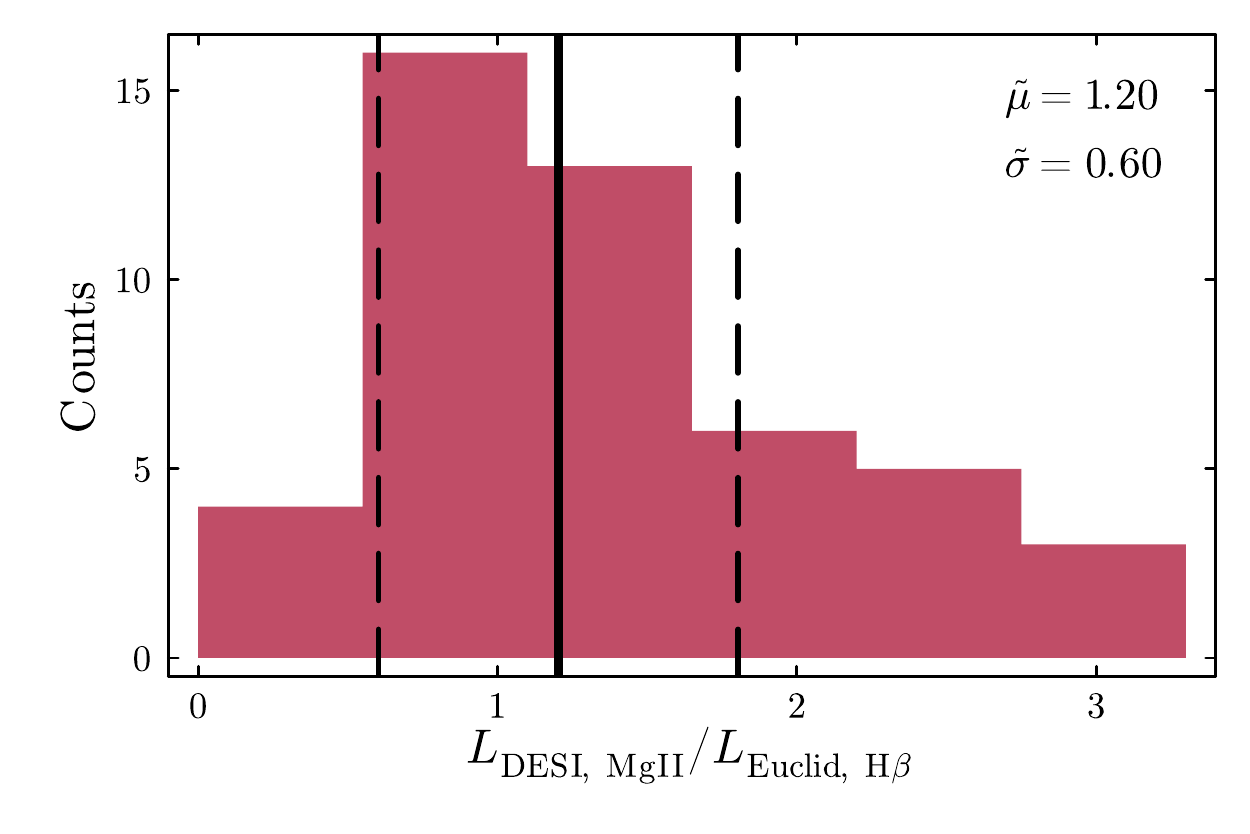}
\caption{Histogram of the ratio between the luminosities of the \ion{Mg}{ii} emission line from DESI spectra and the \hb\,line from \Euclid spectra, for sources in our sample where both spectra are available.}
\label{fig:HbMgII_ratio}
\end{figure}

Another point of comparison is the relative strength of the \ion{Mg}{ii} and \hb\,lines. There is evidence that the broad line components of the \hb\,and \ion{Mg}{ii} lines originate in the same emission region 
\citep{2010MNRAS.409.1033M}. Additionally, \ion{Mg}{ii} is often used as a stand in for the Balmer lines in the estimation of SMBH masses, when the latter are not available. Such comparison is not straightforward for us, as we do not have QSOs with both emission lines in their \Euclid spectra. However, we can compare the \hb\,line luminosity estimate from the \Euclid spectrum and the \ion{Mg}{ii} line from the corresponding DESI spectrum. A total of 50 sources, passing the quality cut and the \hb\,and \ion{Mg}{ii} reliability flags, are available to calculate such line ratios (Fig~\ref{fig:HbMgII_ratio}). 
We obtain a median ratio of 
$(L_{\mathrm{DESI, MgII}}/L_{\Euclid\mathrm{, \hb}})=1.2\pm0.6$.
From the line flux measurements from \cite{2001AJ....122..549V} in their median QSO composite spectrum, we obtain a ratio between the \ion{Mg}{ii} and \hb\,lines of $\sim 1.7$. 
A similar result has also been found by \cite{1991ApJ...373..465F} in their arithmetic mean QSO composite, where the \ion{Mg}{ii} line is $\sim 1.55$ times stronger than \hb. A similar value (of $\sim 1.5$) is also found by \cite{2010MNRAS.409.1033M}.
At first glance, our results are therefore comparable with those of the literature. However, we need a larger sample and better S/N spectra in order to further tighten our constraints.

\begin{figure}[htbp!]
\centering
\includegraphics[angle=0,width=1.0\hsize]{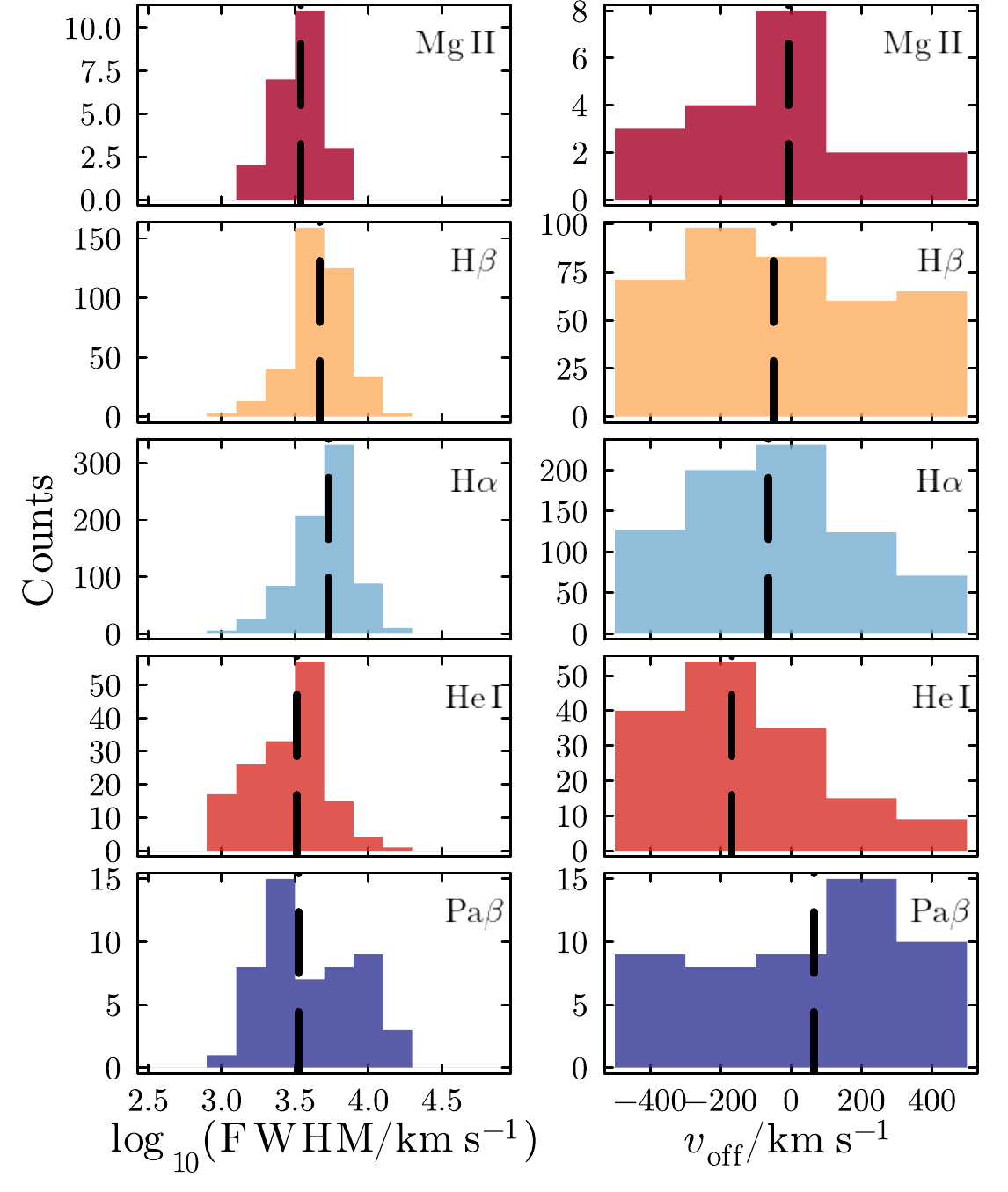}
\caption{\emph{Left panels}: Histograms of FWHM for the five emission lines considered in this work. \emph{Right panels}: Histograms for the
velocity offset associated with each of the five lines.}
\label{fig:FWHM_Voff_Hist}
\end{figure}

We show histograms for the FWHM of \hb, \ion{Mg}{ii}, Pa\,$\beta$, and \ion{He}{i} in Fig.~\ref{fig:FWHM_Voff_Hist}.
Inspection of the FWHM estimates of the \ha\, line shows that the majority of sources have $\mathrm{FWHM_{\rm H\alpha}}$ values within the expected range of $\rm 3 <\logten(FWHM_{\rm H\alpha}/\mathrm{km\,s^{-1}}) < 4$.  
Values for FWHM higher than this range are less reliable, as they may signal problems with the fitting such as confusing continuum with a line, or fitting outflowing components or unexpected emission in the line vicinity as if coming from the line itself. Although there is a `wing' that extends to the extreme values of $\rm \logten(FWHM_{\rm H\alpha}/km \, s^{-1}) > 4$ (Fig. \ref{fig:FWHM_Voff_Hist}), the overall number of sources with such $\mathrm{FWHM_{\rm H\alpha}}$ values is small ($<4\%$), so we expect the majority of the measurements to be trustworthy.
Our \hb\,FWHM estimates are also distributed in a Gaussian-like shape. A small number of sources ($<3\%$) likewise shows $\rm \logten(FWHM_{\rm \hb}/\mathrm{km\,s^{-1}}) >4$. It is possible that these higher values for the \hb\,FWHM are due to unaccounted for additional components present in the $486$--$500\, \text{nm}$ wavelength region, due to the same `red shelf' that may influence the estimation of the BD. 
The \ion{He}{i} and Pa\,$\beta$ histograms show a similar feature. The \ion{Mg}{ii} line histogram may as well, but the low number counts make it unclear if an extension towards higher or lower values is also present. It is possible that the enhancement of the FWHM histogram of the Pa\,$\beta$ line at higher values is also due to unaccounted elements in the Pa\,$\beta$ region, particularly a \ion{Fe}{ii} emission line at $\sim1257\,\text{nm}$. Finally, there is the possibility that these high FWHM sources, for all lines considered, are affected by low-quality spectra that nevertheless pass our quality and reliability flags. Values of $\logten(\mathrm{FWHM/km\, s^{-1}})>4$ should therefore be considered carefully, even when passing the quality and reliability flags.

Comparing the FWHM of the Balmer lines for the sources with both \ha\, and \hb\,emission, we obtain a mean (median) $\logten(\mathrm{FWHM}_{\mathrm{H}\alpha}/\mathrm{km\,s^{-1}})= 3.77\pm0.15$ ($3.76\pm0.13$) and $\logten(\mathrm{FWHM}_{\mathrm{H}\beta}/\mathrm{km\, s^{-1}})= 3.67\pm0.18$ ($3.69\pm0.15$) corresponding to linear means (medians) of $\mathrm{FWHM}_{\mathrm{H_\alpha}}= (5559 \pm 2331) \, \mathrm{km \, s^{-1}}$ [$\mathrm{FWHM}_{\mathrm{H_\alpha}}= (5381 \pm 2332) \, \mathrm{km \, s^{-1}}$] and $\mathrm{FWHM}_{\mathrm{H_\beta}}= (5054 \pm 2123) \, \mathrm{km \, s^{-1}}$ [$\mathrm{FWHM}_{\mathrm{H_\beta}}= (4692 \pm 1747) \, \mathrm{km \, s^{-1}}$]. This results in a mean difference of $\lesssim 1211 \, \mathrm{km\,s^{-1}}$ ($\lesssim 857 \, \mathrm{km\,s^{-1}}$) between our \ha\, and \hb\,lines (taking the larger logarithmic values as reference), in agreement with the range afforded by the standard deviations and normalised MADs of the FWHM of both lines.

Also shown in Fig.~\ref{fig:FWHM_Voff_Hist} are the histograms for the velocity offsets $v_\mathrm{off}$ of our lines. Unlike the FWHM histograms, with clearly defined peaks, the $v_\mathrm{off}$ histograms are much flatter. This is probably due to instrumental limitations: \Euclid's spectral resolution of $R=450$ corresponds to $\sim 670 \, \mathrm{km \, s^{-1}}$. Since we restrict the allowed values for the $v_\mathrm{off}$ to the range of $|v_\mathrm{off}| < 450 \, \mathrm{km \, s^{-1}}$, it is possible that this is why we do not get such peaks. 

\subsection{\label{sc:SMBH_masses_Lbol_discussion} SMBH masses and bolometric luminosities}
\begin{figure}[htbp!]
\centering
\includegraphics[angle=0,width=1.0\hsize]{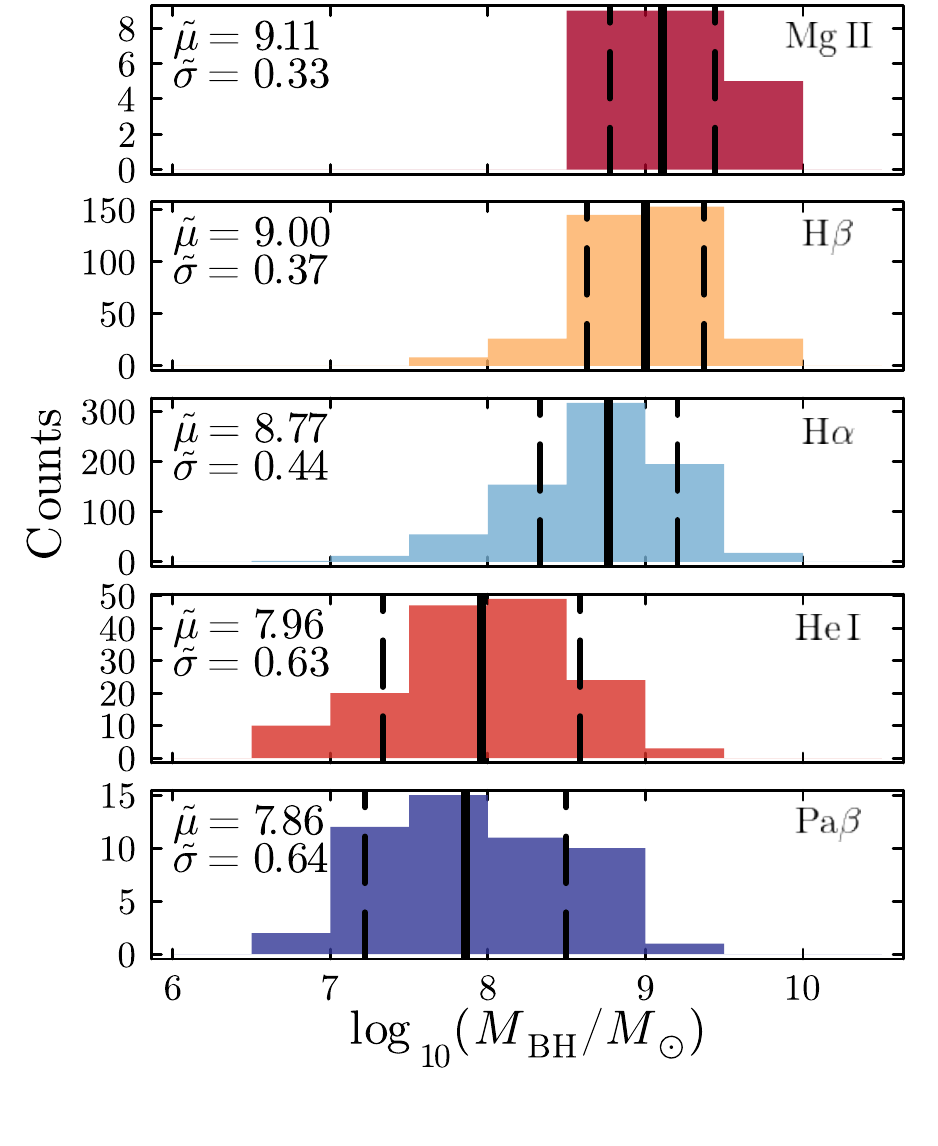}
\caption{Histograms showing the SMBH masses of \Euclid's QSOs, as derived from the broad component of the five emission lines considered in this work.}
\label{fig:BHMass_Hist}
\end{figure}

We use the continuum luminosity and FWHM of the \hb\,and \ion{Mg}{ii} lines to estimate the masses of the SMBHs powering the QSOs for which these lines are available. For the estimation of \ha, Pa\,$\beta$, and \ion{He}{i}-based SMBH masses, we make use of the line luminosity itself instead of the continuum luminosity (Sect.~\ref{sc:BH_quantities}). Table~\ref{tab:QSO_quantities_summary} includes the mean BH mass estimated from our lines, while Fig.~\ref{fig:BHMass_Hist} 
shows the histograms of those black hole masses from each respective line, along with the median values for those masses. Our masses have a mean uncertainty of $\sim 0.3$ dex, as propagated from the luminosities and FWHM used in the estimators, but an additional contribution of 0.3--0.5 dex should be considered to take into account the uncertainties in the single-epoch virial mass calibrations.

\begin{figure}[htbp!]
\centering
\includegraphics[angle=0,width=1.0\hsize]{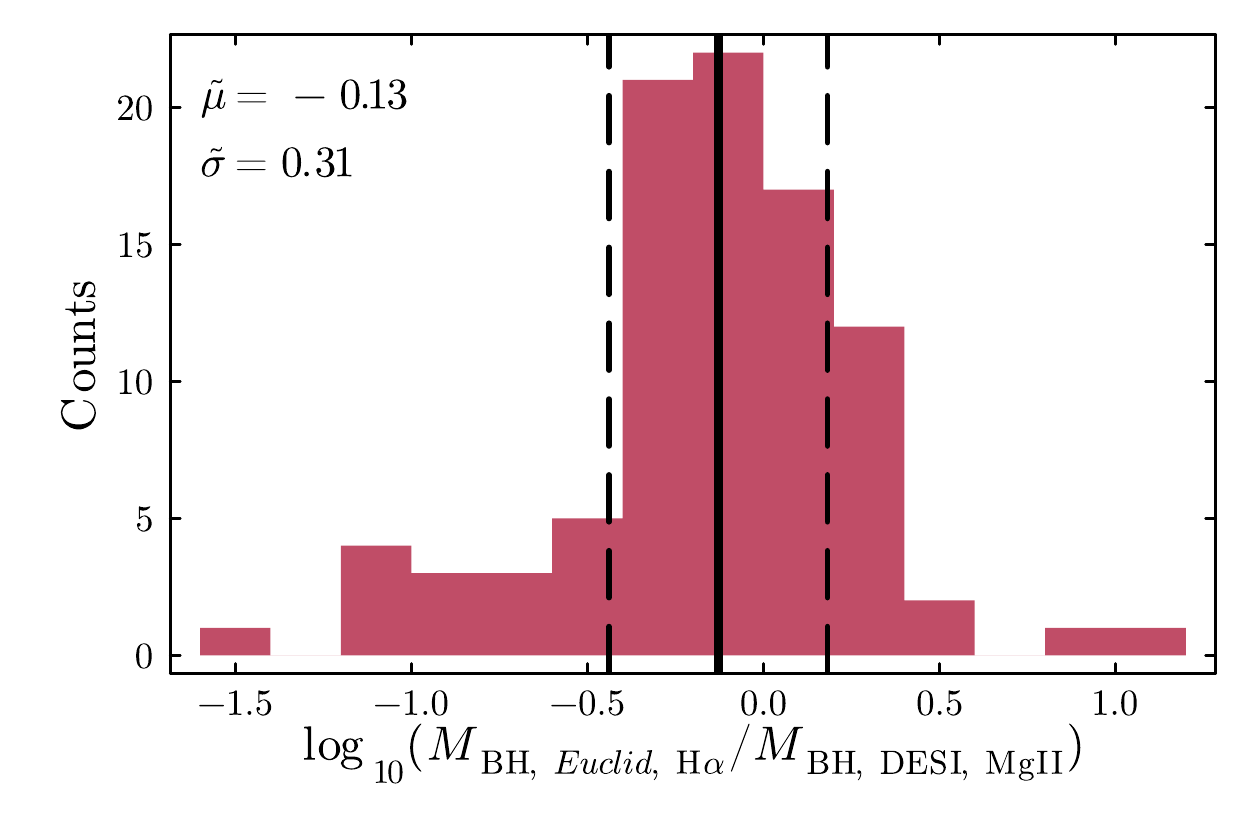}\\
\includegraphics[angle=0,width=1.0\hsize]{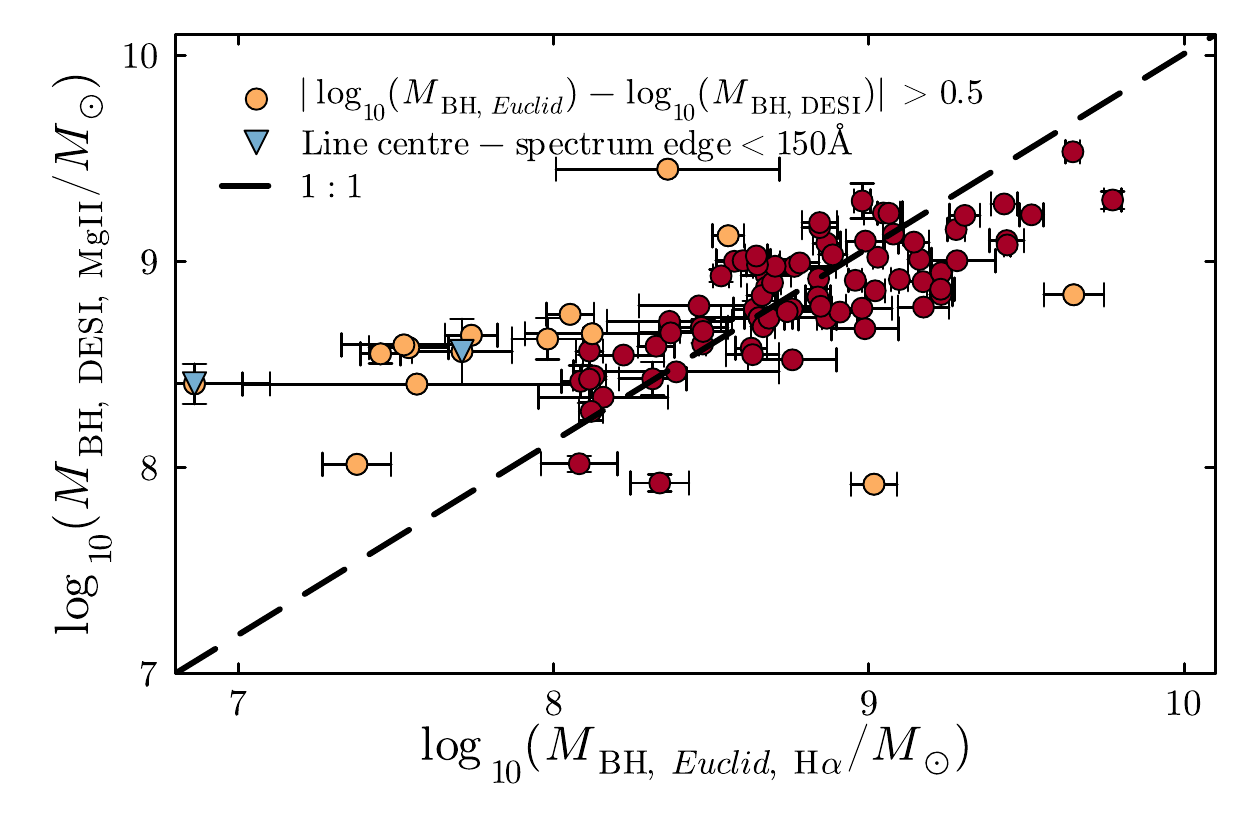}
\caption{\emph{Top panel}: Histogram of the ratio between the BH masses derived with \Euclid's H$\rm \alpha$ emission line and DESI's \ion{Mg}{ii} line. \emph{Bottom panel}: Comparison of the BH masses estimated with the \Euclid \ha\, emission line and the DESI \ion{Mg}{ii} line. Yellow markers are the sources for which the logarithmic BH masses from DESI and \Euclid disagree by more than $0.5$, for sources where both spectra are available. Blue down triangle markers represent the sources where the emission line is too close to the edge of the spectrum, thus leading to a suspicious line fit. Error bars represent the uncertainty of the BH masses without accounting for the  intrinsic uncertainty of the single epoch virial mass estimates.}
\label{fig:EuclidHa_DESIMgII_hist}
\end{figure}

We performed the analysis for the DESI sources in our sample using DESI DR1 spectra, and compared the BH masses obtained from the analysis with our \ha\, results. We specifically targeted sources with \ion{Mg}{ii} lines in the DESI spectra, as this allows a direct comparison of BH masses using the \ha\, line in the \Euclid spectra. The comparison (Fig.~\ref{fig:EuclidHa_DESIMgII_hist}) 
shows a median difference of $-0.13$ dex and a normalised absolute median deviation of $0.3$. This is a good agreement, albeit with somewhat large scatter, in range of what is expected when using single-epoch mass estimate methods \citep[see, e.g., the review of][]{2013BASI...41...61S}. 

In order to try to understand why the agreement is not closer to a 1:1 relation, we conducted an inspection of the fitting for the sources in which the estimated masses disagree the most ($|\logten(M_{\mathrm{BH}, \, \Euclid, \, \mathrm{H}\alpha}) - \logten(M_{\mathrm{BH, \, DESI, \, \ion{Mg}{II}}})|>0.5$, the yellow markers in the bottom panel of Fig.~\ref{fig:EuclidHa_DESIMgII_hist}. We also show the BH mass uncertainty propagated from the luminosities and FWHM used in the estimation, not accounting for the intrinsic uncertainty of single epoch estimators). We found that the disagreement between estimated masses for the two instruments are due to difficulties in fitting the \ha\, line in the \Euclid spectrum. These difficulties are due to low S/N \Euclid spectra, which, nevertheless, pass the quality and reliability flags, and where noise is confused with a legitimate \ha\, emission line, or due to unexpected emission features close to the \ha\, line. The differences can also be due to the emission line being too close to the edge of the spectrum (blue triangle markers). Finally, we find cases where there appears to be no problems with the fitting and the difference in masses is due to genuine varying strengths between the two lines. Despite these setbacks, the majority of our sources agree on the BH masses estimated from the two instruments, and we conclude that our \ha-based BH mass estimates are consistent.
The Pa\,$\beta$ and \ion{He}{i} mass estimations are slightly lower than the remaining lines, as expected due to the lower line luminosities of the low-redshift sources (Table~\ref{tab:QSO_quantities_summary} and Fig.~\ref{fig:Lines_Lum}).

\begin{figure}[htbp!]
\centering
\includegraphics[angle=0,width=1.0\hsize]{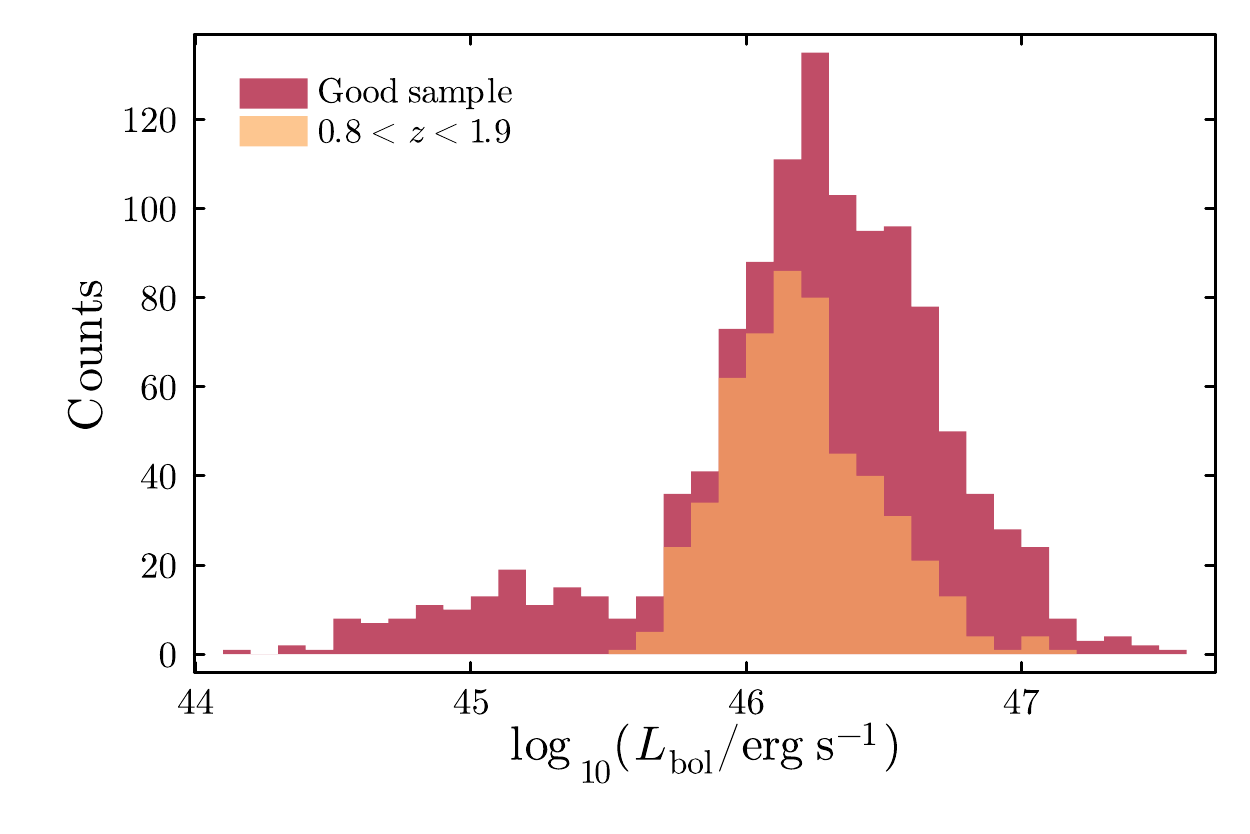}\\
\includegraphics[angle=0,width=1.0\hsize]{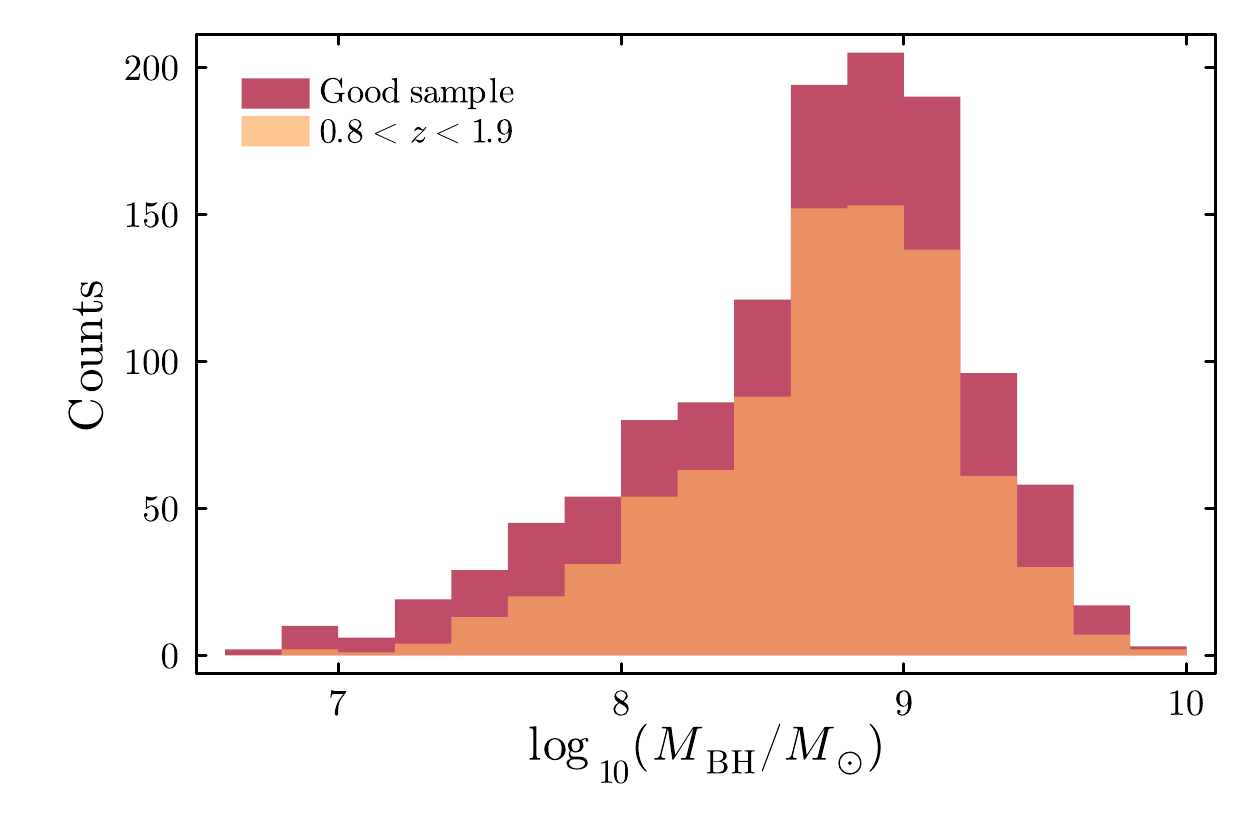}
\caption{\emph{Top panel}: Histogram of the mean bolometric luminosity of the QSOs in our sample. \emph{Bottom panel}: Mean SMBH masses of the QSOs in our sample derived from the available individual lines. The orange histogram in both panels shows the mean BH masses of the QSOs at $0.8<z<1.9$ (\Euclid's cosmological range).}
\label{fig:BOL_Mean_Hist}
\end{figure}

The line luminosities and continuum luminosities estimated from our QSO selection translate (Sect.~\ref{sc:Lbol}) into a mean bolometric luminosity of $\logten(L_{\rm bol}/\mathrm{erg \, s^{-1}}) \sim 46.2 \pm 0.5$ and a mean BH mass of $\logten(M_{\mathrm{BH}}/\si{\solarmass})=8.7 \pm 0.6$. Figure \ref{fig:BOL_Mean_Hist}
shows the histograms of the bolometric luminosities (top panel) and for the BH masses (bottom panel) obtained as a mean of the BH masses estimated from the individual lines. In both figures, we include the histograms for the sample passing the quality and reliability flags, as well as the sources within the $0.8<z<1.9$ range (which specifically targets \ha\, and is of interest for \Euclid's cosmology objectives) also passing those cuts.
Both he bolometric luminosity and BH mass distributions for the good sample show an extended tail towards the lower values. This tail is especially prominent in the bolometric luminosity histogram. The bolometric luminosity tail at the lowest values is composed of the lower-$z$ sources with Pa\,$\beta$ and \ion{He}{i} emission and does not therefore appear in the sample spanning the $0.8<z<1.9$ range. We use an indirect way of estimating the bolometric luminosity (Sect.~\ref{sc:Lbol}) for Pa\,$\beta$ and \ion{He}{i}, when compared to the other lines. We therefore  consider that these bolometric luminosity estimates may be less reliable than the ones from the other lines.
We compared our estimated bolometric luminosities with the SED-derived bolometric luminosities of \citet{Q1-Laloux}, for sources in common between both works. The number of sources is small ($64$) but their values are in good agreement with a mean difference of only $0.04\pm0.19$ dex.

\begin{figure}[htbp!]
\centering
\includegraphics[angle=0,width=1.0\hsize]{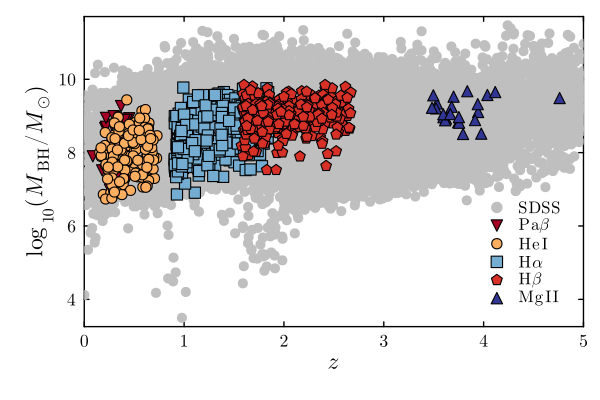}\\
\includegraphics[angle=0,width=1.0\hsize]{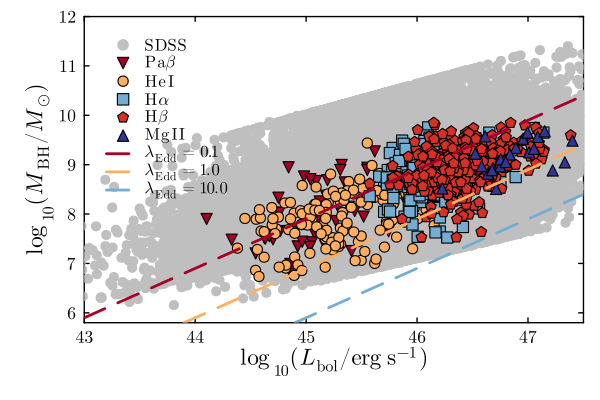}
\caption{\emph{Top panel}: SMBH masses of \Euclid's QSOs derived from the broad component of the Pa\,$\beta$, \ion{He}{i}, \ha, \hb, and \ion{Mg}{ii} emission lines versus redshift. \emph{Bottom panel}: SMBH masses of \Euclid's QSOs derived from the broad component of the same five emission lines as in the top panel versus bolometric luminosity. Dashed lines represent the Eddington ratio limits for $\rm \lambda_\mathrm{Edd}=0.1$, $1$, and $10$. Flat grey markers show the SDSS DR16 QSOs from \cite{Wu_2022}.}
\label{fig:BHM_z}
\end{figure}

As a final check, we also compare our BH masses with the results of \cite{Wu_2022}. The authors applied virial mass methods, similar to those used in our work, and estimated the BH masses of $\sim 750\,000$ AGN from SDSS DR16 using the \hb, \ion{Mg}{ii}, and \ion{C}{iv} emission lines. They obtained a mean BH mass of $\logten (M_{\mathrm{BH}}/\si{\solarmass})=8.2\pm0.5$, $\logten (M_{\mathrm{BH}}/\si{\solarmass})=8.6\pm0.5$, and $\logten (M_{\mathrm{BH}}/\si{\solarmass})=8.7\pm0.5$ using the \hb, \ion{Mg}{ii}, and \ion{C}{iv} lines, respectively. Our own overall mean BH mass stands at $\logten (M_{\mathrm{BH}}/\si{\solarmass})=8.7\pm0.6$, which agrees with the measurements from \cite{Wu_2022}. We also compare the mean bolometric luminosity of \Euclid's QSOs with the mean bolometric luminosity of SDSS DR16 QSOs \citep[][]{Wu_2022}. These comparisons are illustrated in Fig. \ref{fig:BHM_z}. 
which shows the plot of our estimates of the BH mass vs. the redshift (top panel), and the BH masses as a function of the bolometric luminosity (bottom panel) plotted against the background of sources from \cite{Wu_2022}. The QSOs detected by \Euclid have a mean bolometric luminosity of $\logten(L_{\mathrm{bol}}/\mathrm{erg\,s^{-1}})=46.2$, which is approximately $0.4$ dex higher than the mean for the quasars of SDSS DR16  [$\logten(L_\mathrm{{bol}}/\mathrm{erg\,s^{-1}})=45.8$]. This discrepancy is due to the sources selected through the \ha, \hb, and \ion{Mg}{ii} lines which form the bulk of our sample and have their low luminosity sources excluded by our selection criteria (due to low S/N), independently of their redshift.
The average Eddington ratio is $\rm \lambda_\mathrm{Edd} = 0.34 \pm 0.5$. Approximately $3.4\%$ of the good sample show $\rm \lambda_\mathrm{Edd} > 1$. Inspection of the fitting of these sources reveals some cases where spectral features get confused with emission lines, leading to erroneous flagging of such sources as reliable. However, most of our sources share the parameter space occupied by the SDSS sources for our redshift and bolometric luminosity ranges.
The agreements with both SDSS and DESI lead us to believe our BH mass estimates are statistically robust, although some care should be taken with the more extreme cases, such as sources with $\lambda_{\mathrm{Edd}}>1$.

\subsection{\label{sc:composite_compare} QSO continuum of the \Euclid geometrical mean composite}
\noindent
In this section, we compare the averages of the continuum slopes estimated for the individual sources to those measured on our \Euclid geometric composite spectrum (Sect.~\ref{sec:composite_generation}).
\begin{figure}[htbp!]
\centering
\includegraphics[angle=0,width=1.0\hsize]{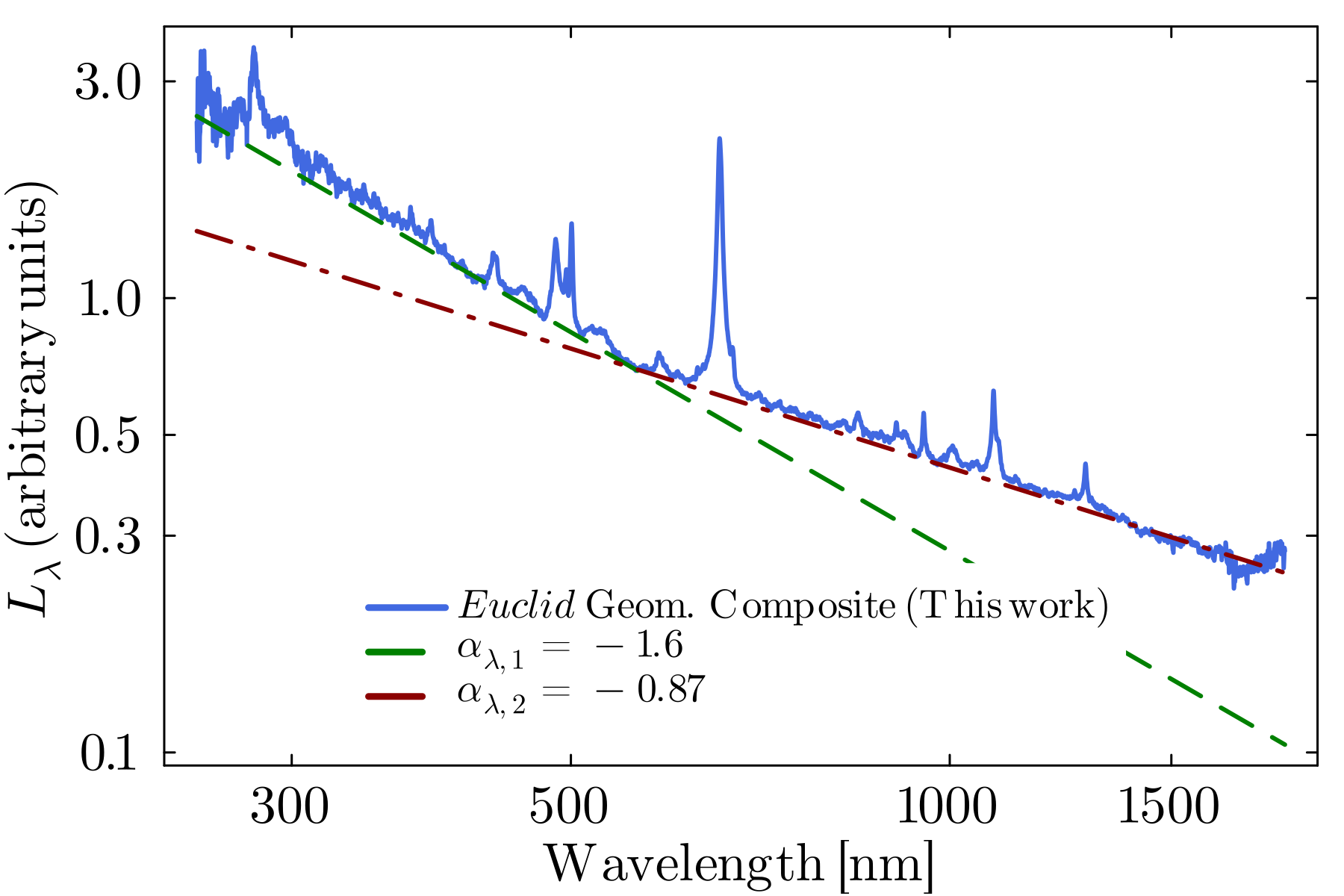}
\caption{Geometric mean composite obtained from \Euclid spectra (in blue, see also Sect.~\ref{sec:composite_generation}) is best-fit by two power laws. 
The wavelength of the spectral break is at a rest-frame wavelength of $\lambda \sim 561.2 \, \text{nm}$.
}
\label{fig:composite_slopes}
\end{figure}
Figure~\ref{fig:composite_slopes} shows our geometric mean composite (in blue) and demonstrates its broad-band continuum to be clearly more complex than a single power law over a sufficiently wide wavelength range. Using a smoothly-broken power-law model results in two different slopes, with the break wavelength at $\lambda \sim 561.2 \, \text{nm}$. The blue side of the composite ($\lambda<561.2 \, \text{nm}$) has $\alpha_{\lambda,1} = -1.6\pm0.05$ and the red side ($\lambda>561.2 \, \text{nm}$) has $\alpha_{\lambda,2}=-0.87\pm0.01$. The slopes obtained from fitting the composite are in agreement with the median values for the QSO continuum slope of our individual sources (Sect.~\ref{sc:QSOalpha_Disc} and Fig.~\ref{fig:QSOcont_Redshift}): briefly, sources at $z>1.9$, corresponding roughly to the composite wavelengths of $\lambda<640 \, \text{nm}$, have a median $\alpha_{\lambda}= -1.58\pm1.04$, in agreement with our composite slope of $\alpha_{\lambda,1} = -1.6\pm0.05$. Sources at $z<0.8$ (thus covering the range redwards of the break wavelength, at $625 \, \text{nm}<\lambda<1830\,\text{nm}$) have a median slope of $-1.0\pm0.4$, also in agreement with the composite's red slope $\alpha_{\lambda,2}=-0.87\pm0.01$. These results are in contrast with those of \cite{2014ApJ...794...75S} and \cite{2015MNRAS.449.4204L}, who could fit their composites with single power laws. However, their composites did not extend much beyond the rest-frame UV and so missed the wavelength break found in our composite. Such a wavelength break, however, is captured by the UV/optical composite of \cite{2001AJ....122..549V} that extends to $\lambda \sim 800\,\text{nm}$. Their composite is best-fit by $\alpha_{\lambda,1} \sim -1.5$ and $\alpha_{\lambda,2} \sim -0.4$, with the break at a somewhat lower wavelength of $\lambda \sim 430 \, \text{nm}$.

\begin{figure}[htbp!]
\centering
\includegraphics[angle=0,width=1.0\hsize]{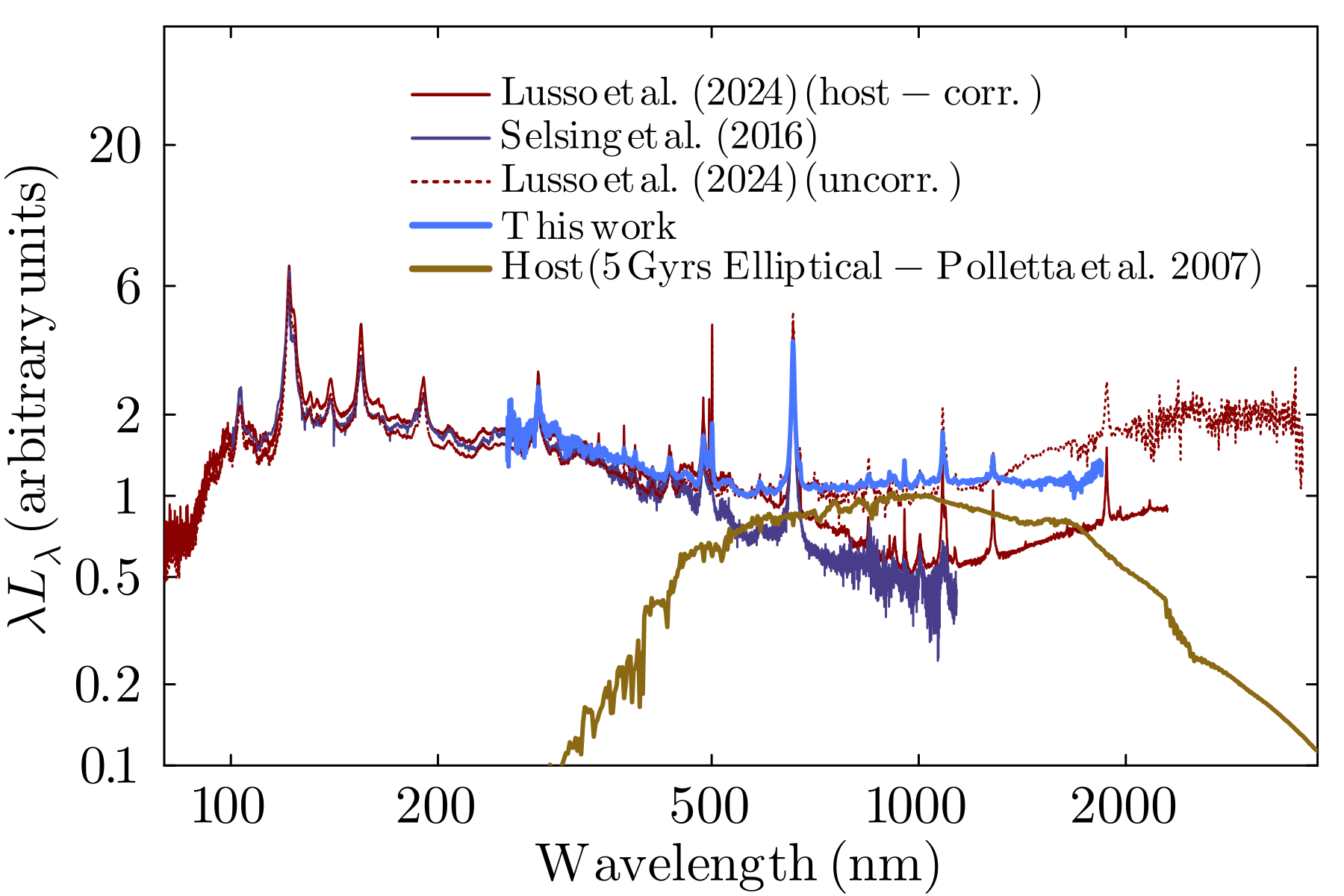}
\caption{Geometric composite obtained from \Euclid spectra (in blue, see also Sect.~\ref{sec:composite_generation}) compared to the host galaxy corrected (dark red continuous line) and host galaxy uncorrected (dark red dotted line) composites of \cite{EP-Lusso}. Also shown is the geometrical host-corrected composite of \cite{Selsing2016} in dark blue and the elliptical host galaxy template \citep[][]{2007ApJ...663...81P} used to fit our sources in this work, scaled to match the \Euclid composite (dark gold).}
\label{fig:composite_comp}
\end{figure}

We also compared our composite to the composites of \cite{EP-Lusso} and \cite{Selsing2016}. The first was built from SDSS in the optical and public NIR spectroscopy (not from \Euclid) and thus covered the rest-frame wavelength range of $70\,\text{nm}<\lambda<3600\,\text{nm}$, while the second uses X-Shooter VLT data covering the rest-frame wavelengths of $100\,\text{nm}<\lambda<1135\,\text{nm}$ (Fig. \ref{fig:composite_comp}).
\cite{EP-Lusso} make available composites both corrected and uncorrected for host galaxy contribution. At wavelengths $\lambda \ga 500 \, \text{nm}$, their host-corrected composite shows a declining luminosity up to $\lambda=1000$~nm and then a rise in flux due to the presence of the hot dust emission from the obscuring torus \citep{Landt2011}. However, the \Euclid composite using the actual data becomes relatively flat at $\lambda \ga 500 \, \text{nm}$ and remains so throughout the sampled wavelength range. A similar flatness is also present in the uncorrected composite of \cite{EP-Lusso}, suggesting that contribution from the host galaxy is the explanation for the flatness.
A factor to consider is that the effect of the host galaxy as the driver of the flatness in the composite slope at $\lambda>500\,\mathrm{nm}$ changes depending on the available aperture. In the works of \cite{EP-Lusso} and \cite{2001AJ....122..549V}, SDSS uses a 2\arcsecond\, to 3\arcsecond\, aperture fibres. This corresponds to 16--25 kpc at the average redshift of our sample ($z\sim1.2$). The half-light radius of a massive elliptical galaxy can reach up to $\sim10$ kpc, with late type galaxies being generally smaller \citep[][]{2014ApJ...788...28V}, with spiral galaxies being generally smaller. The angular size of astronomical sources has a minimum at $z\sim1.5$ \citep{Buchalter_1998} hence, SDSS measurements may miss a sizeable fraction of the host galaxy contribution for lower redshift QSOs, making the central AGN component appear more dominant. The work by \cite{EP-Lusso} constructs their composite by using SDSS data of sources from a redshift range of $z=$~0.08--4.9 (see their table 3 and section 2), therefore their composite may be affected by the underestimation of the host contribution. By contrast, \Euclid uses grism based spectroscopy and so captures the full contribution of the host galaxy, thus making the flatness of the \Euclid composite more obvious when compared with composites using SDSS data.

The host being responsible for the flattening at $\lambda > 500 \, \text{nm}$ is further supported by the work of \cite{Landt2011} and \cite{10.1093/mnras/stt421}, who find similar flattening between $550$ and $1800\, \text{nm}$ for low-luminosity AGN. These authors also attribute the feature as due to the possible large contribution of host galaxy light relative to the AGN emission in the NIR spectra of these sources. 
This highlights an inconsistency between the results for the composite and the results for the individual sources: Fig.~\ref{fig:composite_comp} shows that the host galaxy emission is sufficient to explain the difference between our composite and the corrected composite of \cite{EP-Lusso}, with only a more `residual' difference at $\lambda \ga 1500 \, \text{nm}$. However, as stated in Sect.~\ref{sc:Fitting}, most of the sample yields better fitting results when the host galaxy template is not included in the fit. This is probably due to the lower S/N of the individual spectra not allowing us to properly estimate both the QSO continuum slope and the host galaxy contribution. This also means that there is the possibility that the slopes at $z<0.8$ are biased towards slightly higher values (see the second panel of Fig.~\ref{fig:QSOcont_Redshift}).

Finally, at rest-frame wavelengths of $\lambda \ga 1500 \, \text{nm}$, where the flux of the host galaxy steeply declines and the blackbody hot dust emission starts to dominate, the composites of \cite{EP-Lusso} show an increase in flux. We do not see this increase in our \Euclid composite, which could be a consequence of the understimation of the host galaxy luminosities in \cite{EP-Lusso}. The composite from \cite{Selsing2016} shows similar slopes in the optical wavelengths as our own (up to $500 \, \text{nm}$) but, as with our own composite, the data is not sufficient to verify whether the inversion of slope present in \cite{EP-Lusso} is also present in the \cite{Selsing2016} composite.
Further study on these issues is beyond the scope of the present work, but is something to tackle in a future study, especially with the forthcoming Euclid DR1, which would provide a higher number of sources for the composites (allowing us to reach fainter fluxes through the stacking process), as well as make available deeper individual spectra and the blue grism data.

\section{\label{sc:Conclusion}Conclusions}

We presented the results from the spectroscopic analysis of QSOs detected in the Q1. We selected sources and assigned redshifts using external catalogues such as QUBRICS, DESI, and visually confirmed QSOs from \cite{Q1-SP068}.
In total, we used a sample of $5387$ QSOs and provided estimations of line fluxes, FWHM, velocity offsets, QSO continuum luminosity, slope, and break wavelengths for each source. We also included fitting statistics such as S/N, reduced $\chi^2$, and number of data points used in the fit. These metrics allowed us to establish a simple criterion to select more reliable spectra and measurements based on fit quality. Our main findings are summarised below:
\begin{itemize}
    \item A total of $2410$ sources pass the quality cut. The remaining sample ($55\%$) either has too low S/N, or is affected by artefacts which compromise the reliability of the measurements (Sect.~\ref{sc:Final_sample}).
    \item Our QSOs have a median power-law continuum slope of \mbox{$\alpha=-1.17\pm0.83$}, consistent with values previously reported in the literature and within the expected range for QSOs. However, a proper comparison is difficult, due to the lack of studies targetting the same wavelength range as our work (Sect.~\ref{sc:QSOalpha_Disc}).
    \item We provide measurements (luminosity, central wavelength, FWHM, velocity offset, and respective uncertainties) for the following broad lines: Pa\,$\beta$, \ion{He}{i}, \ha, \hb, and \ion{Mg}{ii}. We also provide estimates for single-epoch BH masses and Eddington ratios based on these lines (Sect.~\ref{sc:overall_line_emission}).
    \item The sources display bolometric luminosities with median values of $\logten(L_{\mathrm{bol}}/{\rm erg\,s^{-1}})=46.21\pm0.52$, as well as a mean Eddington ratio of $\lambda_\mathrm{Edd}=0.34 \pm 0.5$ (Sect.~\ref{sc:SMBH_masses_Lbol_discussion}).
    \item The estimated SMBH masses for our QSOs average around $\logten(M_{\rm BH}/\si{\solarmass})=8.66 \pm 0.57$. We compare SMBH masses derived from \Euclid's \ha\, line with those derived from DESI spectra using the \ion{Mg}{ii} line. The mass estimates are in general agreement, with a typical median difference of $-0.13$ dex (Sect.~\ref{sc:SMBH_masses_Lbol_discussion}).
    \item The geometric composite spectrum obtained from our sources shows a clear flattening in the NIR region (at rest-frame wavelengths $\lambda \sim 550$--$1800\, \text{nm}$), in contradiction with composites from the literature. This is probably due to uncorrected contribution from host galaxy light or hot dust emission (Sect.~\ref{sc:composite_compare}).
\end{itemize}

Our results show that, even though AGN demography is not one of \Euclid's main science goals, \Euclid spectra can be used to achieve outcomes comparable to those of existing studies conducted with other instruments. By using data from external catalogues, we adopt a multi-wavelength approach that enables a more comprehensive characterization of our sources. However, it is essential to conduct careful selection of the spectra due to the low S/N and contamination affecting a significant portion of the Q1 data. Our selection criteria offer a practical way to identify problematic spectra and help resolve these issues going forward.

We make available the resulting catalogue for the $5387$ \Euclid QSOs, as well as the {\tt Julia} script used to run the analysis\footnote{\url{https://github.com/astrocalhau/LinesInTheSky}}.
Appendix \ref{sc:Appendix_B} provides a description of the columns found in the results catalogue.

\begin{acknowledgements}
\AckQone
\AckEC

JC, GC, VA and FR acknowledge the support from the INAF Large Grant ``AGN \& \Euclid: a close entanglement'' Ob. Fu. 01.05.23.01.14. A.F. acknowledges the support from project “VLT-MOONS” CRAM 1.05.03.07, INAF Large Grant 2022 “Dual and binary SMBH in the multi-messenger era” Ob. Fu. 1.05.12.01.13, INAF Mini Grant 2024 ``The pc-scale view of \ion{H}{ii} regions in M33'' Ob. Fu. 1.05.24.07.01.
JC thanks Mar Mezcua and Christoph Saulder for the DESI QSO catalogue used in this study.
MS acknowledges the PhD program in Space Science and Technology at the University of Trento, Cycle XXXIX, with the support of a scholarship financed by the Ministerial Decree no. 118 of 2nd march 2023, based on the NRRP -- funded by the European Union -- NextGenerationEU -- Mission 4 ``Education and Research'', Component 1 ``Enhancement of the offer of educational services: from nurseries to universities'' -- Investment 4.1 “Extension of the number of research doctorates and innovative doctorates for public administration and cultural heritage”. 
M.M. acknowledges support from the Spanish Ministry of Science and Innovation through the project PID2021-124243NBC22. This work was partially supported by the program Unidad de Excelencia Mar\'ia de Maeztu CEX2020-001058-M.

This research used data obtained with the Dark Energy Spectroscopic Instrument (DESI). DESI construction and operations is managed by the Lawrence Berkeley National Laboratory. This material is based upon work supported by the U.S. Department of Energy, Office of Science, Office of High-Energy Physics, under Contract No. DE–AC02–05CH11231, and by the National Energy Research Scientific Computing Center, a DOE Office of Science User Facility under the same contract. Additional support for DESI was provided by the U.S. National Science Foundation (NSF), Division of Astronomical Sciences under Contract No. AST-0950945 to the NSF’s National Optical-Infrared Astronomy Research Laboratory; the Science and Technology Facilities Council of the United Kingdom; the Gordon and Betty Moore Foundation; the Heising-Simons Foundation; the French Alternative Energies and Atomic Energy Commission (CEA); the National Council of Humanities, Science and Technology of Mexico (CONAHCYT); the Ministry of Science and Innovation of Spain (MICINN), and by the DESI Member Institutions: www.desi.lbl.gov/collaborating-institutions. The DESI collaboration is honored to be permitted to conduct scientific research on I’oligam Du’ag (Kitt Peak), a mountain with particular significance to the Tohono O’odham Nation. Any opinions, findings, and conclusions or recommendations expressed in this material are those of the author(s) and do not necessarily reflect the views of the U.S. National Science Foundation, the U.S. Department of Energy, or any of the listed funding agencies. This research uses services or data provided by the SPectra Analysis and Retrievable Catalog Lab (SPARCL) and the Astro Data Lab, which are both part of the Community Science and Data Center (CSDC) program at NSF National Optical-Infrared Astronomy Research Laboratory. NOIRLab is operated by the Association of Universities for Research in Astronomy (AURA), Inc. under a cooperative agreement with the National Science Foundation.

\end{acknowledgements}

\bibliography{Lines, Q1, Euclid}

\clearpage

\appendix
\section{\label{sc:Appendix_A}Example spectra present in the Q1}

\begin{figure*}[htbp!]
\centering
\includegraphics[angle=0,width=1.0\hsize]{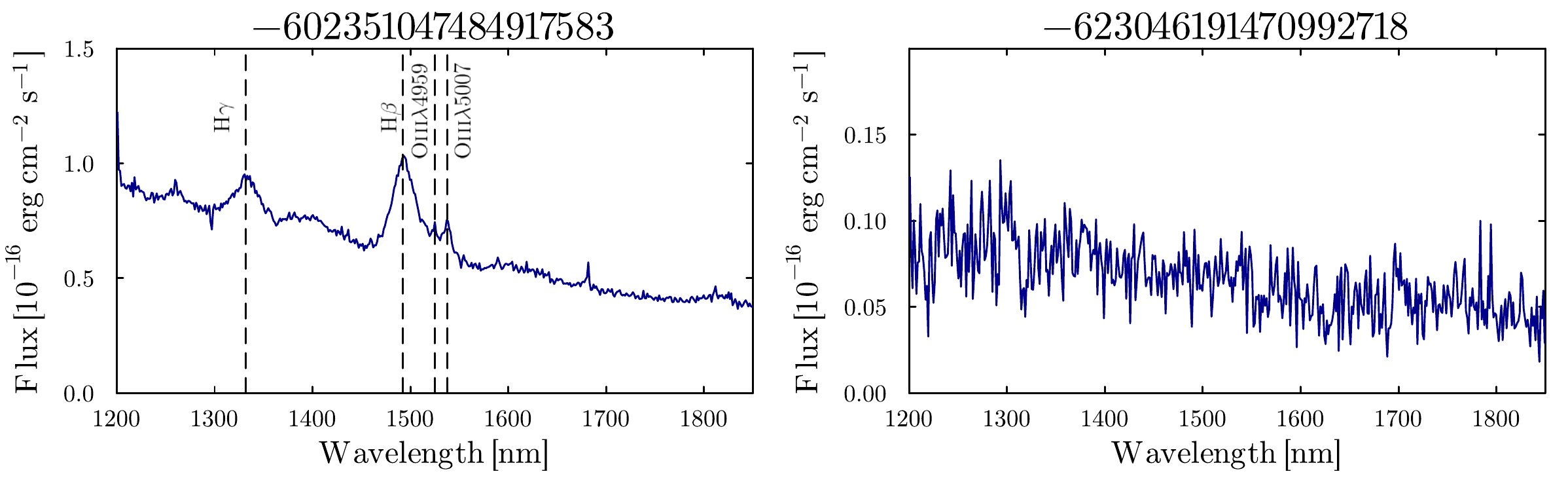}
\caption{Example of good quasar spectra (without artefact or contamination) obtained from \Euclid. Left panel shows a spectrum with overall S/N$=47$, while the right panel shows a spectrum with a much lower S/N$\sim4$.}
\label{fig:Example_spectra}
\end{figure*}

Roughly 55\% of the spectra analysed in this work exhibits problems in the measured flux, either through contamination, artefacts or low S/N. We refer the interested reader to \cite{Q1-TP006} and \cite{Q1-TP007} for more details.
In Fig.~\ref{fig:Example_spectra} 
we show examples of typical \Euclid spectra for QSO, for both high and low S/N cases. The left plot shows a good spectrum with high enough S/N to detect emission lines clearly visible in the figure. The right plot shows a good spectrum with low S/N. All spectra had a wavelength range selection applied during the analysis, in order to filter bad data at the edges. We include the IDs of the sources for each example. A negative ID code corresponds to a source with negative declination. 

\begin{figure*}[htbp!]
\centering
\includegraphics[angle=0,width=1.0\hsize]{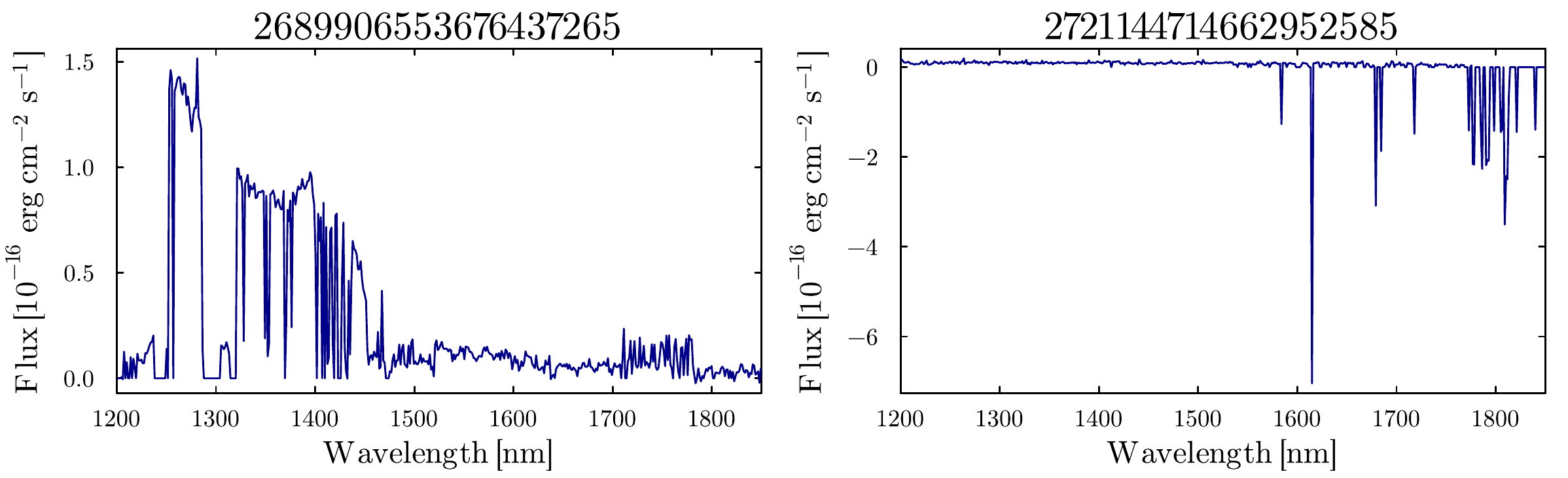}
\caption{Example of the types of bad spectra obtained from \Euclid. The spectrum on the left panel is rejected by the quality cut due to a very high reduced $\chi^2 = 134$ and low number of valid pixels, with $N_{\rm POINTS}=378$. The spectrum on the right panel is rejected due to low number of valid pixels, with $N_{\rm POINTS}=332$.}
\label{fig:Bad_spectra}
\end{figure*}

In Fig.~\ref{fig:Bad_spectra} 
we show two examples of bad spectra present in the our sample. The plot on the left shows a spectrum with irregular, box-like features, with no clear emission line identifiable. On the right, a spectrum with erroneous negative flux is shown, possibly due to over-subtraction. Both sources have low numbers of valid points for fitting, with only 378 and 332 valid pixels for the left and right panel's spectrum respectively, thus failing the quality cut requirement of $N_{\rm POINTS}>450$. The left panel additionally fails the quality cut requirement of $\chi_{\mathrm{red}}^2 >6$, having a $\chi_{\mathrm{red}}^2=134$. The majority of sources that fall outside the quality cut defined in this work, fail to pass the quality cut due to high $\chi_{\mathrm{red}}^2$ values or low number of available points for fitting.

\begin{figure*}[htbp!]
\centering
\includegraphics[angle=0,width=.95\hsize]{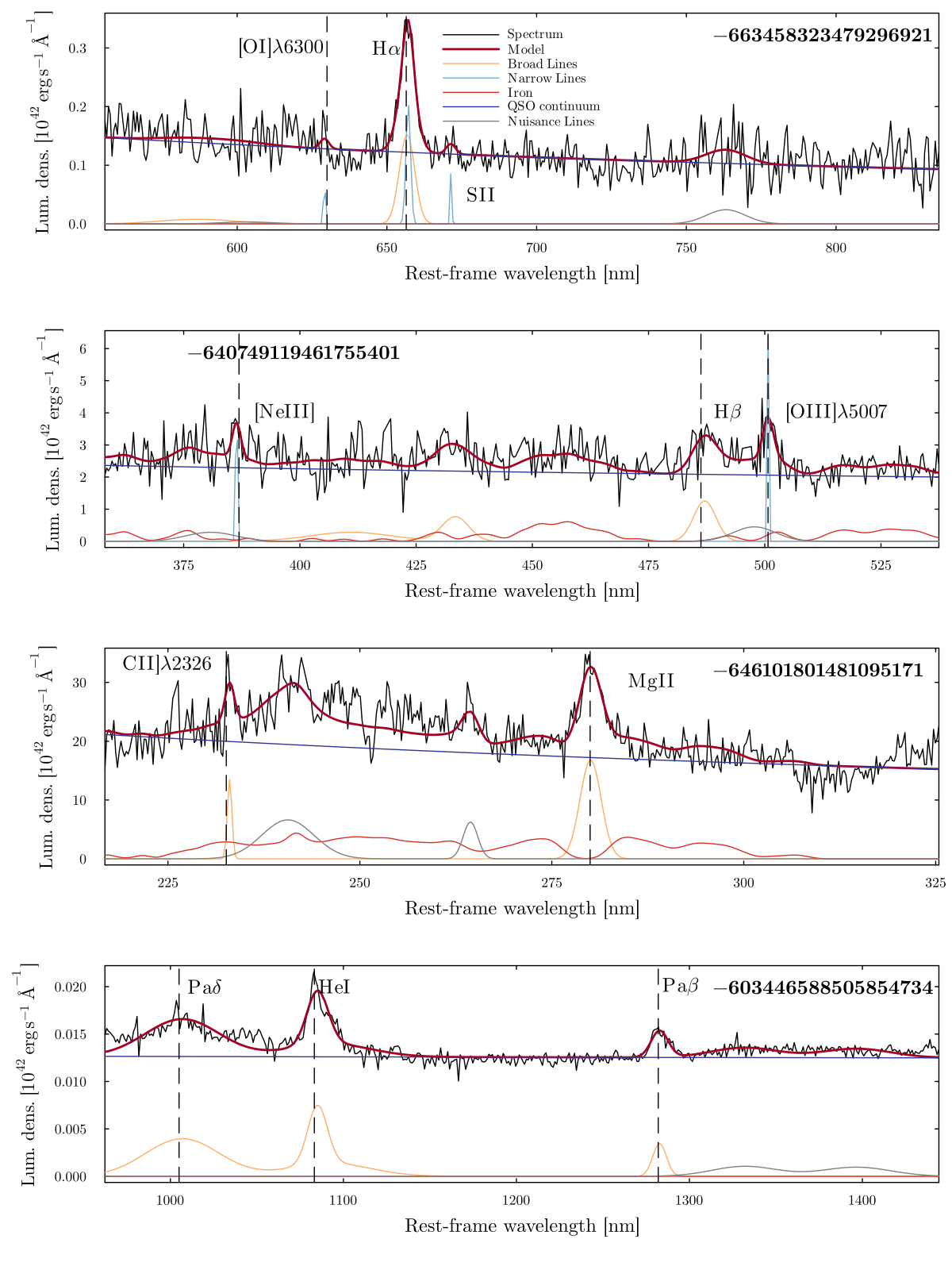}
\caption{Examples of fits using \texttt{QSFit} for sources with H$\rm \alpha$ (first panel), H$\rm \beta$ (second panel), \ion{Mg}{ii} (third panel), and Pa\,$\beta$ and \ion{He}{i} lines (fourth panel), which are used in this analysis for BH mass estimation. The dark red bold line shows the model obtained by \texttt{QSFit} when convolving the spectrum (thin grey line) with a Gaussian template based on the resolution of the spectra $(R\sim450)$. The remaining components are shown before the convolution is applied which is why some lines are thinner than the adopted resolution permits.}
\label{fig:QSFIT_Ha_Hb_MgII}
\end{figure*}

We also show here most of the main components used by \texttt{QSFit} in fitting the spectra of \Euclid QSOs. Figure~\ref{fig:QSFIT_Ha_Hb_MgII} shows the different spectral components fitted to sources with \ha, \hb, and \ion{Mg}{ii} emission.

\section{\label{sc:Appendix_B}Description of quantities present in the results table}

The results of the analysis undertaken in this work are available in table format as an accompanying data product to this paper. In addition to \Euclid IDs, coordinates, and redshifts of the QSOs, we provide information on the fit through statistics like reduced $\chi^2$, number of points used in the fit, and S/N of the source spectrum, as defined in Sect.~\ref{sc:Statistics_Parameters}. We further provide estimations on the QSO continuum luminosity, spectral index, Balmer ratio, and emission line measurements, including luminosity, central wavelength, FWHM, and velocity offsets. Uncertainties are provided for these quantities when applicable. 

The full table contains more than 100 columns and $>5000$ sources. Table \ref{tab:Table_description} describes the columns found in the accompanying catalogue to this paper.

\begin{table*}[htbp!]
\caption{List of columns and their respective description present in the this paper's catalogue. For columns related to emission line quantities, the inclusion of `\_na' and `\_br' in the name refers to the narrow and broad component of the line, respectively. ``[quantity]'' refers to emission lines, Balmer contribution, host galaxy template, iron templates and QSO continuum luminosities.}
\centering
\begin{tabular}{ll}
\hline
\hline
    Column name & Description\\
\hline
\hline
    ID & ID number from Q1\\
    Redshift & Redshift (from the external source)\\
    Hmag & \HE magnitude\\
    Ref\_QUBRICS & article in which a source present in QUBRICS is published\\
    QUBRICS & Boolean column identifying sources present in QUBRICS\\
    DESI & Boolean column identifying sources present in the DESI DR1\\
    FU & Boolean column identifying sources present in F26's visually inspected catalogue\\
    redchisq & reduced $\rm \chi^2$ statistic from \texttt{QSFit}\\
    $N_{\rm POINTS}$ & number of data points used in the fitting of that source's spectrum\\
    SNR & signal-to-noise ratio as derived from the spectra. Not used in this study (Sect. \ref{sc:Statistics_Parameters})\\
    DER\_SNR & signal-to-noise ratio used in this work, as defined by \cite{2008ASPC..394..505S}, Sect. \ref{sc:Statistics_Parameters}\\
    L$3000$ & continuum luminosity at $\lambda=300 \, \text{nm}$ in $10^{42} \, \mathrm{erg\, s^{-1}}$\\
    L$5100$ & continuum luminosity at $\lambda=510 \, \text{nm}$ in $10^{42} \, \mathrm{erg\, s^{-1}}$\\
    $\rm QSOcont\_x0$ & break wavelength estimated for the QSO power-law continuum in $\rm \text{\r{A}}$\\
    $\rm QSOcont\_alpha$ & spectral index estimated for the QSO continuum\\
    $\rm [{quantity}]\_reliable$ & flag for quantities with measurements considered reliable (independent of quality cut flag)\\ 
    $\rm [{quantity}]\_norm$ & luminosity of $\rm [{quantity}]$ estimated by \texttt{QSFit} in $\rm 10^{42}\,erg\,s^{-1}$\\
    $\rm [{quantity}]\_unc$ & uncertainty in the preceding measured quantity in the same units as the parent quantity\\
    $\rm [{line}]\_center$ & central wavelength of emission line in $\rm \text{\r{A}}$\\
    $\rm [{line}]\_fwhm$ & full width at half-maximum of emission line in $\rm km\,s^{-1}$ \\
    $\rm [{line}]\_voff$ & velocity offset of emission line in $\rm km\,s^{-1}$\\
    MBH\_[{line}]\_[{ref}] & BH mass estimated based on a given emission line following the relation from a given reference work (in \si{\solarmass})\\
    MBH\_mean & mean of the BH mass estimated from the individual lines (in \si{\solarmass})\\
    $\rm Lbol\_[{source}]$ & bolometric luminosity as estimated from a given monochromatic luminosity (e.g. L$3000$)\\
    $\rm Lbol\_mean$ & mean bolometric luminosity. Equal to $\rm Lbol\_[source]$ if only one monochromatic luminosity is available\\
    Ledd\_mean & mean eddington luminosity in $\rm erg\,s^{-1}$\\
    Edd\_ratio & Eddington ratio as estimated based on the mean Eddington and bolometric luminosities\\
    good & flag for sources passing the quality cut (independent of reliability flag)\\
\hline
\end{tabular}
\label{tab:Table_description}
\end{table*}

\end{document}